\documentclass{myappolb}
\usepackage{graphicx}

\begin{document}
\headauthor{Frieder Kleefeld}
\headtitle{The metric for (anti-)pseudo-Hermitian 2-dim.\ Hamilton operators}
\title{Identification of the metric for diagonalizable (anti-)pseudo-Hermitian Hamilton operators represented by two-dimensional matrices%
\thanks{This comprehensive article is a recollection of personal notes containing original research work, which should hopefully provide to  the open minded reader a clearer view on some important technical subtleties underlying non-Hermitian Quantum Physics.}%
}
\author{Frieder Kleefeld
\address{Collaborator of the\\ Center of Physics and Engineering of Advanced Materials (CeFEMA), \\
             Instituto Superior T\'{e}cnico (IST), \\
             Physics Building, 3rd floor,  
             Av. Rovisco Pais, 
             P-1049-001 LISBOA,
             Portugal\\
	 Private permanent address:\\
 Frankenstr. 3, 91452 Wilhermsdorf, Germany \\
              e-Mail: frieder.kleefeld@web.de}
}
\maketitle
\begin{abstract}
A general strategy is provided to identify the most general metric for diagonalizable pseudo-Hermitian and anti-pseudo-Hermitian Hamilton operators represented by two-dimensional matrices. It is investigated how a permutation of the eigen-values of the Hamilton operator in the process of its diagonalization influences the metric and how this permutation equivalence affects energy eigen-values.  We try to understand on one hand, how  the metric depends on the normalization of the chosen left and right eigen-basis of the matrix representing the diagonalizable pseudo-Hermitian or anti-pseudo-Hermitian Hamilton operator, on the other hand, whether there has to exist a positive semi-definite metric required to set up a meaningful Quantum Theory even for non-Hermitian Hamilton operators of this type. Using our general strategy we determine the metric with respect to the two elements of the two-dimensional permutation group for various topical examples of  matrices representing two-dimensional Hamilton operators found in the literature assuming on one hand pseudo-Hermiticity, on the other hand anti-pseudo-Hermiticity. The (unnecessary) constraint inferred by C. M. Bender and collegues that the ${\cal C}$-operator of ${\cal PT}$-symmetric Quantum Theory should be an involution (${\cal C}^2=1$) is shown ---  in the unbroken phase of ${\cal PT}$-symmetry --- to require the Hamilton operator to be symmetric.  This inconvenient restriction had been already --- with hesitation --- noted by M.~Znojil and H.~B.~Geyer in 2006 (arxiv:quant-ph/0607104) \cite{Znojil:2006ugs}.
A Hamilton operator proposed by Lee and Wick is used to outline implications of the formalism to higher dimensional Hamilton operators. 
\end{abstract}
\newpage

{\em\small ``Physics needs new ideas. But to have a new idea is a very difficult task: it does not mean to write a few lines in a paper. If you want to be the father of a new idea, you should fully devote your intellectual energy to understand all details and to work out the best way in order to put the new idea under experimental test.

This can take years of work. You should not give up. If you believe that your new idea is a good one, you should work hard and never be afraid to reach the point where a new-comer can, with little effort, find the result you have been working, for so many years, to get.

The new-comer can never take away from you the privilege of having been the first to open a new field with your intelligence, imagination and hard work. Do not be afraid to encourage others to pursue your dream. If it becomes real, the community will never forget that you have been the first to open the field." }\\[0mm]
\mbox{} \hfill {\small (Isidor Isaac Rabi, 1972 \cite{Rabi:1972})} 

\section{Introduction} \label{xxsec1}
In the early days of the development of Quantum Theory it was a  brilliant observation \cite{Heisenberg:1925zz}\cite{Born:1926uzf}\cite{Heisenberg:1927zz}  of Werner Heisenberg during his visit of Helgoland in 1925 which induced major progress in the understanding of Quantum Theory: observables can be represented by matrices which will not commute if they cannot be diagonalized simultaneously.  Since the energy content of a physical system is one of the most profound observables it had been the matrix representing the Hamilton operator which attracted soon special attention, as is was known from classical physics that the Hamilton function~\cite{Hamilton_1834} is responsable for generating the time-evolution of a physical system. In the present article we will consider Hamilton operators which are --- without loss of generality --- represented by two-dimensionial diagonalizable matrices. The spectral expansion of such a matrix can be denoted in the following way:
\begin{equation} H \; = \; \left(\begin{array}{cc} H_{11} & H_{12} \\ H_{21} & H_{22} \end{array}\right) \; = \; X^{-1} \left(\begin{array}{cc} E_1 & 0 \\ 0 & E_2 \end{array}\right)  X \; ,
\end{equation}
which can be rewritten as a right or left eigen-value problem:
\begin{equation} H \, X^{-1} \; = \; X^{-1}  \left(\begin{array}{cc} E_1 & 0 \\ 0 & E_2 \end{array}\right) \; , \quad X\, H  \; = \;  \left(\begin{array}{cc} E_1 & 0 \\ 0 & E_2 \end{array}\right) X  \; .
\end{equation}
The column vectors of the matrix $X^{-1}$ being obviously right eigen-vectors of $H$ form a complete two-dimensional basis which is due to $X \, X^{-1} = 1_2$ (with $1_2$ being the two-dimensional unit matrix) orthogonal to the also complete two-dimensional basis formed by the left eigen-vectors of $H$ being the column vectors of the matrix $X^T$ ($T$ denotes here transposition!). It is well known that the normalization of the eigen-vectors of a matrix is not fixed. Replacing $X$ by $N X$ with $N$ being a diagonal invertible matrix, i.~e.,
\begin{equation}  N \; = \; \left(\begin{array}{cc} N_1 & 0 \\ 0 & N_2 \end{array}\right) \; \end{equation}
\noindent we obtain an also feasible spectral expansion of $H$ for now renormalized right eigen-vectors described by the matrix $(N X)^{-1}$ and the left eigen-vectors described by the matrix $(N X)^T$:
\begin{equation} H \; = \; \left(\begin{array}{cc} H_{11} & H_{12} \\ H_{21} & H_{22} \end{array}\right) \; = \; (NX)^{-1} \left(\begin{array}{cc} E_1 & 0 \\ 0 & E_2 \end{array}\right) N X \; . \label{xxyyzz1} \end{equation}
The spectral expansion of the respective Hermitian conjugate matrix $H^+ = H^{\ast \, T}$ is obviously:
\begin{equation} H^+ \; = \; \left(\begin{array}{cc} H^\ast_{11} & H^\ast_{21} \\ H^\ast_{12} & H^\ast_{22} \end{array}\right) \; = \; (NX)^+ \left(\begin{array}{cc} E^\ast_1 & 0 \\ 0 & E^\ast_2 \end{array}\right) \left((N X)^+\right)^{-1} \; .\label{xhdagger1} \end{equation}
Inspection makes clear that the Hermitian conjugation does not only lead to a complex conjugation of eigen-values, yet also to a new two-dimensional basis of right eigen-vectors  described by the matrix $ \left((N X)^+\right)^{-1}$ and also a new two-dimensional basis of left eigen-vectors described by the matrix $(N X)^\ast$ being orthonormal to the new right eigen-basis.

Since the early days of Quantum Theory it is now general opinion that the eigen-values of matrices representing observable quantities like the energy should be real-valued. This is one of the corner-stones of the foundations of traditional Quantum Theory as formalized e.\ g. by J.\ von Neumann and P.A.M.\ Dirac (see e.\ g.\ \cite{vonNeumann:2019}) and found in nearly all present text books on the field of Quantum Theory. There was the longstanding hope that it is reasonable to constrain the formalism of Quantum Theory to Hermitian matrices which possess --- due to their Hermiticity property $H^+ = H$ --- automatically real-valued eigen-values, yet also an (unnecessary) additional constraint $(NX)^{-1} = (NX)^+ = ((NX)^T)^\ast$ relating the basis of the right eigen-vectors $(NX)^{-1}$ and left eigen-vectors $(NX)^T$ by complex conjugation.  

Several years  ago it came as surprise\footnote{It is quite surprising that the fact was noted as a surprise e.\ g.\ in 1998 \cite{Bender:1998ke}, since the scientific literature was full of problems pointing towards the need of a non-Hermitian generalization of Quantum Theory (see e.\ g.\ \cite{Kleefeld:2002au}\cite{Kleefeld:2004jb}\cite{Kleefeld:2004qs}\cite{Kleefeld:2005hf}).} that there should be taken also into account  operators \cite{Bender:1998ke} and matrices (e.\ g.\ \cite{Bender:2002vv}\cite{Bender:2003gu}\cite{Bender:2005tb}\cite{Bender:2007nj}\cite{Bender:2009mq}\cite{Wang:2010br}\cite{Deng:2012yw}\cite{Wang:2013oqw}) in the formalism of Quantum Theory whose eigen-values are real-valued only in a certain range of the parameter space. An example is the following matrix representing a two-dimensional Hamilton operator discussed in Appendix~\ref{xappenda1} describing two ``complex ghosts" \cite{Nakanishi:2006ct}\cite{Nakanishi:1972pt}\cite{Kleefeld:2002au}\cite{Kleefeld:2004jb}\cite{Kleefeld:2004qs}\cite{Kleefeld:1999}\cite{Kleefeld:2001xd}\cite{Ahmed:2003nn}\cite{Wang:2010br}\cite{LiGe:2016}  with bare mass $m-i \,\varepsilon$ and $m+i\,\varepsilon$ ($m$ and $\varepsilon$ are real-valued) interacting via a complex coupling $\gamma\,$:
\begin{equation} H \; = \;  \left(\begin{array}{cc} m-i \,\varepsilon  & \gamma^\ast \\ \gamma & m+i \,\varepsilon \end{array}\right) \; . \label{compghostx1}
\end{equation}
This matrix $H$ is obviously Hermitian in the limit $\varepsilon=0$. Nontheless are the eigen-values $E_{1,2} = m \mp \sqrt{|\gamma|^2 - \varepsilon^2}$ even real-valued in a wider parameter range, i.~e., for $-|\gamma|\le \varepsilon \le +|\gamma|$, where $H$ turns out to be manifestly non-Hermitian, i.\ e. $\varepsilon\not=0 \; \Rightarrow \; H^+ \not= H$. Outside this parameter range, i.~e., for $|\varepsilon|>|\gamma|$  the eigen-values $E_{1,2} = m \mp i\, \sqrt{\varepsilon^2-|\gamma|^2}$ of $H$  are no more real-valued, yet are related by complex conjugation, i.\ e. $E^\ast_2 =E_1 $. Throughout this work we will consider two-dimensional Hamilton operators $H$ for which the eigen-values $E^\ast_1$, $E^\ast_2$ of $H^+$ are related to the eigen-values $E_1$, $E_2$ of $H$ by a two-dimensional invertible matrix $Q$ via the following identity:
\begin{equation} \left(\begin{array}{cc} E^\ast_1 & 0 \\ 0 & E^\ast_2 \end{array}\right) \; = \; \pm \; Q \left(\begin{array}{cc} E_1 & 0 \\ 0 & E_2 \end{array}\right) Q^{-1} \; .
\end{equation}
By insertion of this identity in the spectral expansion of $H^+$ given by Eq.\ (\ref{xhdagger1}) it is straight forward to show with Eq.\ (\ref{xxyyzz1}) that the matrix representing $H$ is either pseudo-Hermitian ($+$) or anti-pseudo-Hermitian ($-$), i.\ e, that there holds the identity $H^+ = \; \pm \; \eta \, H \, \eta^{-1}$ \cite{Pauli:1943}\cite{Pauli:1941zz}\cite{Pandit:1959}\cite{Nagy:1960}\cite{Nagy:1966}\cite{Znojil:2001mc}\cite{Weigert:2003py}\cite{Weigert:2003pn}\cite{Znojil:2006ugs}\cite{Mostafazadeh:2006} relating $H^+$ and $H$ which contains a two-dimensional invertible matrix $\eta$ --- to be called here metric ---, which is calculated in the following way:
\begin{equation} \eta \; = \;  (NX)^+  \,  Q\; N  X \,  . \label{etaqnx0} \end{equation}
Eq.\ (\ref{etaqnx0}) is the fundamental result of this work. Simple inspection leads to the following important observations:
\begin{itemize}
\item If $Q$ is Hermitian ($Q^+=Q)$, then $\eta$ will be also Hermitian ($\eta^+=\eta$).
\item If $Q$ is positive semi-definite ($Q=Q_+)$, then $\eta$ will be also positive semi-definite ($\eta=\eta_+$). A  particular example is the case $Q=1_2$ (to be called metric in the``trivial phase" throughout this work), i.~e., $\eta_+  =  (NX)^+  N  X$, discussed for $N=1_2$ at various places in the literature.\footnote{Eq.\ (\ref{etaqnx0}) should be distinguished from the usual expression $\eta_+=X^+  X$  found for $N=1_2$ and $Q=1_2$ (``trivial phase") in the literature to calculate some positive semi-definite metric $\eta_+$  providing the non-unitary equivalence between non-Hermitian and Hermitian Hamilton operators  as suggested in 1992 by F.~G.~Scholtz, H.~B.~Geyer and F.~J.~W.~Hahne \cite{Scholtz:1992zz} (see also \cite{Geyer:2007}) and later in  2001 by A.\ Mostafazadeh \cite{Mostafazadeh:2001nr}  (using the notation $\eta_+=O O^+$), L. Solombrino \cite{Solombrino:2002vk} (using the notation $\eta_+=(O O^+)^{-1}$) or P.~D.~Mannheim \cite{Mannheim:2017apd} (using the notation $V=S^+S$), and which was applied to two-dimensional matrices at latest in the year 2003 by Z.\ Ahmed \cite{Ahmed:2003nn}\cite{Ahmed:2004nm} (using the notation $\eta_+=(D D^+)^{-1}$) and A. Mostafazadeh  \cite{Mostafazadeh:2003gz} (using now the notation $\eta_+ = \rho^{\,2}_+$) (see also P.~D. Mannheim \cite{Mannheim:2017apd} in 2018).}
\item The sign of the determinant of $\eta$ coincides with the sign of the determinant of $Q$, as there holds:
\begin{equation} \mbox{det} [\,\eta\, ] \; = \;  \mbox{det} [\,Q\,] \;\, \Big|\,\mbox{det} [N  X]\,\Big|^2 \,  .\end{equation}
\item Eq.\ (\ref{etaqnx0}) displays  and explains --- as has been conjectured e.\ g. in 2009 \cite{Kleefeld:2009vd} ---  clearly the ambiguity of the metric due to its manifest dependence on the undetermined invertible and diagonal matrix $N$ renormalizing the right eigen-vectors of the Hamilton operator $H$ \, (see also \cite{Mostafazadeh:2006}\cite{Znojil:2008}\cite{Znojil:2009}\cite{Shi:2009pc}).\footnote{By now the ambiguity of the positive semidefinite metric $\eta_+$ has been debated mainly in the context of the ambiguity of the ${\cal C}$-operator \cite{Bender:2002vv}\cite{Bender:2005tb}\cite{Weigert:2003py}\cite{Bender:2004ej} in ${\cal PT}$-symmetric Quantum Theory \cite{Bender:2003gu}\cite{Bender:2007nj}\cite{Bender:2019cwm} observed e.\ g.\ by C.~M.~Bender and S.~P.~Klevansky \cite{Bender:2009en} (see also \cite{Mostafazadeh:2005wm}\cite{Znojil:2006ugs}\cite{Mostafazadeh:2006}). The ambiguity had been seemingly fixed by C.~M.~Bender et al.\ by imposing the constraint that the ${\cal C}$-operator should be an involution, i.\ e.\ ${\cal C}^2=1$. Nontheless we demonstrate in Sec.\ \ref{relmetric1} of the manuscript that this constraint does not only fix the renormalizations of the right eigen-vectors to $|N_1|^2 = |N_2|^2=1$, yet --- unfortunately --- requires  also the two-dimensional matrix representing the Hamilton operator to be symmetric ($H_{12}=H_{21}$) being by far too restrictive.} 
\item The --- to our best knowledge --- new aspect of this manuscript is the explicit inclusion and investigation of the matrix $Q$ into the expression Eq.\ (\ref{etaqnx0}) for the calculation of the metric $\eta$ with $Q$ being some ``trivial" ($Q=1$) or ``non-trivial" ($Q\not=1$) element of the permutation group \cite{Chen:2002}. For two dimensions we have $Q\in \{ 1_2 \,, P \}$ with $1_2$ being the two-dimensional unit-matrix and $P$ being the (Hermitian) parity matrix:
\begin{equation} 1_2 \; = \; \left(\begin{array}{cc} 1 & 0 \\ 0 & 1 \end{array}\right)\; , \quad P \; = \;\left(\begin{array}{cc} 0 & 1 \\ 1 & 0 \end{array}\right) \; .\end{equation}
Due to det$[1_2]>0$ and det$[P]<0$ the two elements of the two-dimensional permutation group lead to two fundamentally different phases of the metric. While the former ``trivial" phase $Q=1_2$ allows a positive semi-definite metric $\eta_+$, the metric of the ``non-trivial" phase $Q=P$ must be indefinite in two dimensions. 
\end{itemize}
It is a curious aspect of present scientific research in the field of non-Hermitian Quantum Theory that the ``trivial phase" ($Q=1_2$) (e.g.\ ${\cal PT}$-symmetric Quantum Theories in the so-called phase of ``unbroken" ${\cal PT}$-symmetry) is attracting --- due to the existence of a positive-semidefinite metric $\eta_+$ --- much more attention than the ``non-trivial phase" ($Q\not=1_2$) with det$[Q]<0$ (e.g.\ ${\cal PT}$-symmetric Quantum Theories in the so-called phase of ``broken" ${\cal PT}$-symmetry)  being either avoided or declared ``unphysical", or applied without sufficient due care with respect  to the possible loss of causality, analyticity or Lorentz-invariance by inappropriate use.

We had pointed out on the contrary e.\ g.\ in \cite{Kleefeld:2002au}\cite{Kleefeld:2004jb}\cite{Kleefeld:2004qs}\cite{Kleefeld:1999} that a consistent causal, analytic and Lorentz covariant formulation of Quantum Theory allowing a correspondence principle in the whole complex  space or momentum plane {\em requires} the mere existence of complex ghosts (i.\ e.\ the ``non-trivial" phase $|\varepsilon|>|\gamma|$ in Eq.\ (\ref{compghostx1})) which are obviously governed by some indefinite metric. The only constraint imposed in  \cite{Kleefeld:2002au}\cite{Kleefeld:2004jb}\cite{Kleefeld:2004qs}\cite{Kleefeld:1999} to maintain causality and analyticity is that causal complex ghosts should not interact with anti-causal complex ghosts at all. Inspection of  Eq.\ (\ref{compghostx1}) suggests at least that there might be room for interaction in the presence of causality and absence of analyticity as long as the interaction is sufficently weak, i.\ e., if there holds $|\varepsilon|>|\gamma|$.

The outline of the present manuscript is the following: after the introduction of necessary notations for two-dimensional matrix Hamilton operators in Section \ref{xxsec2} and the concept of generalized parity operators in Section \ref{xxsec3} we develop in Section \ref{xxsec4} in more detail the concept of the metric $\eta$ in the ``trivial" and ``non-trivial" phase. In Section \ref{implhpseudo1} we identify in two-dimensions four different cases for which the metric can be determined: 
\begin{itemize}
\item {\bf Case 1:} $H$ ist pseudo-Hermitian ($H^+ \; = \; + \,\eta \, H \, \eta^{-1}$)\\
\mbox{}$\qquad\quad\;\;\;$ and the phase of the metric $\eta$ is trivial ($Q=1_2$)
\item {\bf Case 2:} $H$ ist pseudo-Hermitian ($H^+ \; = \; + \,\eta \, H \, \eta^{-1}$)\\
\mbox{}$\qquad\quad\;\;\;$ and the phase of the metric $\eta$ is non-trivial ($Q=P$)
\item {\bf Case 3:} $H$ ist anti-pseudo-Hermitian ($H^+ \; = \; -\, \eta \, H \, \eta^{-1}$)\\
\mbox{}$\qquad\quad\;\;\;$ and the phase of the metric $\eta$ is trivial ($Q=1_2$)
\item {\bf Case 4:} $H$ ist anti-pseudo-Hermitian ($H^+ \; = \; -\, \eta \, H \, \eta^{-1}$)\\
\mbox{}$\qquad\quad\;\;\;$ and the phase of the metric $\eta$ is non-trivial ($Q=P$)
\end{itemize}
\noindent After diagonalizing $H$ and $H^+$ in Section \ref{xxsec6} we provide in Section \ref{xmetrici} expressions to calculate $\eta$ for $Q=1_2$ and  in Section \ref{xmetricp} for $Q=P$. Section \ref{xxsec9} is devoted to the calculation of the inverse metric $\eta^{-1}$, while Section \ref{relmetric1} is supposed to clarify for $Q=1_2$ the conditions implied by the --- from our point of view unnecessary --- constraint imposed by C.~M.~Bender and collegues that the ${\cal C}$-operator should be an involution, i.\ e. ${\cal C}^2=1$. In Section \ref{xconvrep1} we provide a summarizing overview to the the four cases identified in Section~\ref{implhpseudo1}. Throughout Section \ref{xxsec12} we apply our methods to calculate the metric for various topical two-dimensional matrix Hamilton operators presently under consideration in scientific literature. The manuscript ends  in Section \ref{xxsec13} with conclusions and some outlook inspired by a non-Hermitian Hamilton operator proposed by T.D.~Lee and C.G.~Wick, several Appendices and a by far not exhaustive list of related references.
  \newpage
\section{Matrix representation of $H$ and $H^+$ in two dimensions} \label{xxsec2}
\noindent Any $2\times2$-matrix $H$ representing a diagonalizable Hamilton operator with eigen-values $E_1$ and $E_2$ can be written in the following way:
\begin{eqnarray}
 H \;  = \;  \left(\begin{array}{cc} H_{11} & H_{12} \\ H_{21} & H_{22} \end{array}\right) & = & \frac{E_1+E_2}{2} \; 1_2 +  \frac{E_1-E_2}{2} \; \vec{n} \cdot \vec{\sigma} \nonumber \\[2mm]
 & = & \frac{1_2+ \vec{n} \cdot \vec{\sigma}}{2} \; E_1 +  \frac{1_2- \vec{n} \cdot \vec{\sigma}}{2} \; E_2 \; , \label{hamop1}
\end{eqnarray}
with $\vec{n}$ being an arbitrary two-dimensional unit vector (i.\ e., $\vec{n}\cdot \vec{n} = 1$), $1_2$ being the two-dimensional unit matrix  and $\vec{\sigma}$ being the well known Pauli matrices taking here the following form:
\begin{equation} \sigma_1 = \left(\begin{array}{cc} 0 & 1 \\ 1 & 0 \end{array}\right) \; , \quad \sigma_2 = \left(\begin{array}{cc} 0 & -i \\ i & 0 \end{array}\right) \; , \quad \sigma_3 = \left(\begin{array}{cc} 1 & 0 \\ 0 & -1 \end{array}\right) \; . \label{paulimat1}
\end{equation}
All three Pauli matrices are Hermitian, i.e., they remain invariant under transposition and subsequent complex conjugation (In our notation: $\sigma^+_1=\sigma_1$, $\sigma^+_2=\sigma_2$, $\sigma^+_3=\sigma_3$). 

The Hermitian conjugate matrix $H^+$ representing the Hermitian conjugate Hamilton operator  with eigen-values $E^\ast_1$ and $E^\ast_2$ is obtained by performing the Hermitian conjugation of Eq.\ (\ref{hamop1}):
\begin{eqnarray}
 H^+ \;  = \;  \left(\begin{array}{cc} H^\ast_{11} & H^\ast_{21} \\ H^\ast_{12} & H^\ast_{22} \end{array}\right) & = & \frac{E^\ast_1+E^\ast_2}{2} \; 1_2 +  \frac{E^\ast_1-E^\ast_2}{2} \; \vec{n}^\ast \cdot \vec{\sigma} \nonumber \\[2mm]
 & = & \frac{1_2+ \vec{n}^\ast \cdot \vec{\sigma}}{2} \; E^\ast_1 +  \frac{1_2- \vec{n}^\ast \cdot \vec{\sigma}}{2} \; E^\ast_2 \; , \label{hamop3}
\end{eqnarray}
while $\vec{n}^\ast$ is also a two-dimensional unit vector  (i.\ e., $\vec{n}^\ast\cdot \vec{n}^\ast = 1$) being obtained by complex conjugation of $\vec{n}$. 
\section{Generalized parity operators}  \label{xxsec3}
\noindent The chosen Pauli matrices Eq.\ (\ref{paulimat1}) respect the well known identities
\begin{eqnarray} \vec{u} \cdot \vec{\sigma} \;\,  \vec{v} \cdot \vec{\sigma} & = & \vec{u} \cdot \vec{v}\;\, 1_2 \; + \;  i\, \vec{\sigma} \cdot [\,\vec{u}\times \vec{v}\; ] \; ,  \label{uvident1} \\[2mm]
 \vec{u} \cdot \vec{\sigma} \; \, \vec{e} \cdot \vec{\sigma}  \; \, \vec{v} \cdot \vec{\sigma} & = & \vec{u} \cdot \vec{e}\;\, \vec{v} \cdot \vec{\sigma} + \vec{v} \cdot \vec{e}\;\, \vec{u} \cdot \vec{\sigma} - \vec{u} \cdot \vec{v}\;\, \vec{e} \cdot \vec{\sigma}   -  i\, \vec{e} \cdot [\,\vec{u}\times \vec{v}\; ] \, 1_2 \, ,  \quad \label{uvident2} 
\end{eqnarray}
for arbitrary two-dimensional vectors $\vec{u}$, $\vec{v}$ and $\vec{e}$. Due to the first identity there holds for any vector $\vec{n}_\perp$ orthogonal to $\vec{n}$ (implying $\vec{n}\cdot \vec{n}_\perp=0$) obviously:
\begin{equation} \vec{n} \cdot\vec{\sigma} \;\, \vec{n}_\perp\cdot \vec{\sigma} \; = \; - \, \vec{n}_\perp \cdot\vec{\sigma} \; \,\vec{n}\cdot \vec{\sigma} \; .
\end{equation}
If we assume $\vec{n}_\perp$ to be a unit vector, i.e.\ $\vec{n}_\perp\cdot\vec{n}_\perp=1$, we can define the following generalized parity operators $P_\pm$ as follows:
\begin{equation} P_\pm \; = \; \pm \; \vec{n}_\perp \cdot \vec{\sigma} \quad \mbox{(with $\vec{n}_\perp\cdot\vec{n}_\perp=1$)}  \; . \label{parop1}
\end{equation}
With the help of the following important properties of the generalized parity operators
\begin{equation} \vec{n}\cdot \vec{\sigma} \;\, P_\pm \; = \; - \, P_\pm \; \,\vec{n}\cdot \vec{\sigma} \; , \qquad  P_\pm\,P_\pm \; = 1_2 \; ,  
\end{equation}
it is now trivial to show by application to  Eq.\ (\ref{hamop1}) that the generalized parity operators $P_\pm$ interchange the eigen-values of the Hamilton operator:
\begin{eqnarray}
P_\pm\; H \; P_\pm & = & \frac{E_2+E_1}{2} \; 1_2 +  \frac{E_2-E_1}{2} \; \vec{n} \cdot \vec{\sigma} \nonumber \\[2mm]
 & = & \frac{1_2+ \vec{n} \cdot \vec{\sigma}}{2} \; E_2 +  \frac{1_2- \vec{n} \cdot \vec{\sigma}}{2} \; E_1 \; . \label{hamop2}
\end{eqnarray}
The respective generalized parity operators $P_\pm^+$  interchanging the eigen-values $E^\ast_1$ and $E^\ast_2$ of $H^+$ are obviously defined in the following way:
\begin{equation} P_\pm^+ \; = \; \pm \; \vec{n}^\ast_\perp \cdot \vec{\sigma} \quad \mbox{(with $\vec{n}^\ast_\perp\cdot\vec{n}^\ast_\perp=1$)}  \; .
\end{equation}
They respect the following identities:
\begin{equation} \vec{n}^\ast\cdot \vec{\sigma} \;\, P_\pm^+ \; = \; - \, P_\pm^+ \; \,\vec{n}^\ast \cdot \vec{\sigma} \; , \qquad  P_\pm^+\, P_\pm^+ \; = 1_2 \; . 
\end{equation}
We would like to stress at this place that for $\vec{n}_\perp\not=\vec{n}^\ast_\perp$ the generalized parity operators are not Hermitian, i.e. there holds $P_\pm \not=P_\pm^+$. For $\vec{n}\not=\vec{n}^\ast$ it is tempting to perform the following choice:
\begin{equation} \vec{n}_\perp \; = \; \frac{\vec{n}\times \vec{n}^\ast}{([\vec{n}\times \vec{n}^\ast]\cdot[\vec{n}\times \vec{n}^\ast])^{1/2}} \; , \quad \vec{n}^\ast_\perp \; = \; \frac{\vec{n}^\ast\times \vec{n}}{([\vec{n}^\ast\times \vec{n}]\cdot[\vec{n}^\ast\times \vec{n}])^{1/2}} \; .
\end{equation}
For this particular choice there holds obviously $P_\pm^+ = P_\mp$.

\section{General form of the metric $\eta$: ``trivial" \& ``non-trivial" phase}  \label{xxsec4}
\noindent The  two-dimensional matrix $\vec{n} \cdot \vec{\sigma}$ with $\vec{n}\cdot\vec{n}=1$  is diagonalizable with eigen-values  $+1$ and $-1$. In a first step we want to outline a very tricky strategy for its diagonalization:
\begin{eqnarray} 
\vec{n} \cdot \vec{\sigma}  & = & n_3 \, \sigma_3 + n_2\, \sigma_2 + n_1 \, \sigma_1  \nonumber \\[2mm]
 & = & \sigma_3\, (n_3 \; 1_2  -i\,  n_2\, \sigma_1 +i\,  n_1 \, \sigma_2)  \nonumber \\[2mm]
 & = & \sigma_3 \left(n_3 \; 1_2  -i\, (n_1^2 + n^2_2)^{1/2}\; \frac{ n_2\, \sigma_1 -\,  n_1 \,\sigma_2}{(n_1^2 + n^2_2)^{1/2}}  \right) \nonumber \\[2mm]
 & = & \sigma_3\, (\cos \alpha \; 1_2  -i\, \sin\alpha \; \,\vec{m} \cdot \vec{\sigma} ) \nonumber \\[2mm]
 & = & \sigma_3\, \exp\left (  -i\, \alpha  \; \vec{m} \cdot \vec{\sigma}  \right) \nonumber  \\[2mm]
 & = & \exp\left (  +i\,  \frac{\alpha}{2} \;  \vec{m} \cdot \vec{\sigma}  \right)  \sigma_3\, \exp\left (  -i\,  \frac{\alpha}{2} \; \vec{m} \cdot \vec{\sigma}  \right) \nonumber  \\[2mm]
 & = & \left(\cos \frac{\alpha}{2} \; 1_2  +i\, \sin\frac{\alpha}{2} \; \vec{m} \cdot \vec{\sigma} \right)  \sigma_3 \left(\cos \frac{\alpha}{2} \; 1_2  -i\, \sin\frac{\alpha}{2} \; \vec{m} \cdot \vec{\sigma} \right) \nonumber \\[2mm]
 & = & \cos \frac{\alpha}{2} \left(1_2  +i\, \tan\frac{\alpha}{2} \; \vec{m} \cdot \vec{\sigma} \right)  \sigma_3 \; \cos \frac{\alpha}{2}  \left(1_2  -i\, \tan\frac{\alpha}{2} \; \vec{m} \cdot \vec{\sigma} \right) \; .\quad \label{nsigtrans1}
\end{eqnarray}
During these steps we performed the following definitions:
\begin{equation} \cos\alpha \; = \; n_3 \;, \quad \sin\alpha \; = \; (n_1^2 + n^2_2)^{1/2} \; , \quad \vec{m} \cdot \vec{\sigma} \; = \; \frac{ n_2\, \sigma_1 -\,  n_1 \,\sigma_2}{(n_1^2 + n^2_2)^{1/2}} \; .\end{equation} 
The vector $\vec{m}$ is obviously a unit vector, as there holds $\vec{m}\cdot \vec{m} = 1$. In a last step we apply the following identities to Eq.\ (\ref{nsigtrans1}):
\begin{eqnarray} \left( \cos \frac{\alpha}{2}\right)^2 & = &  \frac{1}{2} \, (1 + \cos \alpha ) \; = \; \frac{1}{2} \, (1 + n_3 )  \; , \quad \\[2mm]
  \tan \frac{\alpha}{2} & = & \frac{\sin\frac{\alpha}{2}}{\cos\frac{\alpha}{2}} \; = \; \frac{\sin\frac{\alpha}{2}\, \cos\frac{\alpha}{2}}{\left(\cos\frac{\alpha}{2}\right)^2} \; = \; \frac{\sin\alpha}{1 + \cos\alpha} \; , \quad \\[2mm]
 e^{\pm i\alpha} & = & \cos\alpha \pm i \sin \alpha \; = \; n_3 \pm i \; (n_1^2 + n^2_2)^{1/2} \; ,
\end{eqnarray}
to obtain
\begin{equation}  \vec{n} \cdot \vec{\sigma} \;= \; X^{-1} \; \sigma_3 \;  \; X \quad \Leftrightarrow \quad  X \; \vec{n} \cdot \vec{\sigma} \; X^{-1} \;= \; \sigma_3 \; ,  \label{diagmat1}  \end{equation}
with
\begin{eqnarray} 
X & = &  \left(  \left(\cos \frac{\alpha}{2}\right)^2 \right)^{\frac{1}{2}} \;  \left(1_2  -i\, \tan\frac{\alpha}{2} \; \vec{m} \cdot \vec{\sigma} \right) \\[2mm]
  & = &  \frac{1}{\sqrt{2}} \, (1 + \cos \alpha )^{\frac{1}{2}}  \left(1_2  -\, \frac{i\, \sin\alpha}{1 + \cos\alpha} \; \vec{m} \cdot \vec{\sigma} \right)  \\[2mm]
  & = &  \frac{1}{\sqrt{2}} \, (1 + n_3 )^{\frac{1}{2}}  \left(1_2  -\, i\;\frac{n_2\, \sigma_1 -\,  n_1 \,\sigma_2}{1 + n_3}  \right) \; = \;  \frac{1_2  + \sigma_3 \, \vec{n}\cdot \vec{\sigma}}{\sqrt{2} \;  (1 + n_3 )^{\frac{1}{2}}} \; , \label{xxeq1} \\[2mm]
X^{-1} & = &  \left(  \left(\cos \frac{\alpha}{2}\right)^2 \right)^{\frac{1}{2}} \; \left(1_2  +i\, \tan\frac{\alpha}{2} \; \vec{m} \cdot \vec{\sigma} \right) \\[2mm]
  & = &  \frac{1}{\sqrt{2}} \, (1 + \cos \alpha )^{\frac{1}{2}}  \left(1_2  +\, \frac{i\, \sin\alpha}{1 + \cos\alpha} \; \vec{m} \cdot \vec{\sigma} \right) \\[2mm]
  & = &  \frac{1}{\sqrt{2}} \, (1 + n_3 )^{\frac{1}{2}}  \left(1_2  +\, i\;\frac{n_2\, \sigma_1 -\,  n_1 \,\sigma_2}{1 + n_3}  \right) \; = \;  \frac{1_2  + \vec{n}\cdot \vec{\sigma} \; \sigma_3}{\sqrt{2} \;  (1 + n_3 )^{\frac{1}{2}}}\; . \label{xxeq2}
\end{eqnarray}
Note that there holds det$[X] =\;$det$[X^{-1}] = 1$. Eq.\ (\ref{diagmat1}) is not unique. There exists the freedom to multiply this equation from the left with some invertible diagonal matrix $N \;= \;$diag$[N_1,N_2]$ and from the right with the respective inverse matrix $N^{-1}$ with the following result:
\begin{equation} N \, X \; \vec{n} \cdot \vec{\sigma} \; (N\,X)^{-1} \;= \; \sigma_3 \; .  \label{diagmat2} \end{equation}
In defining $Q$ to be a matrix representing an element of the permutation group in two dimensions, i.e.\ $Q\in\{1_2,P\}$ with $P$ being a two-dimensional permutation matrix, and applying it to Eq.\ (\ref{diagmat2}), i.e.,
\begin{equation} Q\, N \, X \; \vec{n} \cdot \vec{\sigma} \; (Q\, N\,X)^{-1} \;= \; Q \; \sigma_3 \; Q^{-1} \; ,  \label{diagmat3} \end{equation}
it gets clear that there exist two fundamentally different phases in diagonalizing the matrix $\vec{n} \cdot \vec{\sigma}$ which are spezified by the order of its eigen-values:
\begin{eqnarray} 1_2\, N \, X \; \vec{n} \cdot \vec{\sigma} \; (1_2\, N\,X)^{-1} & = & 1_2 \; \sigma_3 \; 1_2^{-1} \; = \; + \sigma_3 \; = \; \left(\begin{array}{cc} +1 & 0 \\ 0 & -1 \end{array}\right) \; , \label{pha1} \\[2mm]
 P\; N \, X \; \vec{n} \cdot \vec{\sigma} \; (P\, N\,X)^{-1} & = & P \; \sigma_3 \; P^{-1} \; = \; - \sigma_3 \; = \; \left(\begin{array}{cc} -1 & 0 \\ 0 & +1 \end{array}\right) \; . \label{pha2}
\end{eqnarray}
For later convenience we call the case $Q=1_2$ ``trivial phase" and the case $Q=P$ ``non-trivial phase". After performing the Hermitian conjugation of Eq.\ (\ref{diagmat2}), i.\ e.,
\begin{eqnarray} \lefteqn{((N \, X)^+)^{-1} \; (\vec{n} \cdot \vec{\sigma})^+ \; (N\,X)^+ \;= \; \sigma_3} \nonumber \\[2mm] 
 & \Rightarrow & (\vec{n} \cdot \vec{\sigma})^+  \;= \;  (N\,X)^+ \;  \sigma_3 \; ((N \, X)^+)^{-1} \; ,  \label{diagmat4} \end{eqnarray}
we can determine the matrix $\vec{n}^\ast \cdot\, \vec{\sigma}$  (or, equivalently, \mbox{$(\vec{n} \cdot \vec{\sigma})^+$}) in terms of $\vec{n} \cdot \vec{\sigma}$ by substituting Eqs.\ (\ref{pha1}) and (\ref{pha2}) into Eq.\ (\ref{diagmat4}):
\begin{eqnarray} Q = 1_2 & \Rightarrow &  (\vec{n} \cdot \vec{\sigma})^+  \;= \;  +\, (N\,X)^+    N \, X \;\; \vec{n} \cdot \vec{\sigma} \;\; ((N \, X)^+ N\,X)^{-1}   , \\[2mm]
Q = P & \Rightarrow &  (\vec{n} \cdot \vec{\sigma})^+  \;= \;  -\, (N\,X)^+    P\, N \, X \;\; \vec{n} \cdot \vec{\sigma} \;\; ((N \, X)^+\,P\, N\,X)^{-1}   , \quad \end{eqnarray}
or, equivalently (see also Eq.\ (\ref{etaqnx0}) in Section  \ref{xxsec1}),
\begin{eqnarray} Q = 1_2 & \Rightarrow &  (\vec{n} \cdot \vec{\sigma})^+  \;= \; +\;  \eta \;\; \vec{n} \cdot \vec{\sigma} \;\; \eta^{-1} \quad \mbox{with} \quad \eta \; = \; (N\,X)^+    N \, X \,   , \label{pseudherm1} \\[2mm]
Q = P & \Rightarrow &  (\vec{n} \cdot \vec{\sigma})^+  \;= \; -\;  \eta \;\; \vec{n} \cdot \vec{\sigma} \;\; \eta^{-1} \quad \mbox{with} \quad \eta \; = \;  (N\,X)^+    P\, N \, X \,  .  \quad  \label{pseudherm2} \end{eqnarray}
As pointed out already in Section \ref{xxsec1} we will call the two-dimensional matrix $\eta$ throughout this work ``metric". In the ``trivial phase" ($Q=1_2$)  the matrices $\vec{n} \cdot \vec{\sigma}$ and $(\vec{n} \cdot \vec{\sigma})^+$ are pseudo-Hermitian, as  the matrix $\vec{n} \cdot \vec{\sigma}$ is equivalent to its Hermitian conjugate matrix $(\vec{n} \cdot \vec{\sigma})^+$. In the ``non-trivial phase" ($Q=P$)  the matrices $\vec{n} \cdot \vec{\sigma}$ and $(\vec{n} \cdot \vec{\sigma})^+$ are anti-pseudo-Hermitian, as  the matrices $i\,\vec{n} \cdot \vec{\sigma}$ and $(i\,\vec{n} \cdot \vec{\sigma})^+$ are pseudo-Hermitian.

The calculation of the determinant of the metric $\eta$ making use of Eqs.\ (\ref{pseudherm1}) and (\ref{pseudherm2}) reveals that the sign of det$[\eta]$ coincides with the sign of det$[Q]$ which may be used to identify whether the metric is in the trivial or non-trivial phase, i.\ e., (with $\mbox{det} [\,X \,]=1$ and $\mbox{det} [\,N \,]=N_1\, N_2\,$)
\begin{eqnarray} \mbox{det} [\,\eta \,] & = & \mbox{det} [\,(N\,X)^+    Q\, N \, X\,] \; = \; \Big|\,\mbox{det} [\,N \,]\,\Big|^2\; \Big|\,\mbox{det} [\,X \,]\,\Big|^2 \; \mbox{det} [\,Q \,] \nonumber \\[1mm]
 & = & \Big|N_1|^2 \, \Big|N_2|^2 \;\, \mbox{det} [\,Q \,]  \; \left\{ \begin{array}{l} \mbox{$>0\quad$ for $\quad Q=1_2\; .$} \\ \\ \mbox{$<0\quad $ for $\quad Q=P\; .$} \end{array} \label{identq1}  \right. 
\end{eqnarray}
\section{Implications of $H$ being pseudo-Hermitian or anti-pseudo-Hermitian} \label{implhpseudo1}
\noindent The identities of Eqs.\ (\ref{pseudherm1}) and (\ref{pseudherm2}) can be inserted in the representation of $H^+$ given by Eq.\   (\ref{hamop3}). For the ``trivial phase" ($Q=1_2$)  and ``non-trivial phase" ($Q=P$) we obtain the following result:
\begin{eqnarray}
 Q \; = \; 1_2 & \Rightarrow & H^+ \; = \;  \eta \,\left(\frac{E^\ast_1+E^\ast_2}{2} \; 1_2 +  \frac{E^\ast_1-E^\ast_2}{2} \; \vec{n} \cdot \vec{\sigma}\right) \, \eta^{-1}  \; , \label{xhamop1} \\ [2mm]
 Q \; = \; P & \Rightarrow & H^+ \; = \;  \eta \,\left(\frac{E^\ast_1+E^\ast_2}{2} \; 1_2 +  \frac{E^\ast_2-E^\ast_1}{2} \; \vec{n} \cdot \vec{\sigma}\right) \, \eta^{-1}  \; . \label{xhamop2} 
\end{eqnarray}
The result demonstrates clearly that the application of the elements of the permutation group $Q\in\{1_2,P\}$ does not only permute the eigen-values of the matrix $\vec{n} \cdot \vec{\sigma}$ as shown in Eqs.\ (\ref{pha1}) and (\ref{pha2}). It simultaneously permutes also the order of the eigen-values of the matrix representing the Hermitian conjugate Hamilton operator $H^{+}$ (see Eqs.\ (\ref{xhamop1}) and  (\ref{xhamop2})).
A comparison of Eqs.\ (\ref{xhamop1}) and  (\ref{xhamop2}) with the representation of $H$ in Eq.\   (\ref{hamop1}), i.\ e.,
\begin{equation} H \; = \; \frac{E_1+E_2}{2} \; 1_2 +  \frac{E_1-E_2}{2} \; \vec{n} \cdot \vec{\sigma} \; , \label{hamop10}
\end{equation}
suggests furthermore that the constraint of pseudo-Hermiticity (i.\ e. $H^+ = + \eta \, H \, \eta^{-1}$) or anti-pseudo-Hermiticity (i.\ e. $H^+ = - \eta \, H \, \eta^{-1}$) on the Hamilton operator $H$ imposes constraints on the eigen-values $E_1$ and $E_2$ of the Hamilton operator $H$. For the ``trivial phase" ($Q=1_2$)  and ``non-trivial phase" ($Q=P$) we obtain the following result:
\begin{eqnarray} Q=1_2 \quad \mbox{and} \quad  H^+ = + \eta \, H \, \eta^{-1} & \Rightarrow & E^\ast_1 = E_1 \; , \; E^\ast_2 = E_2 \; , \label{enconstr1} \\[2mm]
Q=P \quad \mbox{and} \quad  H^+ = + \eta \, H \, \eta^{-1} & \Rightarrow &  E^\ast_2 = E_1 \; , \label{enconstr2} \\[2mm]
 Q=1_2 \quad \mbox{and} \quad  H^+ = - \eta \, H \, \eta^{-1} & \Rightarrow & E^\ast_1 = - E_1 \; , \; E^\ast_2 = - E_2 \; ,  \label{enconstr3}\\[2mm]
Q=P \quad \mbox{and} \quad  H^+ = - \eta \, H \, \eta^{-1} & \Rightarrow &  E^\ast_2 = - E_1 \; .  \label{enconstr4} \end{eqnarray}
\subsection{Identification of (anti-)pseudo-Hermiticity and the (non-)trivial phase}
\noindent How can the four classes of Hamilton operators characterized by Eqs.\ (\ref{enconstr1}), (\ref{enconstr2}), (\ref{enconstr3}) and (\ref{enconstr4}) be identified? The pseudo-Hermiticity or anti-pseudo-Hermiticity of $H$ and $H^+$ can be tested by considering their traces and determinants:
\begin{eqnarray} H^+ = + \eta \, H \, \eta^{-1} & \Rightarrow & \mbox{tr}[ H^+] = \mbox{tr} [H] \, , \quad\;\;\;\;\, \mbox{det}[ H^+] = \mbox{det} [H] \, . \label{consid1} \\[2mm]
 H^+ = - \eta \, H \, \eta^{-1} & \Rightarrow & \mbox{tr} [(i\,H)^+] = \mbox{tr} [\, i \, H]  \, , \;\,  \mbox{det}[( i\, H)^+] =  \mbox{det} [\, i\, H] \, .\quad \label{consid2}
\end{eqnarray}
As pointed out in the context of Eqs.\ (\ref{pseudherm1}) and (\ref{pseudherm2}) the ``trivial phase" ($Q=1_2$)  and ``non-trivial phase" ($Q=P$) are related to the pseudo-Hermiticity or anti-pseudo-Hermiticity of the matrix $\vec{n}\cdot \vec{\sigma}$, respectively, or --- equivalently ---  to the pseudo-Hermiticity or anti-pseudo-Hermiticity of the Hermitian conjugate matrix $(\vec{n}\cdot \vec{\sigma})^+$. It is easy to show that there hold the following four identities:
\begin{eqnarray}  &  & \mbox{tr}[ (\vec{n}\cdot \vec{\sigma})^+] = \mbox{tr} [\,\vec{n}\cdot \vec{\sigma}\,] \, , \;\;\;\;\; \mbox{det}[ (\vec{n}\cdot \vec{\sigma})^+] = \mbox{det} [\,\vec{n}\cdot \vec{\sigma}\,] \, , \\[2mm]
  &  & \mbox{tr} [(i\,\vec{n}\cdot \vec{\sigma})^+] = \mbox{tr} [\, i \, \vec{n}\cdot \vec{\sigma}\,]  \, , \;\,  \mbox{det}[( i\, \vec{n}\cdot \vec{\sigma})^+] =  \mbox{det} [\, i\, \vec{n}\cdot \vec{\sigma}\,] \, .\quad
\end{eqnarray}
The reason is that both matrices $\vec{n}\cdot \vec{\sigma}$ and $(\vec{n}\cdot \vec{\sigma})^+$ are simultaneously pseudo-Hermitian and anti-pseudo-Hermitian for all unit vectors $\vec{n}$ and $\vec{n}^{\,\ast}$ while the metric $\eta$ for both scenarios --- pseudo-Hermiticity and anti-pseudo-Hermiticity --- is different. Since the metric $\eta$ is not known from the onset it is therefore not the fastest approach to identify the scenario via calculating the sign of its determinant as described in Eq.\ (\ref{identq1}).

For that reason we will perform here some different approach to identify the scenario by considering the eigen-values $E_1$ and $E_2$ of the Hamilton operator $H$ and the eigen-values $i \, E_1$ and $i\, E_2$ of the Hamilton operator $i\, H$. More precisely we consider the following four quantities:
\begin{eqnarray} E_1+E_2\;\;\;  & =& \mbox{tr}[H] \; = \; H_{11} + H_{22}\; , \label{impquant1} \\[2mm]
(E_1-E_2)^2 & = & (E_1+E_2)^2-4\, E_1E_2 \; = \;   \mbox{tr}[H]^2-4\,  \mbox{det}[H] \nonumber \\[2mm]
 & = & (H_{11} + H_{22})^2 - 4 \, (H_{11}\, H_{22} - H_{12}\, H_{21}) \nonumber \\[2mm]
 & = & (H_{11} - H_{22})^2 + 4 \, H_{12}\, H_{21}  \; , \label{impquant2} \\[1mm]
 & & \nonumber \\
 i\, E_1+i\, E_2\;\;\; & = &  \mbox{tr}[i H] \; = \; i\, \mbox{tr}[H]\; = \; i\, (H_{11} + H_{22}) \; ,\label{impquant3} \\[2mm]
(i\, E_1-i\, E_2)^2 & = &  \mbox{tr}[i H]^2-4\,  \mbox{det}[i H]  \; = \;  4\,  \mbox{det}[H]- \mbox{tr}[H]^2  \nonumber \\[2mm]
 & = & 4 \; i\,H_{12}\; i\,H_{21}+ (i\,H_{11} - i\,H_{22})^2  \; . \label{impquant4}
\end{eqnarray}
A first inspection of Eqs.\ (\ref{enconstr1}), (\ref{enconstr2}), (\ref{enconstr3}) and (\ref{enconstr4}) yields the following two observations:
\begin{itemize} 
\item If $H$ is pseudo-Hermitian (i.\ e.\ $H^+ = + \eta \, H \, \eta^{-1}\,$), the two quantities $E_1+E_2=\mbox{tr}[H]$ and $(E_1-E_2)^2=\mbox{tr}[H]^2-4\,  \mbox{det}[H]$ will be simultaneously real-valued. 
\item If $H$ is anti-pseudo-Hermitian (i.\ e.\ $H^+ = - \eta \, H \, \eta^{-1}\,$) or --- equivalently --- if $\,i H$ is pseudo-Hermitian  (i.\ e.\ $(i H)^+ = + \eta \; i H \, \eta^{-1}\,$), the two quantities $i\, E_1+i\, E_2=i\, \mbox{tr}[H]$ and $(i\, E_1-i\, E_2)^2=4\,  \mbox{det}[H]-\mbox{tr}[H]^2$ will be simultaneously real-valued.
\end{itemize}
A more detailed consideration of Eqs.\ (\ref{enconstr1}), (\ref{enconstr2}), (\ref{enconstr3}) and (\ref{enconstr4}) implies then the following four inequalities:

\begin{eqnarray} Q=1_2 \;  , \;  H^+ = + \eta \, H \, \eta^{-1} & \Rightarrow & (E_1-E_2)^2 > 0  \quad \Leftrightarrow \; \mbox{tr}[H]^2> 4\,  \mbox{det}[H]\; , \nonumber  \\ 
 & & \label{enconstr5} \\
Q=P \; , \;  H^+ = + \eta \, H \, \eta^{-1} & \Rightarrow &  (E_1-E_2)^2 < 0 \quad \Leftrightarrow \; \mbox{tr}[H]^2< 4\,  \mbox{det}[H] \; ,\nonumber  \\ 
 & &  \label{enconstr6} \\
 Q=1_2 \; , \; H^+ = - \eta \, H \, \eta^{-1} & \Rightarrow & (i\, E_1-i\, E_2)^2 > 0 \Leftrightarrow \; \mbox{tr}[H]^2< 4\,  \mbox{det}[H]\; , \nonumber  \\ 
 & &  \label{enconstr7}\\
Q=P \; , \;  H^+ = - \eta \, H \, \eta^{-1} & \Rightarrow &  (i\, E_1-i\, E_2)^2 < 0 \Leftrightarrow \; \mbox{tr}[H]^2> 4\,  \mbox{det}[H] \; . \nonumber  \\ 
 & &  \label{enconstr8} \end{eqnarray}
The --- certainly interesting --- discussion of the exceptional point occuring for $\mbox{tr}[H]^2= 4\,  \mbox{det}[H]$ is beyond the scope of this work.
\newpage
\section{Diagonalization of $H$ and $H^+$}  \label{xxsec6}

\noindent The invertible two-dimensional matrix $\eta$ represents the metric operator for the pseudo-Hermitian Hamilton operator ($H^+ = + \eta \, H \, \eta^{-1}\,$) or anti-pseudo-Hermitian Hamilton operator ($H^+ = - \eta \, H \, \eta^{-1}\,$). The major purpose of this work is to identify the most general term for the metric $\eta$ allowed by the two-dimensional matrices $H$ and $H^+$ and to study its properties.

In order to achieve this we want to diagonalize the matrix $H$  in Eq.\ (\ref{hamop1}) with the help of Eq.\ (\ref{diagmat1}), i.\ e.:
\begin{eqnarray} 
 H  \;  = \;  \left(\begin{array}{cc} H_{11} & H_{12} \\ H_{21} & H_{22} \end{array}\right) & = & \frac{E_1+E_2}{2} \; 1_2 +  \frac{E_1-E_2}{2} \; \vec{n} \cdot \vec{\sigma} \nonumber \\[1mm]
 & = &  \frac{E_1+E_2}{2} \; 1_2 +  \frac{E_1-E_2}{2} \; X^{-1}\; \sigma_3 \; X \nonumber \\[1mm]
 & = & X^{-1} \; \left[  \frac{E_1+E_2}{2} \; 1_2 +  \frac{E_1-E_2}{2} \; \sigma_3 \, \right] \; X \, ,  \label{hamop4}
\end{eqnarray}
with (see Eqs.\ (\ref{xxeq1}) and  (\ref{xxeq2}))
\begin{eqnarray} 
X & = & \frac{(1 + n_3 )^{\frac{1}{2}}}{\sqrt{2}} \;   \frac{1_2  + \sigma_3 \, \vec{n}\cdot \vec{\sigma}}{1 + n_3} \label{yyyy1} \\[2mm]
  & = &  \frac{ (1 + n_3 )^{\frac{1}{2}}}{\sqrt{2}} \ \left(1_2  + \frac{\sigma_3\; (n_1\, \sigma_1 +\,  n_2 \,\sigma_2)}{1 + n_3}  \right)  \; , \label{xxxx1} \\[2mm]
X^{-1} & = &  \frac{(1 + n_3 )^{\frac{1}{2}}}{\sqrt{2}} \;   \frac{1_2  + \vec{n}\cdot \vec{\sigma}\; \sigma_3}{1 + n_3} \label{iiii1} \\[2mm]
  & = & \frac{ (1 + n_3 )^{\frac{1}{2}}}{\sqrt{2}} \ \left(1_2  + \frac{(n_1\, \sigma_1 +\,  n_2 \,\sigma_2)\; \sigma_3}{1 + n_3}  \right) \; ,
 \end{eqnarray}
or, equivalently,
\begin{eqnarray} 
X & = & \left(\frac{\frac{E_1-E_2}{2}+\frac{\mbox{tr}[H\sigma_3]}{2}}{2 \cdot \frac{E_1-E_2}{2}} \right)^{\frac{1}{2}}    \frac{\frac{E_1-E_2}{2} \; 1_2  + \sigma_3\left(H - \frac{\mbox{tr}[H]}{2}\; 1_2\right)}{\frac{E_1-E_2}{2}+\frac{\mbox{tr}[H\sigma_3]}{2}}  \\
  & = & \left(\frac{\frac{E_1-E_2}{2}+\frac{\mbox{tr}[H\sigma_3]}{2}}{2 \cdot \frac{E_1-E_2}{2}} \right)^{\frac{1}{2}}   \left(1_2  + \frac{\sigma_3\left(H - \frac{\mbox{tr}[H]}{2}\; 1_2- \frac{\mbox{tr}[H\sigma_3]}{2} \; \sigma_3\right)}{\frac{E_1-E_2}{2}+\frac{\mbox{tr}[H\sigma_3]}{2}}  \right)  , \nonumber \\[2mm]
 & & \\
X^{-1} & = & \left(\frac{\frac{E_1-E_2}{2}+\frac{\mbox{tr}[H\sigma_3]}{2}}{2 \cdot \frac{E_1-E_2}{2}} \right)^{\frac{1}{2}}    \frac{\frac{E_1-E_2}{2} \; 1_2  + \left(H - \frac{\mbox{tr}[H]}{2}\; 1_2\right)\,\sigma_3}{\frac{E_1-E_2}{2}+\frac{\mbox{tr}[H\sigma_3]}{2}}  \\
  & = & \left(\frac{\frac{E_1-E_2}{2}+\frac{\mbox{tr}[H\sigma_3]}{2}}{2 \cdot \frac{E_1-E_2}{2}} \right)^{\frac{1}{2}}   \left(1_2  + \frac{\left(H - \frac{\mbox{tr}[H]}{2}\; 1_2- \frac{\mbox{tr}[H\sigma_3]}{2} \; \sigma_3\right)\,\sigma_3 }{\frac{E_1-E_2}{2}+\frac{\mbox{tr}[H\sigma_3]}{2}}  \right)  , \nonumber \\[2mm]
 & & 
 \end{eqnarray}
i.\ e.,
\begin{eqnarray} 
X  & = & \left(\frac{\frac{E_1-E_2}{2}+\frac{H_{11}-H_{22}}{2}}{2 \cdot \frac{E_1-E_2}{2}} \right)^{\frac{1}{2}}   \left(1_2  - \frac{\left(\begin{array}{cc} 0 & -\,H_{12} \\ H_{21} & 0 \end{array}\right)}{\frac{E_1-E_2}{2}+\frac{H_{11}-H_{22}}{2}}  \right) \label{xxfin1} \\
 & = & \left(\frac{\frac{E_1-E_2}{2}-\frac{H_{11}-H_{22}}{2}}{2 \cdot \frac{E_1-E_2}{2}\; H_{12}\, H_{21}} \right)^{\frac{1}{2}} \left(\frac{E_1-E_2}{2} \; 1_2  +  \left(\begin{array}{cc} \frac{H_{11}-H_{22}}{2} & H_{12} \\[2mm] -\, H_{21} & \frac{H_{11}-H_{22}}{2} \end{array}\right) \right) , \nonumber \\[2mm]
 & & \\
X^{-1}  & = & \left(\frac{\frac{E_1-E_2}{2}+\frac{H_{11}-H_{22}}{2}}{2 \cdot \frac{E_1-E_2}{2}} \right)^{\frac{1}{2}}   \left(1_2  - \frac{\left(\begin{array}{cc} 0 & H_{12} \\ -\,H_{21} & 0 \end{array}\right)}{\frac{E_1-E_2}{2}+\frac{H_{11}-H_{22}}{2}}  \right)   \label{xxfin2} \\
 & = & \left(\frac{\frac{E_1-E_2}{2}-\frac{H_{11}-H_{22}}{2}}{2 \cdot \frac{E_1-E_2}{2}\; H_{12}\, H_{21}} \right)^{\frac{1}{2}}  \left(\frac{E_1-E_2}{2} \; 1_2  +  \left(\begin{array}{cc} \frac{H_{11}-H_{22}}{2} & -\,H_{12} \\[2mm]  H_{21} & \frac{H_{11}-H_{22}}{2} \end{array}\right) \right) . \nonumber \\[2mm]
 & &
 \end{eqnarray}
The matrix $H^+$ in Eq.\ (\ref{hamop3}) is diagonalized analogously:

\begin{eqnarray} 
 H^+  \;  = \;  \left(\begin{array}{cc} H^\ast_{11} & H^\ast_{21} \\ H^\ast_{12} & H^\ast_{22} \end{array}\right) & = & \frac{E^\ast_1+E^\ast_2}{2} \; 1_2 +  \frac{E^\ast_1-E^\ast_2}{2} \;  \vec{n}^\ast \cdot \vec{\sigma} \nonumber \\[1mm]
  & = & \frac{E^\ast_1+E^\ast_2}{2} \; 1_2 +  \frac{E^\ast_1-E^\ast_2}{2} \;   X^+\; \sigma_3 \; \left(X^+\right)^{-1} \nonumber \\[1mm]
 & = &  X^+\, \left[  \frac{E^\ast_1+E^\ast_2}{2} \; 1_2 +  \frac{E^\ast_1-E^\ast_2}{2} \; \sigma_3 \, \right] \, \left(X^+\right)^{-1} \quad \label{hamop5}
\end{eqnarray}
with
\begin{eqnarray} 
X^+ & = & \frac{(1 + n^\ast_3 )^{\frac{1}{2}}}{\sqrt{2}} \;   \frac{1_2  + \vec{n}^\ast\cdot \vec{\sigma}\;\, \sigma_3}{1 + n^\ast_3} \label{yyyy2} \\[2mm]
  & = &  \frac{ (1 + n^\ast_3 )^{\frac{1}{2}}}{\sqrt{2}} \ \left(1_2  + \frac{(n^\ast_1\, \sigma_1 +\,  n^\ast_2 \,\sigma_2)\; \sigma_3}{1 + n^\ast_3}  \right)  \; , \label{xxxx2} \\[2mm]
\left(X^+\right)^{-1} & = &  \frac{(1 + n^\ast_3 )^{\frac{1}{2}}}{\sqrt{2}} \;   \frac{1_2  + \sigma_3\;\vec{n}^\ast\cdot \vec{\sigma}}{1 + n^\ast_3} \label{iiii2} \\[2mm]
  & = & \frac{ (1 + n^\ast_3 )^{\frac{1}{2}}}{\sqrt{2}} \ \left(1_2  + \frac{\sigma_3\; (n^\ast_1\, \sigma_1 +\,  n^\ast_2 \,\sigma_2)}{1 + n^\ast_3}  \right) \; ,
 \end{eqnarray}
or, equivalently,
\begin{eqnarray} 
X^+ & = & \left(\frac{\frac{E^\ast_1-E^\ast_2}{2}+\frac{\mbox{tr}[H^+\sigma_3]}{2}}{2 \cdot \frac{E^\ast_1-E^\ast_2}{2}} \right)^{\frac{1}{2}}    \frac{\frac{E^\ast_1-E^\ast_2}{2} \; 1_2  +\left(H^+ - \frac{\mbox{tr}[H^+]}{2}\; 1_2\right) \sigma_3}{\frac{E_1-E_2}{2}+\frac{\mbox{tr}[H\sigma_3]}{2}}  \\
  & = & \left(\frac{\frac{E^\ast_1-E^\ast_2}{2}+\frac{\mbox{tr}[H^+\sigma_3]}{2}}{2 \cdot \frac{E^\ast_1-E^\ast_2}{2}} \right)^{\frac{1}{2}} \nonumber \\
 & \cdot &  \left(1_2  + \frac{\left(H^+ - \frac{\mbox{tr}[H^+]}{2}\; 1_2- \frac{\mbox{tr}[H^+\sigma_3]}{2} \; \sigma_3\right)\sigma_3}{\frac{E^\ast_1-E^\ast_2}{2}+\frac{\mbox{tr}[H^+\sigma_3]}{2}}  \right)  ,  \\[2mm]
 & & \nonumber \\
\left(X^+\right)^{-1} & = & \left(\frac{\frac{E^\ast_1-E^\ast_2}{2}+\frac{\mbox{tr}[H^+\sigma_3]}{2}}{2 \cdot \frac{E^\ast_1-E^\ast_2}{2}} \right)^{\frac{1}{2}}    \frac{\frac{E^\ast_1-E^\ast_2}{2} \; 1_2  + \sigma_3\left(H^+ - \frac{\mbox{tr}[H^+]}{2}\; 1_2\right)}{\frac{E^\ast_1-E^\ast_2}{2}+\frac{\mbox{tr}[H^+\sigma_3]}{2}}  \\
  & = & \left(\frac{\frac{E^\ast_1-E^\ast_2}{2}+\frac{\mbox{tr}[H^+\sigma_3]}{2}}{2 \cdot \frac{E^\ast_1-E^\ast_2}{2}} \right)^{\frac{1}{2}} \nonumber \\
 & \cdot &   \left(1_2  + \frac{\sigma_3\left(H^+ - \frac{\mbox{tr}[H^+]}{2}\; 1_2- \frac{\mbox{tr}[H^+\sigma_3]}{2} \; \sigma_3\right)}{\frac{E^\ast_1-E^\ast_2}{2}+\frac{\mbox{tr}[H^+\sigma_3]}{2}}  \right)  , 
\end{eqnarray}
i.\ e.,
\begin{eqnarray} 
X^+  & = & \left(\frac{\frac{E^\ast_1-E^\ast_2}{2}+\frac{H^\ast_{11}-H^\ast_{22}}{2}}{2 \cdot \frac{E^\ast_1-E^\ast_2}{2}} \right)^{\frac{1}{2}}   \left(1_2  - \frac{\left(\begin{array}{cc} 0 & H^\ast_{21} \\ -\, H^\ast_{12} & 0 \end{array}\right)}{\frac{E^\ast_1-E^\ast_2}{2}+\frac{H^\ast_{11}-H^\ast_{22}}{2}}  \right)   \label{xxfin3} \\
 & = & \left(\frac{\frac{E^\ast_1-E^\ast_2}{2}-\frac{H^\ast_{11}-H^\ast_{22}}{2}}{2 \cdot \frac{E^\ast_1-E^\ast_2}{2}\; H^\ast_{12}\, H^\ast_{21}} \right)^{\frac{1}{2}} \nonumber \\
 & \cdot &  \left(\frac{E^\ast_1-E^\ast_2}{2} \; 1_2  +  \left(\begin{array}{cc} \frac{H^\ast_{11}-H^\ast_{22}}{2} & -\,H^\ast_{21} \\[2mm] H^\ast_{12} & \frac{H^\ast_{11}-H^\ast_{22}}{2} \end{array}\right) \right) ,  \\[2mm]
 & & \nonumber \\[2mm]
\left(X^+\right)^{-1}  & = & \left(\frac{\frac{E^\ast_1-E^\ast_2}{2}+\frac{H^\ast_{11}-H^\ast_{22}}{2}}{2 \cdot \frac{E^\ast_1-E^\ast_2}{2}} \right)^{\frac{1}{2}}   \left(1_2  - \frac{\left(\begin{array}{cc} 0 & -\, H^\ast_{21} \\ H^\ast_{12} & 0 \end{array}\right)}{\frac{E^\ast_1-E^\ast_2}{2}+\frac{H^\ast_{11}-H^\ast_{22}}{2}}  \right)   \label{xxfin4} \\
 & = & \left(\frac{\frac{E^\ast_1-E^\ast_2}{2}-\frac{H^\ast_{11}-H^\ast_{22}}{2}}{2 \cdot \frac{E^\ast_1-E^\ast_2}{2}\; H^\ast_{12}\, H^\ast_{21}} \right)^{\frac{1}{2}} \nonumber \\
 & \cdot &  \left(\frac{E^\ast_1-E^\ast_2}{2} \; 1_2  +  \left(\begin{array}{cc} \frac{H^\ast_{11}-H^\ast_{22}}{2} & H^\ast_{21} \\[2mm] -\, H^\ast_{12} & \frac{H^\ast_{11}-H^\ast_{22}}{2} \end{array}\right) \right) . 
\end{eqnarray}
The matrices $H$ and $H^+$ are representing the Hamilton operator and its Hermitian conjugate in the vector spaces spanned by their respective left and right eigen-basis. Although the normalizations of the eigen-vectors are not fixed absolutely, it is well known that the left and right eigen-basis can be made mutually orthogonal. The freedom of normalizing the eigen-vectors is expressed in our work by a suitable two-dimensional diagonal matrix $N$ and its inverse $N^{-1}$, by which the left and right eigen-basis of $H$ are multiplied:
\begin{equation} N\; = \; \left(\begin{array}{cc} N_1 & 0 \\ 0 & N_2 \end{array}\right) \; , \qquad  N^{-1}\; = \; \left(\begin{array}{cc} \frac{1}{N_1} & 0 \\ 0 & \frac{1}{N_2} \end{array}\right)\; .  \label{xnorm1} \end{equation} 
After inserting these renormalization matrices $N$ and $N^{-1}$ the matrix $H$ and $H^+$ in Eq.\ (\ref{hamop4}) and (\ref{hamop5}) take the following form (see Eqs.\ (\ref{xxyyzz1}) and (\ref{xhdagger1})):

\begin{eqnarray} 
H & = &    (N\,X)^{-1}  \left[  \frac{E_1+E_2}{2} \; 1_2 +  \frac{E_1-E_2}{2} \; \sigma_3 \, \right] \; N \,X \; , \label{hamop6} \nonumber\\[2mm]
H^+ & = &  (N\, X)^+\;  \left[  \frac{E^\ast_1+E^\ast_2}{2} \; 1_2 +  \frac{E^\ast_1-E^\ast_2}{2} \; \sigma_3 \, \right] \; \left((N\, X)^+\right)^{-1} \; . \label{hamop7}
\end{eqnarray}
\section{The metric $\eta$ for the trivial phase ($Q=1_2$)} \label{xmetrici}
\noindent  The metric for the trivial phase  (denoted often $\eta_+$) was derived in Eq.\ (\ref{pseudherm1}):
\begin{equation}   \eta \; = \; (N\,X)^+    N \, X \; = \; X^+ \; N^+  N \; X\; .  
\end{equation}
This metric $\eta$ is for $N\not= 0$ clearly positive semi-definite and depending on the chosen renormalization of the eigen-vectors Eqs.(\ref{xnorm1})  of $H$ via the diagonal matrix $N^+ N$ which shall be rewritten here for later convenience:
\begin{eqnarray}
 N^+ N \; = \; \left(\begin{array}{cc} |N_1|^2 & 0 \\ 0 & |N_2|^2 \end{array}\right)  & = & \frac{|N_1|^2+|N_2|^2}{2} \; 1_2 +  \frac{|N_1|^2-|N_2|^2}{2}  \; \sigma_3 \nonumber \\
 & = & \frac{\mbox{tr}[N^+N]}{2} \; 1_2 +  \frac{\mbox{tr}[N^+\sigma_3 \, N]}{2}  \; \sigma_3 \; . \label{nnnident1}
\end{eqnarray}
With the help of this relation we obtain the Hermitian metric $\eta \; = \; \eta_0  + \eta_3 =\eta^+$ for the trivial phase ($Q=1_2$) as a linear superposition of two Hermitian matrices $\eta_0=\eta^+_0$ and $\eta_3=\eta^+_3$ given by
\begin{eqnarray} 
 \eta_0  & = & \frac{|N_1|^2+|N_2|^2}{2} \; X^+  X \quad \;\; \; = \; \frac{\mbox{tr}[N^+N]}{2} \; X^+ \; X \; , \label{etafin0} \\[2mm]
 \eta_3  & = & \frac{|N_1|^2-|N_2|^2}{2}  \; X^+ \; \sigma_3 \; X \; = \; \frac{\mbox{tr}[N^+\sigma_3 \, N]}{2}  \; X^+ \; \sigma_3 \; X \; . \label{etafin3}
\end{eqnarray}
Insertion of Eq.\ (\ref{yyyy1}) and (\ref{yyyy2}) yields
\begin{eqnarray} 
 \eta_0 & = & \frac{|N_1|^2+|N_2|^2}{2} \; \frac{\mbox{\small $\bigcirc\hspace{-3.2mm}\pm$}\; 1}{2\; |1+n_3|} \;  \Big(1_2  +  \vec{n}^\ast \cdot \vec{\sigma} \;\, \sigma_3\Big)\;   1_2  \;\Big(1_2  + \sigma_3 \, \vec{n}\cdot \vec{\sigma}\Big) \; , \\[2mm]
 \eta_3  & = & \frac{|N_1|^2-|N_2|^2}{2} \; \frac{\mbox{\small $\bigcirc\hspace{-3.2mm}\pm$}\; 1}{2\; |1+n_3|} \;  \Big(1_2  +  \vec{n}^\ast \cdot \vec{\sigma} \;\, \sigma_3\Big)\;   \sigma_3  \;\Big(1_2  + \sigma_3 \, \vec{n}\cdot \vec{\sigma}\Big) \;  . 
\end{eqnarray}
Here we applied the identity $\left(|1 + n_3|^2 \right)^{\frac{1}{2}} =  \mbox{\small $\bigcirc\hspace{-3.2mm}\pm$} \; |1 + n_3|$ keeping in mind that a power with an exponent $1/2$ is ambiguous. We use the encircled notation $\mbox{\small $\bigcirc\hspace{-3.2mm}\pm$}$ in order to underline the difference to the notation $\pm$ which we will use elsewhere. 
For $\mbox{tr}[N^+\sigma_3\, N]=|N_1|^2-|N_2|^2=0$ there would obviously hold $\eta = \eta_0$ indicating not only that $\eta_0$ is Hermitian (i.\ e., $\eta_0=\eta^+_0$), yet --- for $N\not= 0$ and the sign choice $\oplus$ --- also positive semi-definite. In defining $\vec{e}_3 =(0,0,1)^T$ the metric $\eta_0$ can be expressed in the following way:
\begin{eqnarray} 
 \eta_0  & = &  \frac{|N_1|^2+|N_2|^2}{2} \; \frac{\mbox{\small $\bigcirc\hspace{-3.2mm}\pm$}\; 1}{2\; |1+n_3|} \;  \Big(\vec{e}_3  +  \vec{n}^\ast   \Big)\!\cdot \vec{\sigma} \; \; \Big(\vec{e}_3  +  \vec{n}   \Big)\!\cdot \vec{\sigma} \; . \; \label{xeta0}
\end{eqnarray}
The second --- also Hermitian --- term $\eta_3=\eta_3^+$  in the metric contributing to the metric for $\mbox{tr}[N^+\sigma_3\, N]=|N_1|^2-|N_2|^2\not=0$, being obviously indefinite due the opposite signs of the eigen-values $+1$ and $-1$ of the Pauli matrix $\sigma_3$ and not changing the positive semi-definiteness of the overall metric $\eta=\eta_0+\eta_3$ is given by:
\begin{eqnarray} 
\eta_3  & = &  \frac{|N_1|^2-|N_2|^2}{2} \;\; \frac{\mbox{\small $\bigcirc\hspace{-3.2mm}\pm$}\; 1}{2\; |1+n_3|} \;  \Big(\vec{e}_3  +  \vec{n}^\ast   \Big)\!\cdot \vec{\sigma} \; \;\vec{e}_3 \cdot \vec{\sigma} \;\; \Big(\vec{e}_3  +  \vec{n}   \Big)\!\cdot \vec{\sigma}\;  . \label{xeta3}
\end{eqnarray}
For convenience we want to rewrite --- with the help of Eqs.\ (\ref{uvident1}), (\ref{uvident2})   --- the expressions for $\eta_0$ (see Eq.\ (\ref{xeta0})) and $\eta_3$ (see Eq.\ (\ref{xeta3})) in the following way:
\newpage
\begin{eqnarray} 
 \eta_0   & = &  \frac{|N_1|^2+|N_2|^2}{2} \; \frac{\mbox{\small $\bigcirc\hspace{-3.2mm}\pm$}\; 1}{2\; |1+n_3|} \;\cdot \nonumber \\[2mm]
 & \cdot & \Big( (\vec{e}_3  +  \vec{n}^\ast   )\!\cdot  (\vec{e}_3  +  \vec{n}  )\; 1_2 + i\, \vec{\sigma} \cdot \left[ \, (\vec{e}_3  +  \vec{n}^\ast   )\!\times  \vec{e}_3  +  \vec{n}  ) \, \right] \Big)  \; ,\quad \\[2mm]
 \eta_3  & = &  \frac{|N_1|^2-|N_2|^2}{2} \; \frac{\mbox{\small $\bigcirc\hspace{-3.2mm}\pm$}\; 1}{2\; |1+n_3|} \; \cdot \nonumber \\[2mm]
 & \cdot & \Big((1  +  n_3^\ast) \;\, (\sigma_3  +  \vec{n} \cdot \vec{\sigma}) + (1  +  n_3) \;\, (\sigma_3  +  \vec{n}^\ast \cdot \vec{\sigma}) \nonumber \\[2mm]
 & & - \,(\vec{e}_3  +  \vec{n}^\ast ) \cdot (\vec{e}_3  +  \vec{n} )\, \sigma_3   -  i\, \vec{e}_3 \cdot [\,\vec{n}^\ast\times \vec{n} \, ] \; 1_2 \Big) . \; 
\end{eqnarray}
Although these expressions can be used to determine $\eta_0$ und $\eta_3$ it is more pragmatic to calculate $\eta_0$ und $\eta_3$ by applying Eq.\ (\ref{xxfin1}) und (\ref{xxfin3}) to Eqs.\ (\ref{etafin0}) and (\ref{etafin3}), i.\ e.,
\begin{eqnarray} 
\eta_0  & = & \mbox{\small $\bigcirc\hspace{-3.2mm}\pm$}\;\;\frac{|N_1|^2+|N_2|^2}{2} \;\;\frac{1}{2} \,  \left|  \frac{\frac{H_{11}-H_{22}}{2}+\frac{E_1-E_2}{2}}{\frac{E_1-E_2}{2}}\right| \; \cdot \nonumber  \\[2mm]
 & \cdot &  \left(1_2  - \frac{\left(\begin{array}{cc} 0 & H^\ast_{21} \\ -\,H^\ast_{12} & 0 \end{array}\right)}{\frac{H^\ast_{11}-H^\ast_{22}}{2}+\frac{E^\ast_1-E^\ast_2}{2}}  \right)  \left(1_2  - \frac{\left(\begin{array}{cc} 0 & -\,H_{12} \\ H_{21} & 0 \end{array}\right)}{\frac{H_{11}-H_{22}}{2}+\frac{E_1-E_2}{2}}  \right)  ,  \label{etazzz0} \\[2mm]
\eta_3  & = & \mbox{\small $\bigcirc\hspace{-3.2mm}\pm$}\;\;\frac{|N_1|^2-|N_2|^2}{2} \;\;\frac{1}{2} \,  \left|  \frac{\frac{H_{11}-H_{22}}{2}+\frac{E_1-E_2}{2}}{\frac{E_1-E_2}{2}}\right| \; \cdot \nonumber  \\[2mm]
 & \cdot &  \left(1_2  - \frac{\left(\begin{array}{cc} 0 & H^\ast_{21} \\ -\,H^\ast_{12} & 0 \end{array}\right)}{\frac{H^\ast_{11}-H^\ast_{22}}{2}+\frac{E^\ast_1-E^\ast_2}{2}}  \right) \left(\begin{array}{cc} 1 & 0 \\ 0 & -1 \end{array}\right) \left(1_2  - \frac{\left(\begin{array}{cc} 0 & -\,H_{12} \\ H_{21} & 0 \end{array}\right)}{\frac{H_{11}-H_{22}}{2}+\frac{E_1-E_2}{2}}  \right)  . \nonumber\\[2mm] 
\label{etazzz3} \end{eqnarray}
The most efficient way to calculate the metric (and its inverse) is to make use of the following representation of the matrices $X$,  $X^+$ (and  $X^{-1}$, $\left(X^+\right)^{-1}$) underlying Eqs.\ (\ref{xxfin1}), (\ref{xxfin3}) (and (\ref{xxfin2}),  (\ref{xxfin4})), respectively:

\begin{eqnarray} X  & = &  \left(\frac{\frac{E_1-E_2}{2} \,(1 + n_3 )}{2\cdot \frac{E_1-E_2}{2}} \right)^{\frac{1}{2}} \; \left(1_2  - \frac{\left(\begin{array}{cc} 0 & -\,H_{12} \\ H_{21} & 0 \end{array}\right)}{\frac{E_1-E_2}{2} \, (1 + n_3)}  \right)  \; , \label{xxxident1} \\[2mm]
 X^{-1}  & = &  \left(\frac{\frac{E_1-E_2}{2} \,(1 + n_3 )}{2\cdot \frac{E_1-E_2}{2}} \right)^{\frac{1}{2}} \; \left(1_2  + \frac{\left(\begin{array}{cc} 0 & -\,H_{12} \\ H_{21} & 0 \end{array}\right)}{\frac{E_1-E_2}{2} \, (1 + n_3)}  \right)  \; ,  \label{xxxident2}\\[2mm]
 X^+  & = &  \left(\frac{\frac{E^\ast_1-E^\ast_2}{2} \,(1 + n^\ast_3 )}{2\cdot \frac{E^\ast_1-E^\ast_2}{2}} \right)^{\frac{1}{2}} \; \left(1_2  - \frac{\left(\begin{array}{cc} 0 & H^\ast_{21} \\ -\, H^\ast_{12} & 0 \end{array}\right)}{\frac{E^\ast_1-E^\ast_2}{2} \, (1 + n^\ast_3)}  \right)  \; ,  \label{xxxident3}\\[2mm]
 \left(X^+\right)^{-1}  & = &  \left(\frac{\frac{E^\ast_1-E^\ast_2}{2} \,(1 + n^\ast_3 )}{2\cdot \frac{E^\ast_1-E^\ast_2}{2}} \right)^{\frac{1}{2}} \; \left(1_2  + \frac{\left(\begin{array}{cc} 0 & H^\ast_{21} \\ -\, H^\ast_{12} & 0 \end{array}\right)}{\frac{E^\ast_1-E^\ast_2}{2}\;  (1 + n^\ast_3)}  \right)  \; . \label{xxxident4}
\end{eqnarray}
Invoking Eqs.\ (\ref{xxxident1}) and (\ref{xxxident3}) to Eqs.\ (\ref{etafin0}) and (\ref{etafin3}) leads to the following identities, which can be used to calculate the metric $\eta_0$ and $\eta_3$  --- to our present experience --- most efficiently:
\begin{eqnarray} 
\eta_0  & = & \frac{|N_1|^2+|N_2|^2}{2} \; X^+  X \nonumber \\[2mm]
 &=  & \mbox{\small $\bigcirc\hspace{-3.2mm}\pm$} \; \; \frac{|N_1|^2+|N_2|^2}{2} \; \;\frac{1}{2\; \left|  \frac{E_1-E_2}{2}\right|} \; \; \frac{1}{\left| \frac{E_1-E_2}{2} \, (1+n_3)\right|} \; \cdot \nonumber\\
 & \cdot & \Bigg\{\left(\begin{array}{cc} \left| \frac{E_1-E_2}{2} \, (1+n_3)\right|^2 + |H_{21}|^2 & 0 \\[2mm] 0 & \left| \frac{E_1-E_2}{2} \, (1+n_3)\right|^2 + |H_{12}|^2 \end{array}\right) \nonumber \\[2mm]
 & & - \,\; \mbox{Re} \left[ \; \frac{E_1-E_2}{2} \, (1+n_3) \; \right] \; \left(\begin{array}{cc} 0 & -(H_{12} -\, H^\ast_{21}) \\[2mm]  H_{21} -\, H^\ast_{12}  & 0 \end{array}\right) \nonumber \\[2mm]
 & & - \, i  \; \mbox{Im} \left[ \; \frac{E_1-E_2}{2} \, (1+n_3) \; \right] \; \left(\begin{array}{cc} 0 & H_{12} + H^\ast_{21} \\[2mm]  -(H_{21} + H^\ast_{12})  & 0 \end{array}\right)\Bigg\} \; , \nonumber \\[2mm]
 & & \\
\eta_3  & = & \frac{|N_1|^2-|N_2|^2}{2} \; X^+ \sigma_3 \,  X \nonumber \\[2mm]
 &=  & \mbox{\small $\bigcirc\hspace{-3.2mm}\pm$} \; \; \frac{|N_1|^2-|N_2|^2}{2} \; \;\frac{1}{2\; \left|  \frac{E_1-E_2}{2}\right|} \; \; \frac{1}{\left| \frac{E_1-E_2}{2} \, (1+n_3)\right|} \; \cdot \nonumber\\
 & \cdot & \Bigg\{\left(\begin{array}{cc} \left| \frac{E_1-E_2}{2} \, (1+n_3)\right|^2 - |H_{21}|^2 & 0 \\[2mm] 0 & -\, \left| \frac{E_1-E_2}{2} \, (1+n_3)\right|^2 + |H_{12}|^2 \end{array}\right) \nonumber \\[2mm]
 & & + \,\; \mbox{Re} \left[ \; \frac{E_1-E_2}{2} \, (1+n_3) \; \right] \; \left(\begin{array}{cc} 0 & H_{12} + H^\ast_{21} \\[2mm]  H_{21} + H^\ast_{12}  & 0 \end{array}\right) \nonumber \\[2mm]
 & & - \, i  \; \mbox{Im} \left[ \; \frac{E_1-E_2}{2} \, (1+n_3) \; \right] \; \left(\begin{array}{cc} 0 & H_{12} - H^\ast_{21} \\[2mm]  H_{21} - H^\ast_{12}  & 0 \end{array}\right)\Bigg\} \; . 
\end{eqnarray}

\section{The metric $\eta$ for the non-trivial phase ($Q=P$)} \label{xmetricp}
\noindent The metric for the non-trivial phase  ($Q=P$) being obviously indefinite due to the indefiniteness of $P$ has been derived in Eq.\ (\ref{pseudherm2}), i.e.,
\begin{equation}   \eta \; = \; (N\,X)^+  P \,  N \, X \; = \; X^+ \, N^+ P \,  N \; X\; . 
\end{equation}
As pointed out in Eq.\ (\ref{pha2})  the two-dimensional permutation matrix $P$ interchanging the eigen-values $+1$ and $-1$ of the Pauli matrix $\sigma_3$ should anti-commute with $\sigma_3$, i.\ e.,
\begin{equation} P \;\, \sigma_3 \;\, P^{-1} \; = \; -\, \sigma_3 \; .
\end{equation}
Hence, we chose the most general matrix $P$ anti-commuting with the Pauli matrix $\sigma_3$ and respecting $P=P^{-1}$. It has the form $P=\vec{e}\cdot \vec{\sigma} = e_1 \,\sigma_1 + e_2\,\sigma_2$ with  $\vec{e}$ being some arbitrary unit vector ($\vec{e}\cdot \vec{e}=1$) with $e_3=0$. With the help of this matrix the metric in the non-trivial phase takes the following form:
\begin{eqnarray} \eta  & = &  X^+\;   N^+ \vec{e} \cdot \vec{\sigma}\, N  \;X  \; , 
\end{eqnarray}
which is Hermitian only for a real-valued unit-vector $\vec{e}$, i.\ e.\ for $\vec{e}=\vec{e}^{\,\ast}$ implying $P=P^+$. In the following we assume $P\not=P^+$, i.\ e.,  $\vec{e}\not=\vec{e}^{\,\ast}$.

The dependence of the metric on the renormalization matrix $N$ (see Eq.\ (\ref{xnorm1})) is here  also manifest in the term $N^+\vec{e} \cdot \vec{\sigma}\, N$, yet even more involved. For later convenience we want to reformulate this term slightly:
\begin{eqnarray}
\lefteqn{ N^+ P\, N\; = \; N^+\vec{e} \cdot \vec{\sigma}\, N\; = } \nonumber \\[3mm]
  & = & \left(\frac{N^\ast_1+N^\ast_2}{2} \; 1_2 +  \frac{N^\ast_1-N^\ast_2}{2}  \; \sigma_3 \right) \vec{e} \cdot \vec{\sigma} \left(\frac{N_1+N_2}{2} \; 1_2 +  \frac{N_1-N_2}{2}  \; \sigma_3 \right) \nonumber \\
  & = & \left(\frac{N^\ast_1+N^\ast_2}{2} \; 1_2 +  \frac{N^\ast_1-N^\ast_2}{2}  \; \sigma_3 \right) \left(\frac{N_1+N_2}{2} \; 1_2 -  \frac{N_1-N_2}{2}  \; \sigma_3 \right) \vec{e} \cdot \vec{\sigma} \nonumber \\
  & = & \left(\frac{N^\ast_2\,N_1+N^\ast_1\,N_2}{2} \; 1_2 +  \frac{N^\ast_2\,N_1-N^\ast_1\,N_2}{2\, i}  \,\; i\, \sigma_3 \right)  \vec{e} \cdot \vec{\sigma} \nonumber \\
 & = & \left( \frac{\mbox{tr}[N^+\sigma_1\,N\,\sigma_1]}{2} \; 1_2 +  \frac{\mbox{tr}[N^+\sigma_1 \, N\,\sigma_2]}{2}  \; i\, \sigma_3 \right)  \vec{e} \cdot \vec{\sigma} \nonumber \\
 & = & \frac{\mbox{tr}[N^+\sigma_1\,N\,\sigma_1]}{2} \;\, \vec{e} \cdot \vec{\sigma} +  \frac{\mbox{tr}[N^+\sigma_1 \, N\,\sigma_2]}{2}  \;\, \vec{e}_\perp \cdot \vec{\sigma}  \nonumber \\
 & = & \frac{\mbox{tr}[N^+\sigma_1\,N\,\sigma_1]}{2} \;\, P \; + \;  \frac{\mbox{tr}[N^+\sigma_1 \, N\,\sigma_2]}{2}  \;\, P_\perp \; . \label{nnnid1}
\end{eqnarray}
Here we have introduced an ``orthogonal" permutation matrix $P_\perp$ by $P_\perp= i\, \sigma_3\; P$, i.\ e.,
\begin{equation} P_\perp \; = \;   i\, \sigma_3\; \vec{e} \cdot \vec{\sigma} \; = \; e_2\, \sigma_1 - e_1 \, \sigma_2 \; = \; \vec{e}_\perp \cdot \vec{\sigma} \; , \end{equation}
with $\vec{e}_\perp=(e_2,-\,e_1,0)^T=\vec{e}\,\times\, \vec{e}_3$ being a unit vector respecting the identities $\vec{e}_\perp\cdot \,\vec{e}_\perp = 1$ and $\vec{e}\,\cdot\, \vec{e}_\perp = 0$. The unit vectors $\vec{e}_\perp$, $\vec{e}$ and $\vec{e}_3$ with $\vec{e}_3 =(0,0,1)^T$ forming a orhonormal basis are related by the following identities:
\begin{equation} \vec{e}_\perp\; =\; \vec{e}\,\times\, \vec{e}_3 \; , \quad  \vec{e}_3\; =\; \vec{e}_\perp\,\times\, \vec{e} \; , \quad  \vec{e}\; =\; \vec{e}_3\,\times\, \vec{e}_\perp \; . \end{equation}
With the help of Eq.\ (\ref{nnnid1}) we obtain the metric $\eta \; = \; \eta_1  + \eta_2$ for the non-trivial phase ($Q=P$) as a linear superposition of two matrices $\eta_1$ and $\eta_2$ given by
\begin{eqnarray} 
 \eta_1  & = & \frac{N^\ast_2\,N_1+N^\ast_1\,N_2}{2} \; X^+  \; \vec{e} \cdot \vec{\sigma} \; X  \;\;\, = \; \frac{\mbox{tr}[N^+\sigma_1\,N\,\sigma_1]}{2} \; X^+\; \vec{e} \cdot \vec{\sigma} \; X  , \label{etafin1} \\[2mm]
 \eta_2  & = & \frac{N^\ast_2\,N_1-N^\ast_1\,N_2}{2\, i}  \; X^+ \; \vec{e}_\perp\! \cdot \vec{\sigma} \; X \; = \; \frac{\mbox{tr}[N^+\sigma_1\,N\,\sigma_2]}{2}  \; X^+ \; \vec{e}_\perp \!\cdot \vec{\sigma} \; X  . \quad \label{etafin2}
\end{eqnarray}
Insertion of Eq.\ (\ref{yyyy1}) and (\ref{yyyy2}) yields
\begin{eqnarray} 
 \eta_1 & = & \frac{N^\ast_2\,N_1+N^\ast_1\,N_2}{2} \; \frac{\mbox{\small $\bigcirc\hspace{-3.2mm}\pm$}\;1}{2\; |1+n_3|} \; \Big(1_2  +  \vec{n}^\ast \cdot \vec{\sigma} \;\, \sigma_3\Big)\;   \vec{e} \cdot \vec{\sigma}  \;\Big(1_2  + \sigma_3 \, \vec{n}\cdot \vec{\sigma}\Big) \nonumber \\[2mm] 
 & = & \frac{N^\ast_2\,N_1+N^\ast_1\,N_2}{2} \; \frac{-\,\mbox{\small $\bigcirc\hspace{-3.2mm}\pm$}\;1}{2\; |1+n_3|} \; \Big(\vec{e}_3  +  \vec{n}^\ast   \Big)\!\cdot \vec{\sigma} \;\;   \vec{e} \cdot \vec{\sigma}  \;\;\Big(\vec{e}_3  +  \vec{n}   \Big)\!\cdot \vec{\sigma} ,  \; \label{xeta1} \\[2mm]
\eta_2 & = & \frac{N^\ast_2\,N_1-N^\ast_1\,N_2}{2\, i}\;  \frac{\mbox{\small $\bigcirc\hspace{-3.2mm}\pm$}\;1}{2\; |1+n_3|} \; \Big(1_2  +  \vec{n}^\ast \cdot \vec{\sigma} \;\, \sigma_3\Big)\;   \vec{e}_\perp \cdot \vec{\sigma}  \;\Big(1_2  + \sigma_3 \, \vec{n}\cdot \vec{\sigma}\Big) \nonumber \\[2mm]
 & = & \frac{N^\ast_2\,N_1-N^\ast_1\,N_2}{2\, i} \; \frac{-\,\mbox{\small $\bigcirc\hspace{-3.2mm}\pm$}\;1}{2\; |1+n_3|} \; \Big(\vec{e}_3  +  \vec{n}^\ast   \Big)\!\cdot \vec{\sigma} \;\;   \vec{e}_\perp \cdot \vec{\sigma}  \;\;\Big(\vec{e}_3  +  \vec{n}   \Big)\!\cdot \vec{\sigma} . \label{xeta2}
\end{eqnarray}
We want to rewrite --- with the help of Eqs.\ (\ref{uvident1}), (\ref{uvident2})  and $\vec{e}_3\cdot \vec{e}=\vec{e}_3\cdot \vec{e}_\perp=0$ --- the expressions for $\eta_1$ (see Eq.\ (\ref{xeta1})) and $\eta_2$ (see Eq.\ (\ref{xeta2})) as follows:
\begin{eqnarray} 
 \eta_1 & = & \frac{N^\ast_2\,N_1+N^\ast_1\,N_2}{2} \;\;  \frac{\mbox{\small $\bigcirc\hspace{-3.2mm}\pm$}\;1}{2\; |1+n_3|} \;\cdot  \nonumber \\ 
& \cdot & \Big(   i\, \vec{e} \cdot [\,(\vec{e}_3  +  \vec{n}^\ast )\times (\vec{e}_3  +  \vec{n} )\, ] \, 1_2 + (\vec{e}_3  +  \vec{n}^\ast ) \cdot (\vec{e}_3  +  \vec{n} )\;\, \vec{e} \cdot \vec{\sigma} \nonumber \\
 & & - \,  \vec{n}  \cdot \vec{e}\;\; (\vec{e}_3  +  \vec{n}^\ast ) \cdot \vec{\sigma} -  \vec{n}^\ast  \cdot \vec{e}\;\; (\vec{e}_3  +  \vec{n} ) \cdot \vec{\sigma}   \, \Big)   \\[2mm]
 & = & \frac{N^\ast_2\,N_1+N^\ast_1\,N_2}{2} \;\;  \frac{\mbox{\small $\bigcirc\hspace{-3.2mm}\pm$}\;1}{2\; |1+n_3|} \;\cdot  \nonumber \\ 
& \cdot & \Big(   i\, [\,\vec{e}_3 \times \vec{e}_\perp \, ] \cdot [\,(\vec{e}_3  +  \vec{n}^\ast )\times (\vec{e}_3  +  \vec{n} )\, ] \, 1_2 + (\vec{e}_3  +  \vec{n}^\ast ) \cdot (\vec{e}_3  +  \vec{n} )\;\, \vec{e} \cdot \vec{\sigma} \nonumber \\
 & & - \,  \vec{n}  \cdot \vec{e}\;\; (\vec{e}_3  +  \vec{n}^\ast ) \cdot \vec{\sigma} -  \vec{n}^\ast  \cdot \vec{e}\;\; (\vec{e}_3  +  \vec{n} ) \cdot \vec{\sigma}   \, \Big)   \\[2mm]
 & = & \frac{N^\ast_2\,N_1+N^\ast_1\,N_2}{2} \;\;  \frac{\mbox{\small $\bigcirc\hspace{-3.2mm}\pm$}\;1}{2\; |1+n_3|} \;\cdot  \nonumber \\
& \cdot & \Big(    i\, \Big( (1+n^\ast_3) \; \vec{n} \cdot \vec{e} - (1+n_3) \; \vec{n}^\ast \cdot \vec{e}\Big)  \, 1_2 + (\vec{e}_3  +  \vec{n}^\ast ) \cdot (\vec{e}_3  +  \vec{n} )\;\, \vec{e} \cdot \vec{\sigma} \nonumber \\
 & & - \,  \vec{n}  \cdot \vec{e}\;\; (\vec{e}_3  +  \vec{n}^\ast ) \cdot \vec{\sigma} -  \vec{n}^\ast  \cdot \vec{e}\;\; (\vec{e}_3  +  \vec{n} ) \cdot \vec{\sigma}   \, \Big)  , \\[2mm]
\eta_2 & = & \frac{N^\ast_2\,N_1-N^\ast_1\,N_2}{2\, i} \;\;  \frac{\mbox{\small $\bigcirc\hspace{-3.2mm}\pm$}\;1}{2\; |1+n_3|} \;\cdot  \nonumber \\
& \cdot & \Big(   i\, \vec{e}_\perp \cdot [\,(\vec{e}_3  +  \vec{n}^\ast )\times (\vec{e}_3  +  \vec{n} )\, ] \, 1_2 + (\vec{e}_3  +  \vec{n}^\ast ) \cdot (\vec{e}_3  +  \vec{n} )\;\, \vec{e}_\perp \cdot \vec{\sigma} \nonumber \\ 
 & & - \,  \vec{n}  \cdot \vec{e}_\perp\; (\vec{e}_3  +  \vec{n}^\ast ) \cdot \vec{\sigma} -  \vec{n}^\ast  \cdot \vec{e}_\perp\; (\vec{e}_3  +  \vec{n} ) \cdot \vec{\sigma}   \, \Big)   \\[2mm]
 & = & \frac{N^\ast_2\,N_1-N^\ast_1\,N_2}{2\, i} \;\;  \frac{\mbox{\small $\bigcirc\hspace{-3.2mm}\pm$}\;1}{2\; |1+n_3|} \;\cdot  \nonumber \\ 
& \cdot & \Big(   i\, \Big( (1+n_3) \; \vec{n}^\ast \cdot \vec{e} - (1+n^\ast_3) \; \vec{n} \cdot \vec{e}\Big)  \, 1_2 + (\vec{e}_3  +  \vec{n}^\ast ) \cdot (\vec{e}_3  +  \vec{n} )\;\, \vec{e}_\perp \cdot \vec{\sigma} \nonumber \\
 & & - \, \vec{n}  \cdot \vec{e}_\perp\; (\vec{e}_3  +  \vec{n}^\ast ) \cdot \vec{\sigma} -   \vec{n}^\ast  \cdot \vec{e}_\perp\; (\vec{e}_3  +  \vec{n} ) \cdot \vec{\sigma}   \, \Big)  . 
\end{eqnarray}
Although these expressions can be used to determine $\eta_1$ und $\eta_2$ it is more pragmatic to calculate $\eta_1$ und $\eta_2$ by applying Eq.\ (\ref{xxfin1}) und (\ref{xxfin3}) to Eqs.\ (\ref{etafin1}) and (\ref{etafin2}), i.\ e.,
\newpage
\begin{eqnarray} 
\eta_1  & = & \mbox{\small $\bigcirc\hspace{-3.2mm}\pm$}\;\; \frac{N^\ast_2\,N_1+N^\ast_1\,N_2}{2} \;\;\frac{1}{2} \,  \left|  \frac{\frac{E_1-E_2}{2}+\frac{H_{11}-H_{22}}{2}}{\frac{E_1-E_2}{2}}\right| \; \cdot \nonumber  \\[2mm]
 & \cdot &  \left(1_2  - \frac{\left(\begin{array}{cc} 0 & H^\ast_{21} \\ -\,H^\ast_{12} & 0 \end{array}\right)}{\frac{E^\ast_1-E^\ast_2}{2}+\frac{H^\ast_{11}-H^\ast_{22}}{2}}  \right)  \vec{e} \cdot \vec{\sigma} \left(1_2  - \frac{\left(\begin{array}{cc} 0 & -\,H_{12} \\ H_{21} & 0 \end{array}\right)}{\frac{E_1-E_2}{2}+\frac{H_{11}-H_{22}}{2}}  \right)  , \nonumber \\[2mm]
 & & \label{etazzz1} \\[2mm]
\eta_2  & = & \mbox{\small $\bigcirc\hspace{-3.2mm}\pm$}\;\;\frac{N^\ast_2\,N_1-N^\ast_1\,N_2}{2\, i} \;\;\frac{1}{2} \,  \left|  \frac{\frac{E_1-E_2}{2}+\frac{H_{11}-H_{22}}{2}}{\frac{E_1-E_2}{2}}\right| \; \cdot \nonumber  \\[2mm]
 & \cdot &  \left(1_2  - \frac{\left(\begin{array}{cc} 0 & H^\ast_{21} \\ -\,H^\ast_{12} & 0 \end{array}\right)}{\frac{E^\ast_1-E^\ast_2}{2}+\frac{H^\ast_{11}-H^\ast_{22}}{2}}  \right)  \vec{e}_\perp \cdot \vec{\sigma} \left(1_2  - \frac{\left(\begin{array}{cc} 0 & -\,H_{12} \\ H_{21} & 0 \end{array}\right)}{\frac{E_1-E_2}{2}+\frac{H_{11}-H_{22}}{2}}  \right)  , \nonumber\\[2mm] 
\label{etazzz2} \end{eqnarray}
with $e^2_1+e^2_2=1$ (suggesting $e_1=\cos\varphi$ and $e_2=\sin\varphi$ with $\varphi$ being eventually complex-valued) and
\begin{eqnarray}  \vec{e} \cdot \vec{\sigma} & = &   \left(\begin{array}{cc} 0 & e_1 - i \, e_2 \\ e_1 + i \, e_2 & 0 \end{array}\right) \; = \;\left(\begin{array}{cc} 0 & e^{-i\,\varphi} \\ e^{+i\,\varphi} & 0 \end{array}\right) \nonumber \\[2mm]
 & = & \cos\varphi  \; \left(\begin{array}{cc} 0 & 1 \\ 1 & 0 \end{array}\right) + \sin\varphi  \; \left(\begin{array}{cc} 0 & -i \\ i & 0 \end{array}\right) \; = \; \vec{e}_\perp \cdot \vec{\sigma} \; i \, \sigma_3 \;  \,  ,  \label{vvee1} \\[2mm] 
 \vec{e}_\perp \cdot \vec{\sigma} & = &   \left(\begin{array}{cc} 0 & e_2 + i \, e_1 \\ e_2 - i \, e_1 & 0 \end{array}\right)  \; = \;\left(\begin{array}{cc} 0 & -i\,e^{-i\,\varphi} \\ +i\, e^{+i\,\varphi} & 0 \end{array}\right) \nonumber \\[2mm]
 & = & \sin\varphi  \; \left(\begin{array}{cc} 0 & 1 \\ 1 & 0 \end{array}\right) - \cos\varphi  \; \left(\begin{array}{cc} 0 & -i \\ i & 0 \end{array}\right) \; = \; i \,\sigma_3 \; \vec{e} \cdot \vec{\sigma} \, . \label{vvee2} \end{eqnarray}
With the help of these relations Eqs.\ (\ref{etafin1}) and (\ref{etafin2}) can be rewritten as follows:
\begin{eqnarray} 
 \eta_1  & = &  \frac{N^\ast_2\,N_1+N^\ast_1\,N_2}{2} \; \Big( \cos\varphi \; \; X^+ \sigma_1 \, X + \sin\varphi \;  X^+ \sigma_2 \, X \Big)  , \label{xetafin1} \\[2mm]
 \eta_2  & = & \frac{N^\ast_2\,N_1-N^\ast_1\,N_2}{2\, i}  \;\Big( \sin\varphi \; \; X^+ \sigma_1 \, X - \cos\varphi \;  X^+ \sigma_2 \, X \Big)   . \quad \label{xetafin2}
\end{eqnarray}
The most efficient way to determine $\eta_1$ and $\eta_2$ is to insert  the following identities for $X^+ \sigma_1 \, X$ and $X^+ \sigma_2 \, X$ (obtained with the help of Eqs.\ (\ref{xxxident1}) and (\ref{xxxident3})) into Eqs.\ (\ref{xetafin1}) and (\ref{xetafin2}):
\begin{eqnarray} 
\lefteqn{X^+ \sigma_1 \, X \; =} \nonumber \\[2mm]
  & =  & \mbox{\small $\bigcirc\hspace{-3.2mm}\pm$} \; \; \frac{1}{\left|  \frac{E_1-E_2}{2}\right|} \; \; \frac{1}{\left| \frac{E_1-E_2}{2} \, (1+n_3)\right|} \; \cdot \nonumber\\
 & \cdot & \Bigg\{\frac{1}{2}\left(\begin{array}{cc} 0 & \left| \frac{E_1-E_2}{2} \, (1+n_3)\right|^2  -H_{12}\,H^\ast_{21}  \\[2mm] \left| \frac{E_1-E_2}{2} \, (1+n_3)\right|^2  -H_{21}\,H^\ast_{12} & 0 \end{array}\right) \nonumber \\[2mm]
 & & - \,\; \mbox{Re} \left[ \; \frac{E_1-E_2}{2} \, (1+n_3) \; \right] \; \left(\begin{array}{cc} \mbox{Re}[ \, H_{21}\, ] & 0 \\[2mm]  0  & -\, \mbox{Re}[ \, H_{12}\, ]  \end{array}\right) \nonumber \\[2mm]
 & &   -\; \mbox{Im} \left[ \; \frac{E_1-E_2}{2} \, (1+n_3) \; \right] \; \left(\begin{array}{cc} \mbox{Im}[ \, H_{21}\, ] & 0 \\[2mm]  0  & -\, \mbox{Im}[ \, H_{12}\, ]  \end{array}\right)\Bigg\} \; ,  \\[2mm]
\lefteqn{X^+ \sigma_2 \, X \; =} \nonumber \\[2mm]
  & =  & \mbox{\small $\bigcirc\hspace{-3.2mm}\pm$} \; \; \frac{1}{\left|  \frac{E_1-E_2}{2}\right|} \; \; \frac{1}{\left| \frac{E_1-E_2}{2} \, (1+n_3)\right|} \; \cdot \nonumber\\
 & \cdot & \Bigg\{\frac{i}{2}\left(\begin{array}{cc} 0 & -\,\left| \frac{E_1-E_2}{2} \, (1+n_3)\right|^2  -H_{12}H^\ast_{21}  \\[2mm] \left| \frac{E_1-E_2}{2} \, (1+n_3)\right|^2  +H_{21}H^\ast_{12} & 0 \end{array}\right) \nonumber \\[2mm]
 & & - \,\; \mbox{Re} \left[ \; \frac{E_1-E_2}{2} \, (1+n_3) \; \right] \; \left(\begin{array}{cc} \mbox{Im}[ \, H_{21}\, ] & 0 \\[2mm]  0  & \mbox{Im}[ \, H_{12}\, ]  \end{array}\right) \nonumber \\[2mm]
 & &   +\; \mbox{Im} \left[ \; \frac{E_1-E_2}{2} \, (1+n_3) \; \right] \; \left(\begin{array}{cc} \mbox{Re}[ \, H_{21}\, ] & 0 \\[2mm]  0  & \mbox{Re}[ \, H_{12}\, ]  \end{array}\right)\Bigg\} \; . 
\end{eqnarray}

\section{Determination of the inverse metric $\eta^{-1}$}  \label{xxsec9}
\subsection{The inverse metric $\eta^{-1}$ for the trivial phase ($Q=1_2$)}
\noindent Inspection of Eqs.\ (\ref{yyyy1}), (\ref{iiii1}), (\ref{yyyy2}) and (\ref{iiii2}) yields:
\begin{eqnarray}
X^{-1} & = & \sigma_3 \; X \; \sigma_3 \; , \quad\;\;\;\;\, X \; = \; \sigma_3 \; X^{-1} \; \sigma_3 \; ,  \\[2mm]
\left(X^+\right)^{-1} & = & \sigma_3 \; X^+ \; \sigma_3 \; , \quad X^+ \; = \; \sigma_3 \; \left(X^+\right)^{-1} \; \sigma_3 \; . 
\end{eqnarray}
With the help of these relations we deduce the following identities:
\begin{eqnarray}
\left( X^+  X\right)^{-1} & = & X^{-1}\,  \left(X^+\right)^{-1} \nonumber \\[2mm]
 & = & +\, \sigma_3\,  X  X^+ \sigma_3 \;\;\;\; = \; + \, \sigma_3\,  \left( X^+  X\right)^+ \sigma_3 \; , \label{xyident0} \\[2mm]
\left( X^+ \sigma_1\, X\right)^{-1} & = & X^{-1}\, \sigma_1 \left(X^+\right)^{-1} \nonumber \\[2mm]
 & = & -\, \sigma_3\,  X \sigma_1 X^+ \sigma_3 \; = \; - \, \sigma_3\,  \left( X^+ \sigma_1\, X\right)^+ \sigma_3 \; , \label{xyident1} \\[2mm]
\left( X^+ \sigma_2\, X\right)^{-1} & = & X^{-1}\, \sigma_2 \left(X^+\right)^{-1} \nonumber \\[2mm]
 & = & -\, \sigma_3\,  X \sigma_2 X^+ \sigma_3 \; = \; - \, \sigma_3\,  \left( X^+ \sigma_2\, X\right)^+ \sigma_3 \; , \label{xyident2} \\[2mm]
\left( X^+ \sigma_3\, X\right)^{-1} & = & X^{-1}\, \sigma_3 \left(X^+\right)^{-1} \nonumber \\[2mm]
 & = & +\, \sigma_3\,  X \sigma_3 X^+ \sigma_3 \; = \; + \,\sigma_3\,  \left( X^+ \sigma_3 \, X \right)^+ \sigma_3 \; . \label{xyident3}
\end{eqnarray}
Since the metric for the trivial phase ($Q=1_2$) is given due to Eqs.\ (\ref{etafin0}) and (\ref{etafin3}) by the following expression:
\begin{equation} \eta \; = \; \eta_0 + \eta_3 \; = \; \frac{|N_1|^2+|N_2|^2}{2} \; X^+  X \; + \; \frac{|N_1|^2-|N_2|^2}{2} \; X^+ \sigma_3\,  X \; ,
\end{equation}
we make the follwing ansatz for the respective inverse metric $\eta^{-1}$ with yet unknown variables $c_0$ and $c_3\;$:
\begin{equation} \eta^{-1}  \; = \; c_0 \; \left(X^+  X\right)^{-1} \; + \; c_3 \; \left(X^+ \sigma_3\,  X\right)^{-1} \; . \label{etainv1}
\end{equation}
The solution of the identity $\eta \; \eta^{-1} \; = \; 1_2$ is given by the solution of the following system of linear equations:
\begin{equation} \left(\begin{array}{cc} \frac{|N_1|^2+|N_2|^2}{2}   & \frac{|N_1|^2-|N_2|^2}{2} \\[2mm] \frac{|N_1|^2-|N_2|^2}{2} & \frac{|N_1|^2+|N_2|^2}{2}  \end{array}\right)\left(\begin{array}{c} c_0 \\ c_3  \end{array}\right) \; = \; \left(\begin{array}{c} 1 \\ 0  \end{array}\right) \; ,
\end{equation}
i.\ e.,
\begin{equation}  \left(\begin{array}{c} c_0 \\ c_3  \end{array}\right) \; = \; \frac{1}{|N_1 N_2|^2} \left(\begin{array}{c} +\, \frac{|N_1|^2+|N_2|^2}{2} \\[2mm]  -\, \frac{|N_1|^2-|N_2|^2}{2} \end{array}\right) \; .
\end{equation}
Inserting this solution in Eq.\ (\ref{etainv1}) we obtain with the help of Eqs.\ (\ref{xyident0}) and (\ref{xyident3}) for the inverse metric in the trivial phase  ($Q=1_2$):

\begin{eqnarray} \eta^{-1}  & = & \frac{1}{|N_1 N_2|^2} \left( \frac{|N_1|^2+|N_2|^2}{2} \; \left(X^+  X\right)^{-1}  -  \frac{|N_1|^2-|N_2|^2}{2} \; \left(X^+ \sigma_3\,  X\right)^{-1} \right) \nonumber \\[2mm]
 & & \label{invtrivppp1} \\
& = & \frac{1}{|N_1 N_2|^2} \, \; \sigma_3 \, \left( \eta^+_0 - \eta^+_3 \right) \, \sigma_3 \; = \; \frac{1}{|N_1 N_2|^2} \, \; \sigma_3 \, \left( \eta_0 - \eta_3 \right) \, \sigma_3 \; . \quad \label{etainv2}
\end{eqnarray}
The most efficient way to calculate the inverse metric is to apply the following expressions to Eq.\ (\ref{invtrivppp1}) obtained making use of Eqs.\ (\ref{xxxident2}) and (\ref{xxxident4}):

\begin{eqnarray}\lefteqn{\left(X^+  X\right)^{-1}  \; = \; X^{-1} \left(X^+\right)^{-1} \; =} \nonumber \\[2mm]
 &=  & \mbox{\small $\bigcirc\hspace{-3.2mm}\pm$} \; \; \frac{1}{2\; \left|  \frac{E_1-E_2}{2}\right|} \; \; \frac{1}{\left| \frac{E_1-E_2}{2} \, (1+n_3)\right|} \; \cdot \nonumber\\
 & \cdot & \Bigg\{\left(\begin{array}{cc} \left| \frac{E_1-E_2}{2} \, (1+n_3)\right|^2 + |H_{21}|^2 & 0 \\[2mm] 0 & \left| \frac{E_1-E_2}{2} \, (1+n_3)\right|^2 + |H_{12}|^2 \end{array}\right) \nonumber \\[2mm]
 & & + \,\; \mbox{Re} \left[ \; \frac{E_1-E_2}{2} \, (1+n_3) \; \right] \; \left(\begin{array}{cc} 0 & -(H_{12} -\, H^\ast_{21}) \\[2mm]  H_{21} -\, H^\ast_{12}  & 0 \end{array}\right) \nonumber \\[2mm]
 & & + \, i  \; \mbox{Im} \left[ \; \frac{E_1-E_2}{2} \, (1+n_3) \; \right] \; \left(\begin{array}{cc} 0 & H_{12} + H^\ast_{21} \\[2mm]  -(H_{21} + H^\ast_{12})  & 0 \end{array}\right)\Bigg\} \; , \nonumber \\[2mm]
 & & \\
\lefteqn{\left(X^+ \sigma_3\,  X\right)^{-1}  \; = \;  X^{-1} \; \sigma_3 \,  \left(X^+\right)^{-1} \; =} \nonumber \\[2mm]
 &=  & \mbox{\small $\bigcirc\hspace{-3.2mm}\pm$} \; \;\frac{1}{2\; \left|  \frac{E_1-E_2}{2}\right|} \; \; \frac{1}{\left| \frac{E_1-E_2}{2} \, (1+n_3)\right|} \; \cdot \nonumber\\
 & \cdot & \Bigg\{\left(\begin{array}{cc} \left| \frac{E_1-E_2}{2} \, (1+n_3)\right|^2 - |H_{21}|^2 & 0 \\[2mm] 0 & -\, \left| \frac{E_1-E_2}{2} \, (1+n_3)\right|^2 + |H_{12}|^2 \end{array}\right) \nonumber \\[2mm]
 & & - \,\; \mbox{Re} \left[ \; \frac{E_1-E_2}{2} \, (1+n_3) \; \right] \; \left(\begin{array}{cc} 0 & H_{12} + H^\ast_{21}) \\[2mm]  H_{21} + H^\ast_{12}  & 0 \end{array}\right) \nonumber \\[2mm]
 & & + \, i  \; \mbox{Im} \left[ \; \frac{E_1-E_2}{2} \, (1+n_3) \; \right] \; \left(\begin{array}{cc} 0 & H_{12} - H^\ast_{21} \\[2mm]  H_{21} - H^\ast_{12}  & 0 \end{array}\right)\Bigg\} \; .
\end{eqnarray}
\subsection{The inverse metric $\eta^{-1}$ for the non-trivial phase ($Q=P$)}
\noindent The determination of the inverse metric $\eta^{-1}$ for the non-trivial phase ($Q=P$) is more involved. Again we start with the expression of the metric for the non-trivial phase in Eqs.\  (\ref{etafin1}) and (\ref{etafin2}) in combination with Eqs.\ (\ref{vvee1}) and (\ref{vvee2}), i.\ e.,
\begin{eqnarray} \eta & = & \eta_1 + \eta_2 \nonumber \\[3mm] 
 & = &  \frac{N^\ast_2\,N_1+N^\ast_1\,N_2}{2} \; X^+  \; \vec{e} \cdot \vec{\sigma} \; X \; + \;  \frac{N^\ast_2\,N_1-N^\ast_1\,N_2}{2\, i} \; X^+  \; \vec{e}_\perp \cdot \vec{\sigma} \;  X \nonumber \\ 
 & = &  \frac{N^\ast_2\,N_1+N^\ast_1\,N_2}{2} \;\Big(\, \cos\varphi \; \, X^+  \sigma_1 \, X \; + \;\sin\varphi \; \, X^+  \sigma_2 \, X \, \Big) \nonumber \\[2mm]
 & + &  \frac{N^\ast_2\,N_1-N^\ast_1\,N_2}{2\, i} \; \Big( \,\sin\varphi \; \, X^+  \sigma_1 \, X \; - \;\cos\varphi \; \, X^+  \sigma_2 \, X \, \Big)  \; , \label{qqqq1}
\end{eqnarray}
which can be rewritten slightly in the following way:
\begin{eqnarray} \eta & = & \left(  \frac{N^\ast_2\,N_1+N^\ast_1\,N_2}{2} \;\cos\varphi + \frac{N^\ast_2\,N_1-N^\ast_1\,N_2}{2\, i} \;\sin\varphi\right) \;  X^+  \sigma_1 \, X \nonumber \\[2mm]
 & + & \left(  \frac{N^\ast_2\,N_1+N^\ast_1\,N_2}{2} \;\sin\varphi - \frac{N^\ast_2\,N_1-N^\ast_1\,N_2}{2\, i} \;\cos\varphi\right) \;  X^+  \sigma_2 \, X\; . \quad
\end{eqnarray}
Again we make the follwing ansatz for the respective inverse metric $\eta^{-1}$ with yet unknown variables $c_1$ and $c_2\;$:
\begin{equation} \eta^{-1}  \; = \; c_1 \; \left(X^+  X\right)^{-1} \; + \; c_2 \; \left(X^+ \sigma_3\,  X\right)^{-1} \; . \label{etainv3}
\end{equation}
The solution of the identity $\eta \; \eta^{-1} \; = \; 1_2$ is given by the solution of the following system of linear equations:
\begin{equation} \left(\begin{array}{cc} \frac{N^\ast_2\,N_1+N^\ast_1\,N_2}{2}   & \frac{N^\ast_2\,N_1-N^\ast_1\,N_2}{2\, i} \\[2mm] \frac{N^\ast_2\,N_1-N^\ast_1\,N_2}{2\, i} & -\, \frac{N^\ast_2\,N_1+N^\ast_1\,N_2}{2}  \end{array}\right)\left(\begin{array}{cc} \cos\varphi   & \sin\varphi \\[2mm] \sin\varphi & -\, \cos\varphi  \end{array}\right)\left(\begin{array}{c} c_1 \\ c_2  \end{array}\right)  = \left(\begin{array}{c} 1 \\ 0  \end{array}\right)  ,
\end{equation}
i.\ e.,
\begin{equation}  \left(\begin{array}{c} c_1 \\ c_2  \end{array}\right) \; = \; \frac{1}{|N_1 N_2|^2} \left(\begin{array}{cc} \cos\varphi   & \sin\varphi \\[2mm] \sin\varphi & -\, \cos\varphi  \end{array}\right)\left(\begin{array}{c} \frac{N^\ast_2\,N_1+N^\ast_1\,N_2}{2} \\[2mm] \frac{N^\ast_2\,N_1-N^\ast_1\,N_2}{2\, i} \end{array}\right) \; .
\end{equation}
Inserting this solution in Eq.\ (\ref{etainv3}) we obtain for the inverse metric in the non-trivial phase  ($Q=P$) the following intermediate result:

\begin{eqnarray} \lefteqn{\eta^{-1} \; = \; \frac{1}{|N_1 N_2|^2} \; \cdot} \nonumber \\[2mm]
 & \cdot & \Bigg(\left(  \frac{N^\ast_2\,N_1+N^\ast_1\,N_2}{2} \;\cos\varphi + \frac{N^\ast_2\,N_1-N^\ast_1\,N_2}{2\, i} \;\sin\varphi\right) \left( X^+  \sigma_1 \, X\right)^{-1} \nonumber \\[2mm]
 &  & +\, \left(  \frac{N^\ast_2\,N_1+N^\ast_1\,N_2}{2} \;\sin\varphi - \frac{N^\ast_2\,N_1-N^\ast_1\,N_2}{2\, i} \;\cos\varphi\right) \left( X^+  \sigma_2 \, X\right)^{-1} \Bigg) . \nonumber \\[2mm]
\label{etainv4} \end{eqnarray}
The final step is to perform the Hermitian conjugation of the expression for $\eta_1$ and $\eta_2$ as given in Eq.\ (\ref{qqqq1}) and to apply Eqs.\ (\ref{xyident1}) and (\ref{xyident2}):
\begin{eqnarray} 
\eta^+_1 & = &  -\, \frac{N^\ast_2\,N_1+N^\ast_1\,N_2}{2} \;\cdot \nonumber\\[2mm]
 &  & \cdot \; \sigma_3 \, \Big(\, \cos\varphi^\ast \, \left(X^+  \sigma_1 \, X\right)^{-1}  + \sin\varphi^\ast \, \left(X^+  \sigma_2 \, X\right)^{-1} \, \Big) \, \sigma_3 \; , \\[2mm]
\eta^+_2 & = &  -\, \frac{N^\ast_2\,N_1-N^\ast_1\,N_2}{2\, i} \;\cdot \nonumber\\[2mm]
 &  & \cdot \;  \sigma_3 \, \Big( \,\sin\varphi^\ast  \, \left(X^+  \sigma_1 \, X\right)^{-1}  - \cos\varphi^\ast  \, \left( X^+  \sigma_2 \, X\right)^{-1} \, \Big) \, \sigma_3 \; , 
\end{eqnarray}
or, equivalently,
\begin{eqnarray} 
\lefteqn{\left(X^+  \sigma_1 \, X\right)^{-1} \; =} \nonumber \\[2mm]
 & =  & - \; \sigma_3 \, \left(\, \cos\varphi^\ast \, \frac{2\, \eta^+_1}{N^\ast_2\,N_1+N^\ast_1\,N_2}  + \sin\varphi^\ast \, \frac{2\, i\,  \eta^+_2}{N^\ast_2\,N_1-N^\ast_1\,N_2} \, \right) \, \sigma_3  , \\[2mm]
\lefteqn{\left(X^+  \sigma_2 \, X\right)^{-1} \; =} \nonumber \\[2mm]
 & =  & - \;  \sigma_3 \, \left( \,\sin\varphi^\ast  \, \frac{2\, \eta^+_1}{N^\ast_2\,N_1+N^\ast_1\,N_2}  - \cos\varphi^\ast  \, \frac{2\, i\,  \eta^+_2}{N^\ast_2\,N_1-N^\ast_1\,N_2} \, \right) \, \sigma_3  . 
\end{eqnarray}
Insertion of these expressions into Eq.\ (\ref{etainv4}) leads then to the final result for the inverse metric $\eta^{-1}$ in the non-trivial phase ($Q=P$):
\newpage
\begin{eqnarray}
\eta^{-1} & = & -\;  \frac{1}{|N_1 N_2|^2} \; \sigma_3 \, \Bigg( \cosh(2\, \mbox{Im}[\varphi]) \Big(\eta^+_1+\eta^+_2\Big) +   i\,\sinh(2\, \mbox{Im}[\varphi])  \; \cdot \nonumber \\[2mm]
 & & \cdot \; \left(\frac{N^\ast_2\,N_1-N^\ast_1\,N_2}{i\, (N^\ast_2\,N_1+N^\ast_1\,N_2)} \; \eta^+_1-\frac{i\,(N^\ast_2\,N_1+N^\ast_1\,N_2)}{N^\ast_2\,N_1-N^\ast_1\,N_2}\;\eta^+_2\right)\Bigg) \, \sigma_3 . \nonumber \\
\end{eqnarray}
Note that there holds
\begin{eqnarray} \cosh(2\, \mbox{Im}[\varphi]) & = & \cos(\varphi-\varphi^\ast) \; = \; \cos\varphi^\ast \; \cos\varphi \; +  \sin\varphi^\ast \; \sin\varphi \; , \\[2mm]
 i\, \sinh(2\, \mbox{Im}[\varphi]) & = & \sin(\varphi-\varphi^\ast) \; = \; \cos\varphi^\ast \;\, \sin\varphi \; -  \sin\varphi^\ast \; \cos\varphi \; . \end{eqnarray}
The result for $\eta^{-1}$ in the non-trivial phase ($Q=P$)  simplifies significantly for $\varphi$ being real-valued, i.\ e.,
\begin{equation} \mbox{Im}[\varphi]\;  \rightarrow \; 0   \quad \Rightarrow \quad \eta^{-1} \; \rightarrow \; -\;  \frac{1}{|N_1 N_2|^2} \; \sigma_3 \,  \Big(\eta^+_1+\eta^+_2\Big)  \, \sigma_3 . 
\end{equation}
The most efficient way to calculate the inverse metric is to apply the following expressions to Eq.\ (\ref{etainv4}) obtained making use of Eqs.\ (\ref{xxxident2}) and (\ref{xxxident4}):

\begin{eqnarray} 
\lefteqn{(X^+ \sigma_1 \, X)^{-1} \; = \; X^{-1}\,  \sigma_1  \left(X^+\right)^{-1} \; =} \nonumber \\[2mm]
  & =  & \mbox{\small $\bigcirc\hspace{-3.2mm}\pm$} \; \; \frac{1}{\left|  \frac{E_1-E_2}{2}\right|} \; \; \frac{1}{\left| \frac{E_1-E_2}{2} \, (1+n_3)\right|} \; \cdot \nonumber\\
 & \cdot & \Bigg\{\frac{1}{2}\left(\begin{array}{cc} 0 & \left| \frac{E_1-E_2}{2} \, (1+n_3)\right|^2  -H_{12}\,H^\ast_{21}  \\[2mm] \left| \frac{E_1-E_2}{2} \, (1+n_3)\right|^2  -H_{21}\,H^\ast_{12} & 0 \end{array}\right) \nonumber \\[2mm]
 & & + \,\; \mbox{Re} \left[ \; \frac{E_1-E_2}{2} \, (1+n_3) \; \right] \; \left(\begin{array}{cc} \mbox{Re}[ \, H_{21}\, ] & 0 \\[2mm]  0  & -\, \mbox{Re}[ \, H_{12}\, ]  \end{array}\right) \nonumber \\[2mm]
 & &   +\; \mbox{Im} \left[ \; \frac{E_1-E_2}{2} \, (1+n_3) \; \right] \; \left(\begin{array}{cc} \mbox{Im}[ \, H_{21}\, ] & 0 \\[2mm]  0  & -\, \mbox{Im}[ \, H_{12}\, ]  \end{array}\right)\Bigg\} \; ,  \\[2mm]
 & & \nonumber \\[30mm]
\lefteqn{(X^+ \sigma_2 \, X)^{-1} \; = \; X^{-1}\,  \sigma_2  \left(X^+\right)^{-1} \; =} \nonumber \\[2mm]
  & =  & \mbox{\small $\bigcirc\hspace{-3.2mm}\pm$} \; \; \frac{1}{\left|  \frac{E_1-E_2}{2}\right|} \; \; \frac{1}{\left| \frac{E_1-E_2}{2} \, (1+n_3)\right|} \; \cdot \nonumber\\
 & \cdot & \Bigg\{\frac{i}{2}\left(\begin{array}{cc} 0 & -\,\left| \frac{E_1-E_2}{2} \, (1+n_3)\right|^2  -H_{12}H^\ast_{21}  \\[2mm] \left| \frac{E_1-E_2}{2} \, (1+n_3)\right|^2  +H_{21}H^\ast_{12} & 0 \end{array}\right) \nonumber \\[2mm]
 & & + \,\; \mbox{Re} \left[ \; \frac{E_1-E_2}{2} \, (1+n_3) \; \right] \; \left(\begin{array}{cc} \mbox{Im}[ \, H_{21}\, ] & 0 \\[2mm]  0  & \mbox{Im}[ \, H_{12}\, ]  \end{array}\right) \nonumber \\[2mm]
 & &   -\; \mbox{Im} \left[ \; \frac{E_1-E_2}{2} \, (1+n_3) \; \right] \; \left(\begin{array}{cc} \mbox{Re}[ \, H_{21}\, ] & 0 \\[2mm]  0  & \mbox{Re}[ \, H_{12}\, ]  \end{array}\right)\Bigg\} \; . 
\end{eqnarray}

\section{Relating the metric $\eta$ and the ${\cal CP}$-operator of C.\ M.\ Bender et al.\ by consideration of expectation values} \label{relmetric1}
\noindent The time evolution of a quantum mechanical system in the Schr\"odinger-picture is contained in the states $\left|\psi(t)\right>$ and $\left<\psi(t)\right|=\left|\psi(t)\right>^+$ determined by the time-dependent Schr\"odinger equation and its Hermitian conjugate, respectively, i.\ e.:
\begin{eqnarray} i\, \hbar \;\frac{d}{dt} \,  \left|\psi(t)\right> & = & H\; \left|\psi(t)\right> \; , \\[2mm]
- \,  i\, \hbar \;\frac{d}{dt} \,  \left<\psi(t)\right| & = & \left<\psi(t)\right| \; H^+\; .
\end{eqnarray}
Hence, the time-derivative of any expectation value $\left<\psi(t)\right| A(t) \left|\psi(t)\right>$ of some --- eventually manifestly time-dependent --- operator $A(t)$ can be determined in the spirit of theorems of Ehrenfest as follows:
\begin{eqnarray} \lefteqn{\frac{d}{dt} \left<\psi(t)\right| A(t) \left|\psi(t)\right> \; =} \nonumber \\[2mm] & = &  \left<\psi(t)\right| \frac{\partial A(t)}{\partial t}  \left|\psi(t)\right> \; + \; \frac{1}{ i\, \hbar} \; \left<\psi(t)\right| \left[ \, A\, H - H^+ A\,\right] \left|\psi(t)\right> \; .
\end{eqnarray}
In order to maintain the expectation value stationary for any time-independent operator $A$ there has to hold:
\begin{equation} 0 \; = \; A\, H - H^+ A \; = \; \left\{ \begin{array}{l} \eta \; \Big( \eta^{-1} A\, H -  H\, \eta^{-1}  A \Big) \quad \mbox{for} \quad H^+ = + \,\eta \, H \, \eta^{-1} \, , \\[2mm] \eta \; \Big( \eta^{-1} A\, H +  H\, \eta^{-1}  A \Big) \quad \mbox{for} \quad H^+ = -\, \eta \, H \, \eta^{-1} \, , \end{array} \right.
\end{equation}
implying
\begin{eqnarray} \Big[ \; \eta^{-1} A \; , \, H \; \Big] \; = \; \eta^{-1} A\, H -  H\, \eta^{-1}  A \; =\; 0 & \mbox{for} & H^+ = + \,\eta \, H \, \eta^{-1} \, , \\[2mm]
 \Big\{ \, \eta^{-1} A \; , \,  H \; \Big\} \; = \; \eta^{-1} A\, H +  H\, \eta^{-1}  A \; =\; 0 & \mbox{for} & H^+ = - \,\eta \, H \, \eta^{-1} \, . \quad
\end{eqnarray}
Hence,  we have
\begin{equation} \eta^{-1} A \; = \; B \quad \Leftrightarrow \quad A \; = \eta \, B \; \label{etaab1}
\end{equation}
with $B$ being some arbitrary operator commuting with the Hamilton operator (i.\ e., $[ B, H] = 0$) for a pseudo-Hermitian Hamilton operator ($H^+ = + \,\eta \, H \, \eta^{-1}$) and anti-commuting with the Hamilton operator (i.\ e., $\{ B, H\} = 0$) for an anti-pseudo-Hermitian Hamilton operator ($H^+ = - \,\eta \, H \, \eta^{-1}$). 

Since the expectation value of the transposed ${\cal CP}$-operator with respect to the ${\cal CPT}$-inner product suggested by C.~M.~Bender, D.~C.~Brody and H.~F.~Jones in \cite{Bender:2002vv} (see also \cite{Bender:2005tb}\cite{Bender:2004ej}\cite{Bender:2009en}) corresponding to some time-independent Hamilton operator  should  be time-independent there holds in our notation ($T$ denotes here the transposition, NOT the time-conjugation):
\begin{equation} \frac{d}{dt} \left<\psi(t)\right| ({\cal CP})^T \left|\psi(t)\right> \; = \; 0 \; .
\end{equation} 
Consequently, there holds due to Eq.\ (\ref{etaab1}):
\begin{eqnarray} \eta^{-1} ({\cal CP})^T \; = \; B & \Leftrightarrow & ({\cal CP})^T \; = \eta \, B \nonumber \\[2mm] 
 & \Leftrightarrow & {\cal CP} \; = (\eta \, B)^T\; = \; B^T \eta^T \nonumber \\[2mm] 
 & \Leftrightarrow & {\cal C} \; = (\eta \, B)^T\, {\cal P}^{-1}\; = \; B^T \eta^T \, {\cal P}^{-1}\; . \label{etaab2}
\end{eqnarray}
According to C.~M.~Bender et al.\ the ${\cal C}$-operator should be an involution, i.\ e., there holds ${\cal C}^{\,2}= 1_2$ or, equivalently, ${\cal C}= {\cal C}^{-1}$, implying with ${\cal P}^{\,2}= 1_2\; $:
\begin{equation} (\eta \, B)^T\, {\cal P}^{-1} \; = \; {\cal P} \left( (\eta \, B)^T\,\right)^{-1} \quad \Rightarrow \quad (\eta \, B)^{-1}  \; = \; {\cal P}^T \; \eta \, B\;\, {\cal P}^T  \; . \label{xxxinvol1}
\end{equation}
It is --- yet --- unclear whether this symmetry relation between $(\eta \, B)^{-1}$ and $\eta \, B$ can be always realized by selecting some specific normalizations $N_1$ and $N_2$ of the eigen-vectors of an eventually asymmetric Hamilton operator. Nonetheless we will provide here a definite answer to this question for Case 1 ($H$ is pseudo-Hermitian ($H^+ = + \eta \, H \, \eta^{-1}\,$) and the phase of the metric is trivial ($Q=1_2$)) as specified in Eqs.\ (\ref{enconstr1}) and (\ref{enconstr5}), as the ${\cal CPT}$-inner product suggested by C.~M.~Bender, D.~C.~Brody and H.~F.~Jones is used --- yet --- only under these conditions. One condition between $N_1$ and $N_2$ resulting from $\cal C$ being an involution is obtained by taking the determinant of Eq.\ (\ref{xxxinvol1}) and invoking Eq.\ (\ref{identq1}):
\begin{eqnarray} 1 & = & \mbox{det} [ \, \eta \, B\;\, {\cal P}^T  \, \eta \, B\;\, {\cal P}^T\, ] \nonumber \\[1mm]
 & = & \Big( \mbox{det} [ \, \eta\,]\,\mbox{det} [ \, B \, ] \, \mbox{det} [ \, P \,] \Big)^2 \; = \; \Big( \mbox{det} [ \, \eta\,]\Big)^2 \; \Big(\mbox{det} [ \, B \, ] \Big)^2\nonumber \\[1mm]
 \Rightarrow  \; 1 & = & \Big( |N_1|^2 \,|N_2|^2 \Big)^2 \; \Big(\mbox{det} [ \, B \, ] \Big)^2\nonumber \\[2mm]
 \Rightarrow  \; 1 & = & |N_1|^2 \,|N_2|^2  \quad \mbox{for} \quad \Big(\mbox{det} [ \, B \, ] \Big)^2 = 1  \; . \label{nnid12} \end{eqnarray}

With the help of Eqs.\  (\ref{pseudherm1}), (\ref{xxxident1}), (\ref{xxxident2}), (\ref{xxxident3}), (\ref{xxxident4}) and (\ref{nnnident1}) we can denote the metric $\eta$ and its inverse $\eta^{-1}$ for a pseudo-Hermitian Hamilton operator ($H^+ = + \eta \, H \, \eta^{-1}\,$) in the trivial phase ($Q=1_2$) in the following way:
\begin{eqnarray} 
 \eta  & = & X^+ \; N^+ \; N \; X \; = \; \mbox{$\bigcirc\hspace{-3.4mm}\pm$} \;\; \frac{|1+n_3|}{2} \;  \Bigg( \left(\begin{array}{cc} |N_1|^2 & 0 \\ 0 & |N_2|^2 \end{array}\right)  \nonumber \\
 & &   + \; \frac{ \left(\begin{array}{cc} 0 & H^\ast_{21} \\ -\,H^\ast_{12} & 0 \end{array}\right)\left(\begin{array}{cc} |N_1|^2 & 0 \\ 0 & |N_2|^2 \end{array}\right)\left(\begin{array}{cc} 0 & -\,H_{12} \\ H_{21} & 0 \end{array}\right)}{\left|\frac{E_1-E_2}{2} \, (1+n_3)\right|^2}   \nonumber \\
 & &   - \; \frac{ \left(\begin{array}{cc} 0 & H^\ast_{21} \\ -\,H^\ast_{12} & 0 \end{array}\right)\left(\begin{array}{cc} |N_1|^2 & 0 \\ 0 & |N_2|^2 \end{array}\right)}{\frac{E^\ast_1-E^\ast_2}{2} \,(1+n^\ast_3)}    \nonumber \\
 & &   - \; \frac{\left(\begin{array}{cc} |N_1|^2 & 0 \\ 0 & |N_2|^2 \end{array}\right)\left(\begin{array}{cc} 0 & -\,H_{12} \\ H_{21} & 0 \end{array}\right)}{\frac{E_1-E_2}{2} \, (1+n_3)} \; \Bigg) \nonumber \\[2mm]
  & = &  \mbox{$\bigcirc\hspace{-3.4mm}\pm$} \;\; \frac{|1+n_3|}{2} \;  \Bigg( \left(\begin{array}{cc} |N_1|^2 & 0 \\ 0 & |N_2|^2 \end{array}\right)    + \frac{\left(\begin{array}{cc}  |H_{21}|^2 \, |N_2|^2 & 0 \\ 0 &  |H_{12}|^2 \, |N_1|^2 \end{array}\right)}{\left|\frac{E_1-E_2}{2} \,(1+n_3)\right|^2}   \nonumber \\
 & &   - \; \frac{ \left(\begin{array}{cc} 0 & H^\ast_{21} \,  |N_2|^2  \\ -\,H^\ast_{12} \,  |N_1|^2 & 0 \end{array}\right)}{\frac{E^\ast_1-E^\ast_2}{2} \,(1+n^\ast_3)}    \;   - \; \frac{\left(\begin{array}{cc} 0 & -\,H_{12} \, |N_1|^2 \\ H_{21} \, |N_2|^2 & 0 \end{array}\right)}{\frac{E_1-E_2}{2} \,(1+n_3)} \;  \Bigg)  \; , \nonumber \\[2mm]
 & & \label{wwweta1}\\[10mm] 
 \eta^{-1}  & = &  X^{-1} \; (N^+  N)^{-1} \; (X^+)^{-1} \; = \; \mbox{$\bigcirc\hspace{-3.4mm}\pm$} \;\; \frac{|1+n_3|}{2} \;  \Bigg( \left(\begin{array}{cc} \frac{1}{|N_1|^2} & 0 \\ 0 & \frac{1}{|N_2|^2} \end{array}\right)  \nonumber \\
 & &   + \; \frac{ \left(\begin{array}{cc} 0 & H^\ast_{21} \\ -\,H^\ast_{12} & 0 \end{array}\right)\left(\begin{array}{cc} \frac{1}{|N_1|^2} & 0 \\ 0 & \frac{1}{|N_2|^2} \end{array}\right)\left(\begin{array}{cc} 0 & -\,H_{12} \\ H_{21} & 0 \end{array}\right)}{\left|\frac{E_1-E_2}{2} \,(1+n_3)\right|^2}   \nonumber \\
 & &   + \; \frac{ \left(\begin{array}{cc} 0 & H^\ast_{21} \\ -\,H^\ast_{12} & 0 \end{array}\right)\left(\begin{array}{cc} \frac{1}{|N_1|^2} & 0 \\ 0 & \frac{1}{|N_2|^2} \end{array}\right)}{\frac{E^\ast_1-E^\ast_2}{2} \,(1+n^\ast_3)}    \nonumber \\[2mm]
 & &   + \; \frac{\left(\begin{array}{cc} \frac{1}{|N_1|^2} & 0 \\ 0 & \frac{1}{|N_2|^2} \end{array}\right)\left(\begin{array}{cc} 0 & -\,H_{12} \\ H_{21} & 0 \end{array}\right)}{\frac{E_1-E_2}{2} \,(1+n_3)} \; \Bigg) \nonumber \\
  & = &  \mbox{$\bigcirc\hspace{-3.4mm}\pm$} \;\; \frac{|1+n_3|}{2} \;  \Bigg( \left(\begin{array}{cc} \frac{1}{|N_1|^2} & 0 \\ 0 &  \frac{1}{|N_2|^2} \end{array}\right)    + \frac{\left(\begin{array}{cc}  \frac{|H_{21}|^2}{|N_2|^2} & 0 \\ 0 & \frac{|H_{12}|^2}{|N_1|^2} \end{array}\right)}{\left|\frac{E_1-E_2}{2} \,(1+n_3)\right|^2}   \nonumber \\
 & &   + \; \frac{ \left(\begin{array}{cc} 0 & \frac{H^\ast_{21}}{|N_2|^2}  \\ -\,\frac{H^\ast_{12}}{|N_1|^2} & 0 \end{array}\right)}{\frac{E^\ast_1-E^\ast_2}{2} \,(1+n^\ast_3)}    \;   + \; \frac{\left(\begin{array}{cc} 0 & -\,\frac{H_{12}}{|N_1|^2} \\ \frac{H_{21}}{|N_2|^2} & 0 \end{array}\right)}{\frac{E_1-E_2}{2} \,(1+n_3)} \;  \Bigg)  \; . \label{uuuetainv1}
\end{eqnarray}
According to C.~M.~Bender, P.~Meisinger and  Q.\ Wang \cite{Bender:2003gu} the most general parity operator should take the following for:
\begin{equation} \mbox{$\cal P$} \; = \;  \mbox{$\cal P$}^T \; = \;  \mbox{$\cal P$}^{-1} \; = \;  \left(\begin{array}{cc} \cos\phi & \;\;\;\sin \phi \\ \sin\phi &  -\sin\phi \end{array}\right) \; .
\end{equation}
The condition of C.~M.~Bender et al., that the ${\cal C}$-operator should be an involution, i.\ e., ${\cal C}^{\,2}= 1_2$, implies due to Eq.\ (\ref{xxxinvol1}) for $B=1_2$ the identity $\eta^{-1}  \; = \; {\cal P}^T \, \eta \, {\cal P}^T\;= \; {\cal P} \, \eta \, {\cal P}$ . The identification of the expressions for $\eta^{-1}$  given by in Eq.\ (\ref{uuuetainv1}) and ${\cal P} \, \eta \, {\cal P}$ with $\eta$ given by Eq.\ (\ref{wwweta1}) requires with $|N_1|^2\, |N_2|^2=1$ (see Eq.~(\ref{nnid12})) the following two identities:
\begin{eqnarray} {\cal P} \, \left(\begin{array}{cc} |N_1|^2 & 0 \\ 0 & |N_2|^2 \end{array}\right) \, {\cal P} & = & \left(\begin{array}{cc} |N_2|^2 & 0 \\ 0 & |N_1|^2 \end{array}\right)  \; , \\[2mm]
 {\cal P} \, \left(\begin{array}{cc} 0 & -\,H_{12} \, |N_1|^2 \\ H_{21} \, |N_2|^2 & 0 \end{array}\right) \, {\cal P} & = & \left(\begin{array}{cc} 0 & H_{12} \,|N_2|^2 \\ -\, H_{21}\, |N_1|^2 & 0 \end{array}\right)   ,\quad \quad
\end{eqnarray}
or, equivalently,
\begin{eqnarray}  \Big(\,|N_1|^2 \; - \; |N_2|^2\,\Big) \; \sin\phi \,\cos\phi & = & 0 \; , \\[1mm]
|N_1|^2\; (\cos \phi)^2 \; + \; |N_2|^2\; (\sin\phi)^2 & = & |N_2|^2 \; ,\\[1mm]
|N_1|^2\; (\sin \phi)^2 \; + \; |N_2|^2\; (\cos\phi)^2 & = & |N_1|^2 \; , \\[1mm]
 \Big(\,H_{12} \, |N_1|^2 \; - \; H_{21}\, |N_2|^2\,\Big) \; \sin\phi \,\cos\phi & = & 0 \; , \\[1mm]
H_{12} \, |N_1|^2\; (\cos \phi)^2 \; + \; H_{21} \, |N_2|^2\; (\sin\phi)^2 & = & H_{12} \, |N_2|^2 \; ,\\[1mm]
H_{12} \, |N_1|^2\; (\sin \phi)^2 \; + \; H_{21} \, |N_2|^2\; (\cos\phi)^2 & = & H_{21}\, |N_1|^2 \; .
\end{eqnarray}
Consequently, the constraint that the ${\cal C}$-operator should be an involution (i.~e., ${\cal C}^{\,2}= 1_2$) while keeping simultaneously the angle $\phi$ arbitrary implies for $B=1_2$ and $|N_1|^2\, |N_2|^2=1$ (see Eq.~(\ref{nnid12})) the following identities:
\begin{equation} |N_1|^2 = |N_2|^2 \; = \; 1 \; , \quad H_{12} \; = \; H_{21} \; . \label{cinvol1}
\end{equation}
Hence, the unnecessary \cite{Znojil:2006ugs} constraint ${\cal C}^2=1_2$ requires $H$ to be symmetric. Unfortunately it excludes the class of non-symmetric Hermitian Hamilton operators being well allowed in traditional Hermitian Quantum Theory.
\section{Convenient representations for the (anti-)pseudo-Hermitian Hamilton operator and the respective metric} \label{xconvrep1}
\noindent At this place we want to provide --- for all four cases specified in Eqs.\ (\ref{enconstr1}), (\ref{enconstr2}), (\ref{enconstr3}), (\ref{enconstr4}) or, equivalently, Eqs.\ (\ref{enconstr5}), (\ref{enconstr6}), (\ref{enconstr7}), (\ref{enconstr8}) ---  the most convenient expressions for the matrix $H$ representing the Hamilton operator, for its eigen-values $E_1$ and $E_2$, for the respective matrix $\vec{n}\cdot \vec{\sigma}$ and the resulting metric $\eta\,$:
\begin{itemize} 
\item {\bf Case 1:} $H$ is pseudo-Hermitian ($H^+ = + \eta \, H \, \eta^{-1}\,$), \\
\mbox{} $\qquad\quad\;\;$ the phase of the metric is trivial ($Q=1_2$)\\[2mm]
There hold the following properties:
\begin{itemize} \item The eigen-values $E_1$ and $E_2$ of $H$ are real-valued, i.e. there holds: $E_1=E^\ast_1$, $E_2=E^\ast_2$ (see also Eq.\ (\ref{enconstr1})). 
\item The quantities $E_1+E_2=\mbox{tr}[H]=H_{11}+H_{22}$ and $(E_1-E_2)^2=\mbox{tr}[H]^2+4\,  \mbox{det}[H]=(H_{11}-H_{22})^2+ 4\, H_{12}\, H_{21}$ are ---  due to the pseudo-Hermiticity of $H$  --- real-valued.
\item  According to  Eq.\ (\ref{enconstr5}) there holds in the trivial phase ($Q=1_2$) the inequality $\mbox{tr}[H]^2> 4\,  \mbox{det}[H]$, or, equivalently, $(H_{11}-H_{22})^2\; >\;-\, 4\, H_{12}\, H_{21}$.
\end{itemize}
The square root of $(E_1-E_2)^2=\mbox{tr}[H]^2-4\,  \mbox{det}[H]=(H_{11}-H_{22})^2+ 4\, H_{12}\, H_{21}> 0$ (see Eq.\ (\ref{impquant2})) is performed most conveniently in the following way:
\begin{eqnarray} \frac{E_1-E_2}{2} & = & \pm \; \frac{\sqrt{\mbox{tr}[H]^2-4\,  \mbox{det}[H] }}{2} \nonumber \\[2mm]
  & = & \pm \; \sqrt{\left(\frac{H_{11}-H_{22}}{2}\right)^2+ H_{12}\, H_{21}} \; . 
\end{eqnarray}
Taking this expression for $E_1-E_2$ and the identity $E_1+E_2=\mbox{tr}[H]=H_{11}+H_{22}$ we can use the following identities (see also Eq.\ (\ref{hamop1}))
\begin{eqnarray}  H  & = &  \frac{E_1+E_2}{2} \; 1_2 +  \frac{E_1-E_2}{2} \; \vec{n} \cdot \vec{\sigma} \label{usefid1} \\
  E_1 & = &   \frac{E_1+E_2}{2} \; + \;  \frac{E_1-E_2}{2} \; , \label{usefid2} \\
 E_2 & = & \frac{E_1+E_2}{2} \; - \;  \frac{E_1-E_2}{2} \;  , \label{usefid3}
\end{eqnarray}
to determine the most convenient representation for the Hamilton operator and its corresponding eigen-values $E_1$ and $E_2$:
\begin{eqnarray}  H  & = &  \frac{\mbox{tr}[H]}{2} \; 1_2 \; \pm \; \frac{\sqrt{\mbox{tr}[H]^2-4\,  \mbox{det}[H] }}{2} \; \vec{n} \cdot \vec{\sigma} \; , \label{xrep1} \\[2mm]
 E_1 & = & \frac{\mbox{tr}[H]}{2} \; \pm \;  \frac{\sqrt{\mbox{tr}[H]^2-4\,  \mbox{det}[H] }}{2}\; , \label{xrep2} \\[2mm]
 E_2 & = & \frac{\mbox{tr}[H]}{2} \; \mp \;  \frac{\sqrt{\mbox{tr}[H]^2-4\,  \mbox{det}[H] }}{2} \;  , \label{xrep3} 
\end{eqnarray}
or, equivalently,
\begin{eqnarray}  H  & = &  \frac{H_{11}+H_{22}}{2} \; 1_2 \; \pm \;  \sqrt{\left(\frac{H_{11}-H_{22}}{2}\right)^2+ H_{12}\, H_{21}} \;\; \vec{n} \cdot \vec{\sigma} \; , \quad \label{zrep1} \\[2mm]
 E_1 & = & \frac{H_{11}+H_{22}}{2} \; \pm \;   \sqrt{\left(\frac{H_{11}-H_{22}}{2}\right)^2+ H_{12}\, H_{21}}\; , \label{zrep2} \\[2mm]
 E_2 & = & \frac{H_{11}+H_{22}}{2} \; \mp \;   \sqrt{\left(\frac{H_{11}-H_{22}}{2}\right)^2+ H_{12}\, H_{21}} \;  . \label{zrep3} 
\end{eqnarray}
Eqs.\ (\ref{xrep1}) and (\ref{zrep1}) can be solved for the matrix $\vec{n} \cdot \vec{\sigma}$ to determine the most convenient expression for the matrix $\vec{n} \cdot \vec{\sigma}$ for a pseudo-Hermitian Hamilton operator in the trivial phase:
\begin{eqnarray} \vec{n} \cdot \vec{\sigma} & = & \pm\; \frac{H - \frac{\mbox{tr}[H]}{2} \; 1_2}{\sqrt{\frac{\mbox{tr}[H]^2}{4}-  \mbox{det}[H] }}  \; = \; \pm\; \frac{H - \frac{H_{11}+H_{22}}{2} \; 1_2}{\sqrt{\left(\frac{H_{11}-H_{22}}{2}\right)^2+ H_{12}\, H_{21}}} \; . \nonumber \\
 \label{nsig1}
\end{eqnarray}
With the help of these results we can denote on the basis of Eqs.\ (\ref{etazzz0})  and (\ref{etazzz3}) the following convenient representation for the metric $\eta=\eta_0+\eta_3$ for a two-dimensional pseudo-Hermitian Hamilton operator in the trivial phase ($Q=1_2$), i.\ e., 
\begin{eqnarray} 
\eta_0  & = & \mbox{\small $\bigcirc\hspace{-3.2mm}\pm$}\;\;\frac{|N_1|^2+|N_2|^2}{2} \;\;\frac{1}{2} \,  \left|  \frac{ \frac{H_{11}-H_{22}}{2}\pm \; \sqrt{\left(\frac{H_{11}-H_{22}}{2}\right)^2+ H_{12}\, H_{21}}}{\pm \;\sqrt{\left(\frac{H_{11}-H_{22}}{2}\right)^2+ H_{12}\, H_{21}}}\right| \; \cdot \nonumber  \\[2mm]
 & \cdot &  \left(1_2 \; - \; \frac{\left(\begin{array}{cc} 0 & H^\ast_{21} \\ -\,H^\ast_{12} & 0 \end{array}\right)}{\frac{H^\ast_{11}-H^\ast_{22}}{2}\pm \; \sqrt{\left(\frac{H^\ast_{11}-H^\ast_{22}}{2}\right)^2+ H^\ast_{21}\, H^\ast_{12}}}  \right)  \nonumber\\[2mm]
 & \cdot & \left(1_2  \; - \; \frac{\left(\begin{array}{cc} 0 & -\,H_{12} \\ H_{21} & 0 \end{array}\right)}{\frac{H_{11}-H_{22}}{2}\pm \; \sqrt{\left(\frac{H_{11}-H_{22}}{2}\right)^2+ H_{12}\, H_{21}}}  \right)  ,  \\[10mm]
\eta_3  & = & \mbox{\small $\bigcirc\hspace{-3.2mm}\pm$}\;\;\frac{|N_1|^2-|N_2|^2}{2} \;\;\frac{1}{2} \,  \left|  \frac{\frac{H_{11}-H_{22}}{2}\pm \; \sqrt{\left(\frac{H_{11}-H_{22}}{2}\right)^2+ H_{12}\, H_{21}}}{\pm \;\sqrt{\left(\frac{H_{11}-H_{22}}{2}\right)^2+ H_{12}\, H_{21}}}\right| \; \cdot \nonumber  \\[2mm]
 & \cdot &  \left(1_2 \; - \; \frac{\left(\begin{array}{cc} 0 & H^\ast_{21} \\ -\,H^\ast_{12} & 0 \end{array}\right)}{\frac{H^\ast_{11}-H^\ast_{22}}{2}\pm \; \sqrt{\left(\frac{H^\ast_{11}-H^\ast_{22}}{2}\right)^2+ H^\ast_{21}\, H^\ast_{12}}}  \right) \left(\begin{array}{cc} 1 & 0 \\ 0 & -1 \end{array}\right) \nonumber\\[2mm]
 & \cdot & \left(1_2 \; - \; \frac{\left(\begin{array}{cc} 0 & -\,H_{12} \\ H_{21} & 0 \end{array}\right)}{\frac{H_{11}-H_{22}}{2}\pm \; \sqrt{\left(\frac{H_{11}-H_{22}}{2}\right)^2+ H_{12}\, H_{21}}}  \right) . 
\end{eqnarray}
\item {\bf Case 2:} $H$ is pseudo-Hermitian ($H^+ = + \eta \, H \, \eta^{-1}\,$), \\
\mbox{} $\qquad\quad\;\;$ the phase of the metric is non-trivial ($Q=P$)\\[2mm]
There hold the following properties:
\begin{itemize} \item The eigen-values $E_1$ and $E_2$ of $H$ form a complex conjugate pair, i.e. there holds: $E_1=E^\ast_2$, $E_2=E^\ast_1$ (see also Eq.\ (\ref{enconstr2})). 
\item The quantities $E_1+E_2=\mbox{tr}[H]=H_{11}+H_{22}$ and $(E_1-E_2)^2=\mbox{tr}[H]^2-4\,  \mbox{det}[H]=(H_{11}-H_{22})^2+ 4\, H_{12}\, H_{21}$ are ---  due to the pseudo-Hermiticity of $H$  --- real-valued.
\item According to  Eq.\ (\ref{enconstr6}) there holds in the non-trivial phase \mbox{($Q=P$)} the inequality $\mbox{tr}[H]^2< 4\,  \mbox{det}[H]$, or, equivalently, $(H_{11}-H_{22})^2\; <\;-\, 4\, H_{12}\, H_{21}$.
\end{itemize}
The square root of $(E_1-E_2)^2=\mbox{tr}[H]^2-4\,  \mbox{det}[H]=(H_{11}-H_{22})^2+ 4\, H_{12}\, H_{21}< 0$ (see Eq.\ (\ref{impquant2})) is performed most conveniently in the following way:
\begin{eqnarray} \frac{E_1-E_2}{2} & = & \pm \, i\; \frac{\sqrt{4\,  \mbox{det}[H] -\mbox{tr}[H]^2}}{2}  \nonumber \\[2mm]
  & = & \pm \; i\;\, \sqrt{\left(\frac{i H_{11}-i H_{22}}{2}\right)^2+ i H_{12}\, i H_{21}} \; .
\end{eqnarray}
Taking this expression for $E_1-E_2$ and the identity $E_1+E_2=\mbox{tr}[H]=H_{11}+H_{22}$ we can use the identities Eqs. (\ref{usefid1}), (\ref{usefid2}) and (\ref{usefid3}) to determine the most convenient representation for the Hamilton operator and its corresponding eigen-values $E_1$ and $E_2$:\footnote{The application of the generalized parity operators $P_\pm$ (see Eq.\ (\ref{parop1})) to this pseudo-Hermitian Hamilton operator in the non-trivial phase interchanging the eigen-values $E_1$ and $E_2$  is equivalent to a complex conjugation of  its eigen-values $E_1$ and $E_2$. Keeping this in mind one can understand why the non-trivial phase yielding complex conjugate pairs of eigen-values with non-vanishing imaginary part is called in the literature phase of ``broken parity-time  (${\cal PT}$) symmetry". The phase of ``unbroken parity-time symmetry" implies in the literature that both eigen-values  $E_1$ and $E_2$ are real-valued. Unfortunately this situation does not not distinguish between the ``trivial phase" ($Q=1_2$) and the ``non-trivial phase" ($Q=P$). }
\begin{eqnarray}  H  & = &  \frac{\mbox{tr}[H]}{2} \; 1_2 \; \pm \; i\; \frac{\sqrt{4\,  \mbox{det}[H] -\mbox{tr}[H]^2}}{2} \; \vec{n} \cdot \vec{\sigma} \; , \label{xrep4} \\[2mm]
 E_1 & = & \frac{\mbox{tr}[H]}{2} \; \pm \, i\;  \frac{\sqrt{4\,  \mbox{det}[H] -\mbox{tr}[H]^2}}{2}\; , \label{xrep5} \\[2mm]
 E_2 & = & \frac{\mbox{tr}[H]}{2} \; \mp \, i \;  \frac{\sqrt{4\,  \mbox{det}[H] -\mbox{tr}[H]^2}}{2} \;  , \label{xrep6} 
\end{eqnarray}
or, equivalently,
\begin{eqnarray}  H  & = &  \frac{H_{11}+H_{22}}{2} \; 1_2 \; \pm \; i\;\, \sqrt{\left(\frac{i H_{11}-i H_{22}}{2}\right)^2+ i H_{12}\, i H_{21}} \;\; \vec{n} \cdot \vec{\sigma} \; , \quad  \nonumber \\[2mm]
 & &  \label{zrep4} \\
 E_1 & = & \frac{H_{11}+H_{22}}{2} \; \pm \; i\;\, \sqrt{\left(\frac{i H_{11}-i H_{22}}{2}\right)^2+ i H_{12}\, i H_{21}}\; , \label{zrep5} \\[2mm]
 E_2 & = & \frac{H_{11}+H_{22}}{2} \; \mp \; i\;\, \sqrt{\left(\frac{i H_{11}-i H_{22}}{2}\right)^2+ i H_{12}\, i H_{21}} \;  . \label{zrep6} 
\end{eqnarray}
Eqs.\ (\ref{xrep4}) and (\ref{zrep4}) can be solved for the matrix $\vec{n} \cdot \vec{\sigma}$ to determine the most convenient expression for the matrix $\vec{n} \cdot \vec{\sigma}$ for a pseudo-Hermitian Hamilton operator in the non-trivial phase:
\begin{eqnarray} \vec{n} \cdot \vec{\sigma} & = &  \mp\, i \; \frac{H - \frac{\mbox{tr}[H]}{2} \; 1_2}{\sqrt{\mbox{det}[H]-\frac{\mbox{tr}[H]^2}{4}}}  \; = \; \mp\, i \;  \frac{H - \frac{H_{11}+H_{22}}{2} \; 1_2}{\sqrt{\left(\frac{i H_{11}-i H_{22}}{2}\right)^2+ i H_{12}\, i H_{21}}} \; . \nonumber \\
 \label{nsig2}
\end{eqnarray}
With the help of these results we can denote on the basis of Eqs.\ (\ref{etazzz1})  and (\ref{etazzz2}) the following convenient representation for the metric $\eta=\eta_1+\eta_2$ for a two-dimensional pseudo-Hermitian Hamilton operator in the non-trivial phase ($Q=P$), i.\ e., 

\begin{eqnarray} 
\eta_1 & = & \mbox{\small $\bigcirc\hspace{-3.2mm}\pm$}\;\;\frac{N^\ast_2\,N_1+N^\ast_1\,N_2}{2} \;\frac{1}{2}   \left|  \frac{ \frac{H_{11}-H_{22}}{2}\pm  i\, \sqrt{\left(\frac{i H_{11}-i H_{22}}{2}\right)^2+ i H_{12}\, i H_{21}}}{\pm \, i\, \sqrt{\left(\frac{i H_{11}-i H_{22}}{2}\right)^2+ i H_{12}\, i H_{21}}}\right| \cdot \nonumber  \\[2mm]
 & \cdot &  \left(1_2  -  \frac{\left(\begin{array}{cc} 0 & H^\ast_{21} \\ -\,H^\ast_{12} & 0 \end{array}\right)}{\frac{H^\ast_{11}-H^\ast_{22}}{2}\mp  i \, \sqrt{\left(\frac{i H_{11}-i H_{22}}{2}\right)^2+ i H_{12}\, i H_{21}}}  \right)  \; \vec{e} \cdot \vec{\sigma} \nonumber\\[2mm]
 & \cdot & \left(1_2  \; - \; \frac{\left(\begin{array}{cc} 0 & -\,H_{12} \\ H_{21} & 0 \end{array}\right)}{\frac{H_{11}-H_{22}}{2}\pm  i\, \sqrt{\left(\frac{i H_{11}-i H_{22}}{2}\right)^2+ i H_{12}\, i H_{21}}}  \right)  ,  \\[20mm]
 & & \nonumber \\
 & & \nonumber \\
\eta_2 & = & \mbox{\small $\bigcirc\hspace{-3.2mm}\pm$}\;\;\frac{N^\ast_2\,N_1-N^\ast_1\,N_2}{2\, i} \;\frac{1}{2}   \left|  \frac{ \frac{H_{11}-H_{22}}{2}\pm  i\, \sqrt{\left(\frac{i H_{11}-i H_{22}}{2}\right)^2+ i H_{12}\, i H_{21}}}{\pm  \, i\, \sqrt{\left(\frac{i H_{11}-i H_{22}}{2}\right)^2+ i H_{12}\, i H_{21}}}\right| \cdot \nonumber  \\[2mm]
 & \cdot &  \left(1_2  -  \frac{\left(\begin{array}{cc} 0 & H^\ast_{21} \\ -\,H^\ast_{12} & 0 \end{array}\right)}{\frac{H^\ast_{11}-H^\ast_{22}}{2}\mp  i \, \sqrt{\left(\frac{i H_{11}-i H_{22}}{2}\right)^2+ i H_{12}\, i H_{21}}}  \right)  \; \vec{e} \cdot \vec{\sigma}_\perp \nonumber\\[2mm]
 & \cdot & \left(1_2  \; - \; \frac{\left(\begin{array}{cc} 0 & -\,H_{12} \\ H_{21} & 0 \end{array}\right)}{\frac{H_{11}-H_{22}}{2}\pm  i\, \sqrt{\left(\frac{i H_{11}-i H_{22}}{2}\right)^2+ i H_{12}\, i H_{21}}}  \right)  ,   
 \end{eqnarray}
while $\vec{e} \cdot \vec{\sigma}$ and $\vec{e} \cdot \vec{\sigma}_\perp$ are given by Eqs.\ (\ref{vvee1}) and (\ref{vvee2}), respectively.
 
\item {\bf Case 3:} $H$ is anti-pseudo-Hermitian ($H^+ = - \,\eta \, H \, \eta^{-1}$) implying\\
\mbox{} $\qquad\quad\;\;$ $i H$ to be pseudo-Hermitian  ($(i H)^+ = + \,\eta \; i H \; \eta^{-1}$), \\
\mbox{} $\qquad\quad\;\;$ the phase of the metric is trivial ($Q=1_2$)\\[2mm]
There hold the following properties:
\begin{itemize} \item The eigen-values $E_1$ and $E_2$ of $H$ are purely imaginary, i.e. there holds: $i E_1=(i\,E_1)^\ast$, $i E_2=(i\,E_2)^\ast$ (see also Eq.\ (\ref{enconstr3})). 
\item The quantities $i E_1+i E_2= i \, \mbox{tr}[H]=i\,H_{11}+i\,H_{22}$ and $(i E_1-i E_2)^2=4\,  \mbox{det}[H]-\mbox{tr}[H]^2=(i\,H_{11}-i\,H_{22})^2+ 4\, i\, H_{12}\, i\, H_{21}$ are ---  due to the anti-pseudo-Hermiticity of $H$  --- real-valued.
\item According to  Eq.\ (\ref{enconstr7}) there holds in the trivial phase ($Q=1_2$) the inequality $\mbox{tr}[H]^2< 4\,  \mbox{det}[H]$, or, equivalently, $(H_{11}-H_{22})^2\; <\;-\, 4\, H_{12}\, H_{21}$.
\end{itemize}
The square root of $(i E_1-i E_2)^2=4\,  \mbox{det}[H]-\mbox{tr}[H]^2=(i\,H_{11}-i\,H_{22})^2+ 4\, i\, H_{12}\, i\, H_{21}> 0$ (see Eq.\ (\ref{impquant4})) is performed most conveniently in the following way:
\begin{eqnarray} \frac{i E_1-i E_2}{2} & = & \mp \; \frac{\sqrt{4\,  \mbox{det}[H]-\mbox{tr}[H]^2}}{2} \nonumber \\[2mm]
  & = & \mp \;\, \sqrt{\left(\frac{i H_{11}-i H_{22}}{2}\right)^2+ i H_{12}\, i H_{21}}\; ,
\end{eqnarray}
or, equivalently,

\begin{eqnarray} \frac{E_1- E_2}{2} & = & \pm \, i \; \frac{\sqrt{4\,  \mbox{det}[H]-\mbox{tr}[H]^2}}{2} \nonumber \\[2mm]
  & = & \pm \; i\;\, \sqrt{\left(\frac{i H_{11}-i H_{22}}{2}\right)^2+ i H_{12}\, i H_{21}} \; .
\end{eqnarray}
Taking this expression for $E_1-E_2$ and the identity $E_1+E_2=\mbox{tr}[H]=H_{11}+H_{22}$ we can use the identities Eqs. (\ref{usefid1}), (\ref{usefid2}) and (\ref{usefid3}) to determine the most convenient representation for the Hamilton operator and its corresponding eigen-values $E_1$ and $E_2$:
\begin{eqnarray}  H  & = &  \frac{\mbox{tr}[H]}{2} \; 1_2 \; \pm \, i\; \frac{\sqrt{4\,  \mbox{det}[H]-\mbox{tr}[H]^2}}{2} \; \vec{n} \cdot \vec{\sigma} \; , \label{xrep7} \\[2mm]
 E_1 & = & \frac{\mbox{tr}[H]}{2} \; \pm \, i \;  \frac{\sqrt{4\,  \mbox{det}[H]-\mbox{tr}[H]^2}}{2}\; , \label{xrep8} \\[2mm]
 E_2 & = & \frac{\mbox{tr}[H]}{2} \; \mp \, i \;  \frac{\sqrt{4\,  \mbox{det}[H]-\mbox{tr}[H]^2}}{2} \;  , \label{xrep9} 
\end{eqnarray}
or, equivalently,
\begin{eqnarray}  H  & = &  \frac{H_{11}+H_{22}}{2} \; 1_2 \; \pm \; i\;\, \sqrt{\left(\frac{i H_{11}-i H_{22}}{2}\right)^2+ i H_{12}\, i H_{21}} \;\; \vec{n} \cdot \vec{\sigma} \; , \quad  \nonumber \\[2mm]
 & &  \label{zrep7} \\
 E_1 & = & \frac{H_{11}+H_{22}}{2} \; \pm \; i\;\, \sqrt{\left(\frac{i H_{11}-i H_{22}}{2}\right)^2+ i H_{12}\, i H_{21}}\; , \label{zrep8} \\[2mm]
 E_2 & = & \frac{H_{11}+H_{22}}{2} \; \mp \; i\;\, \sqrt{\left(\frac{i H_{11}-i H_{22}}{2}\right)^2+ i H_{12}\, i H_{21}} \;  . \label{zrep9} 
\end{eqnarray}
Eqs.\ (\ref{xrep7}) and (\ref{zrep7}) can be solved for the matrix $\vec{n} \cdot \vec{\sigma}$ to determine the most convenient expression for the matrix $\vec{n} \cdot \vec{\sigma}$ for a anti-pseudo-Hermitian Hamilton operator in the trivial phase:
\begin{eqnarray} \vec{n} \cdot \vec{\sigma} & = &  \mp\, i \; \frac{H - \frac{\mbox{tr}[H]}{2} \; 1_2}{\sqrt{\mbox{det}[H]-\frac{\mbox{tr}[H]^2}{4}}}  \; = \; \mp\, i \;  \frac{H - \frac{H_{11}+H_{22}}{2} \; 1_2}{\sqrt{\left(\frac{i H_{11}-i H_{22}}{2}\right)^2+ i H_{12}\, i H_{21}}} \; . \nonumber \\
 \label{nsig3}
\end{eqnarray}
With the help of these results we can denote on the basis of Eqs.\ (\ref{etazzz0})  and (\ref{etazzz3}) the following convenient representation for the metric $\eta=\eta_0+\eta_3$ for a two-dimensional anti-pseudo-Hermitian Hamilton operator in the trivial phase ($Q=1_2$), i.\ e., 
\begin{eqnarray} 
\eta_0  & = & \mbox{\small $\bigcirc\hspace{-3.2mm}\pm$}\;\;\frac{|N_1|^2+|N_2|^2}{2}  \;\frac{1}{2}   \left|  \frac{ \frac{H_{11}-H_{22}}{2}\pm  i\, \sqrt{\left(\frac{i H_{11}-i H_{22}}{2}\right)^2+ i H_{12}\, i H_{21}}}{\pm \, i\, \sqrt{\left(\frac{i H_{11}-i H_{22}}{2}\right)^2+ i H_{12}\, i H_{21}}}\right| \cdot \nonumber  \\[2mm]
 & \cdot &  \left(1_2  -  \frac{\left(\begin{array}{cc} 0 & H^\ast_{21} \\ -\,H^\ast_{12} & 0 \end{array}\right)}{\frac{H^\ast_{11}-H^\ast_{22}}{2}\mp  i \, \sqrt{\left(\frac{i H_{11}-i H_{22}}{2}\right)^2+ i H_{12}\, i H_{21}}}  \right)  \nonumber\\[2mm]
 & \cdot & \left(1_2  \; - \; \frac{\left(\begin{array}{cc} 0 & -\,H_{12} \\ H_{21} & 0 \end{array}\right)}{\frac{H_{11}-H_{22}}{2}\pm  i\, \sqrt{\left(\frac{i H_{11}-i H_{22}}{2}\right)^2+ i H_{12}\, i H_{21}}}  \right)  ,  \\[2mm]
 & & \nonumber \\
\eta_3  & = & \mbox{\small $\bigcirc\hspace{-3.2mm}\pm$}\;\;\frac{|N_1|^2-|N_2|^2}{2} \;\frac{1}{2}   \left|  \frac{ \frac{H_{11}-H_{22}}{2}\pm  i\, \sqrt{\left(\frac{i H_{11}-i H_{22}}{2}\right)^2+ i H_{12}\, i H_{21}}}{\pm  \, i\, \sqrt{\left(\frac{i H_{11}-i H_{22}}{2}\right)^2+ i H_{12}\, i H_{21}}}\right| \cdot \nonumber  \\[2mm]
 & \cdot &  \left(1_2  -  \frac{\left(\begin{array}{cc} 0 & H^\ast_{21} \\ -\,H^\ast_{12} & 0 \end{array}\right)}{\frac{H^\ast_{11}-H^\ast_{22}}{2}\mp  i \, \sqrt{\left(\frac{i H_{11}-i H_{22}}{2}\right)^2+ i H_{12}\, i H_{21}}}  \right)  \left(\begin{array}{cc} 1 & 0 \\ 0 & -1 \end{array}\right) \nonumber\\[2mm]
 & \cdot & \left(1_2  \; - \; \frac{\left(\begin{array}{cc} 0 & -\,H_{12} \\ H_{21} & 0 \end{array}\right)}{\frac{H_{11}-H_{22}}{2}\pm  i\, \sqrt{\left(\frac{i H_{11}-i H_{22}}{2}\right)^2+ i H_{12}\, i H_{21}}}  \right)  .
\end{eqnarray}
\newpage
\item {\bf Case 4:} $H$ is anti-pseudo-Hermitian ($H^+ = - \,\eta \, H \, \eta^{-1}$) implying\\
\mbox{} $\qquad\quad\;\;$ $i H$ to be pseudo-Hermitian  ($(i H)^+ = + \,\eta \; i H \; \eta^{-1}$), \\
\mbox{} $\qquad\quad\;\;$ the phase of the metric is non-trivial ($Q=P$)\\[2mm]
There hold the following properties:
\begin{itemize} \item The eigen-values $i E_1$ and $i E_2$ of $i H$ form complex conjugate pairs, i.e. there holds: $i E_1=(i\,E_2)^\ast$, $i E_2=(i\,E_1)^\ast$ (see also Eq.\ (\ref{enconstr4})). 
\item The quantities $i E_1+i E_2= i\, \mbox{tr}[H]=i\,H_{11}+i\,H_{22}$ and $(i E_1-i E_2)^2=4\,  \mbox{det}[H]-\mbox{tr}[H]^2=(i\,H_{11}-i\,H_{22})^2+ 4\, i\, H_{12}\, i\, H_{21}$ are ---  due to the anti-pseudo-Hermiticity of $H$  --- real-valued.
\item According to  Eq.\ (\ref{enconstr8}) there holds in the non-trivial phase \mbox{($Q=P$)} the inequality $\mbox{tr}[H]^2> 4\,  \mbox{det}[H]$, or, equivalently, $(H_{11}-H_{22})^2\; >\;-\, 4\, H_{12}\, H_{21}$.
\end{itemize}
The square root of $(i E_1-i E_2)^2=4\,  \mbox{det}[H]-\mbox{tr}[H]^2=(i\,H_{11}-i\,H_{22})^2+ 4\, i\, H_{12}\, i\, H_{21}< 0$ (see Eq.\ (\ref{impquant4})) is performed most conveniently in the following way:
\begin{eqnarray} \frac{i E_1-i E_2}{2} & = & \pm \; i \; \frac{\sqrt{\mbox{tr}[H]^2-4\,  \mbox{det}[H] }}{2} \nonumber \\[2mm]
  & = & \pm \; i\; \sqrt{\left(\frac{H_{11}-H_{22}}{2}\right)^2+ H_{12}\, H_{21}} \; ,
\end{eqnarray}
or, equivalently,

\begin{eqnarray} \frac{E_1-E_2}{2} & = & \pm \; \frac{\sqrt{\mbox{tr}[H]^2-4\,  \mbox{det}[H] }}{2} \nonumber \\[2mm]
  & = & \pm \; \sqrt{\left(\frac{H_{11}-H_{22}}{2}\right)^2+ H_{12}\, H_{21}} \; . 
\end{eqnarray}
Taking this expression for $E_1-E_2$ and the identity $E_1+E_2=\mbox{tr}[H]=H_{11}+H_{22}$ we can use the identities Eqs. (\ref{usefid1}), (\ref{usefid2}) and (\ref{usefid3}) to determine the most convenient representation for the Hamilton operator and its corresponding eigen-values $E_1$ and $E_2$:
\begin{eqnarray}  H  & = &  \frac{\mbox{tr}[H]}{2} \; 1_2 \; \pm \; \frac{\sqrt{\mbox{tr}[H]^2-4\,  \mbox{det}[H]}}{2} \; \vec{n} \cdot \vec{\sigma} \; , \label{xrep10} \\[2mm]
 E_1 & = & \frac{\mbox{tr}[H]}{2} \; \pm  \;  \frac{\sqrt{\mbox{tr}[H]^2-4\,  \mbox{det}[H]}}{2}\; , \label{xrep11} \\[2mm]
 E_2 & = & \frac{\mbox{tr}[H]}{2} \; \mp  \;  \frac{\sqrt{\mbox{tr}[H]^2-4\,  \mbox{det}[H]}}{2} \;  . \label{xrep12} 
\end{eqnarray}
or, equivalently,
\begin{eqnarray}  H  & = &  \frac{H_{11}+H_{22}}{2} \; 1_2 \; \pm \;  \sqrt{\left(\frac{H_{11}-H_{22}}{2}\right)^2+ H_{12}\, H_{21}} \;\; \vec{n} \cdot \vec{\sigma} \; , \quad \label{zrep10} \\[2mm]
 E_1 & = & \frac{H_{11}+H_{22}}{2} \; \pm \;   \sqrt{\left(\frac{H_{11}-H_{22}}{2}\right)^2+ H_{12}\, H_{21}}\; , \label{zrep11} \\[2mm]
 E_2 & = & \frac{H_{11}+H_{22}}{2} \; \mp \;   \sqrt{\left(\frac{H_{11}-H_{22}}{2}\right)^2+ H_{12}\, H_{21}} \;  . \label{zrep12} 
\end{eqnarray}
Eqs.\ (\ref{xrep10}) and  (\ref{zrep10}) can be solved for the matrix $\vec{n} \cdot \vec{\sigma}$ to determine the most convenient expression for the matrix $\vec{n} \cdot \vec{\sigma}$ for a anti-pseudo-Hermitian Hamilton operator in the non-trivial phase:
\begin{eqnarray} \vec{n} \cdot \vec{\sigma} & = & \pm\; \frac{H - \frac{\mbox{tr}[H]}{2} \; 1_2}{\sqrt{\frac{\mbox{tr}[H]^2}{4}-  \mbox{det}[H] }}  \; = \; \pm\; \frac{H - \frac{H_{11}+H_{22}}{2} \; 1_2}{\sqrt{\left(\frac{H_{11}-H_{22}}{2}\right)^2+ H_{12}\, H_{21}}} \; . \nonumber \\
 \label{nsig4}
\end{eqnarray}
With the help of these results we can denote on the basis of Eqs.\ (\ref{etazzz1})  and (\ref{etazzz2}) the following convenient representation for the metric $\eta=\eta_1+\eta_2$ for a two-dimensional anti-pseudo-Hermitian Hamilton operator in the non-trivial phase ($Q=P$), i.\ e., 
\begin{eqnarray} 
\eta_1  & = & \mbox{\small $\bigcirc\hspace{-3.2mm}\pm$}\;\;\frac{N^\ast_2\,N_1+N^\ast_1\,N_2}{2} \,  \left|  \frac{ \frac{H_{11}-H_{22}}{2}\pm \; \sqrt{\left(\frac{H_{11}-H_{22}}{2}\right)^2+ H_{12}\, H_{21}}}{\pm \;\sqrt{\left(\frac{H_{11}-H_{22}}{2}\right)^2+ H_{12}\, H_{21}}}\right| \; \cdot \nonumber  \\[2mm]
 & \cdot &  \left(1_2 \; - \; \frac{\left(\begin{array}{cc} 0 & H^\ast_{21} \\ -\,H^\ast_{12} & 0 \end{array}\right)}{\frac{H^\ast_{11}-H^\ast_{22}}{2}\pm \; \sqrt{\left(\frac{H^\ast_{11}-H^\ast_{22}}{2}\right)^2+ H^\ast_{21}\, H^\ast_{12}}}  \right)  \vec{e} \cdot \vec{\sigma} \nonumber\\[2mm]
 & \cdot & \left(1_2  \; - \; \frac{\left(\begin{array}{cc} 0 & -\,H_{12} \\ H_{21} & 0 \end{array}\right)}{\frac{H_{11}-H_{22}}{2}\pm \; \sqrt{\left(\frac{H_{11}-H_{22}}{2}\right)^2+ H_{12}\, H_{21}}}  \right)  ,  \\[2mm]
\eta_2  & = & \mbox{\small $\bigcirc\hspace{-3.2mm}\pm$}\;\;\frac{N^\ast_2\,N_1-N^\ast_1\,N_2}{2\, i} \;\;\frac{1}{2} \,  \left|  \frac{\frac{H_{11}-H_{22}}{2}\pm \; \sqrt{\left(\frac{H_{11}-H_{22}}{2}\right)^2+ H_{12}\, H_{21}}}{\pm \;\sqrt{\left(\frac{H_{11}-H_{22}}{2}\right)^2+ H_{12}\, H_{21}}}\right| \; \cdot \nonumber  \\[2mm]
 & \cdot &  \left(1_2 \; - \; \frac{\left(\begin{array}{cc} 0 & H^\ast_{21} \\ -\,H^\ast_{12} & 0 \end{array}\right)}{\frac{H^\ast_{11}-H^\ast_{22}}{2}\pm \; \sqrt{\left(\frac{H^\ast_{11}-H^\ast_{22}}{2}\right)^2+ H^\ast_{21}\, H^\ast_{12}}}  \right)  \vec{e} \cdot \vec{\sigma}_\perp \nonumber\\[2mm]
 & \cdot & \left(1_2 \; - \; \frac{\left(\begin{array}{cc} 0 & -\,H_{12} \\ H_{21} & 0 \end{array}\right)}{\frac{H_{11}-H_{22}}{2}\pm \; \sqrt{\left(\frac{H_{11}-H_{22}}{2}\right)^2+ H_{12}\, H_{21}}}  \right) ,
\end{eqnarray}
while $\vec{e} \cdot \vec{\sigma}$ and $\vec{e} \cdot \vec{\sigma}_\perp$ are given by Eqs.\ (\ref{vvee1}) and (\ref{vvee2}), respectively.

 \end{itemize}
Interestingly, the expressions for the pseudo-Hermitian Hamilton operator in the trivial phase Eqs.\ (\ref{xrep1}), (\ref{xrep2}), (\ref{xrep3}), (\ref{nsig1}) coincide with the respective expressions for the anti-pseudo-Hermitian Hamilton operator in the non-trivial phase Eqs.\ (\ref{xrep10}), (\ref{xrep11}), (\ref{xrep12}), (\ref{nsig4}). The same coincidence appears also between the expressions for the pseudo-Hermitian Hamilton operator in the non-trivial phase Eqs.\ (\ref{xrep4}), (\ref{xrep5}), (\ref{xrep6}), (\ref{nsig2}) and the respective expressions for the anti-pseudo-Hermitian Hamilton operator in the trivial phase Eqs.\ (\ref{xrep7}), (\ref{xrep8}), (\ref{xrep9}), (\ref{nsig3}).

\section{Selective applications}  \label{xxsec12}
In this section we want to apply our formalism to some topical Hamilton operators described by two-dimensional matrices found in the literature. The results obtained show that the formalism developed here is rather general containing --- in most cases --- the results found in the respective literature as a limit. Contrary to what has been done in the literature we determine the respective metric $\eta$ for the Hamilton operator $H$ not only for Case 1, yet also --- in most examples --- for the Cases 2, 3 and 4.
\newpage
\subsection{The metric for a two-dimensional Hamilton operator describing two interacting complex ghosts} 
In Appendix \ref{xappenda1} we summarize results in calculating the metric $\eta$ for the following two-dimensional Hamilton operator $H$ describing two complex ghosts \cite{Nakanishi:2006ct}\cite{Nakanishi:1972pt}\cite{Kleefeld:2002au}\cite{Kleefeld:2004jb}\cite{Kleefeld:2004qs}\cite{Kleefeld:1999}\cite{Ahmed:2003nn}\cite{Wang:2010br}\cite{LiGe:2016} with mass $m-i \,\varepsilon$ and $m+i\,\varepsilon$ ( $m$ and $\varepsilon$ are real-valued) interacting via a complex coupling $\gamma = \gamma_1 + i  \, \gamma_2$ (with $\gamma_1$ and $\gamma_2$ being real-valued) and the complex conjugate coupling $\gamma^\ast = \gamma_1 - i  \, \gamma_2$:
\begin{equation} H \; = \;  \left(\begin{array}{cc} m-i \,\varepsilon  & \gamma^\ast \\ \gamma & m+i \,\varepsilon \end{array}\right) \; .
\end{equation}
A discussion of important features of the Hamilton operator is found in the context of Eq.\  (\ref{compghostx1}) in Section \ref{xxsec1}. Pointing to Appendix \ref{xappenda1} for all other cases we want to present at this place only some results for Case 1, in which the real-valued eigen-values of the Hamilton operator are determined by:
\begin{equation}  E_1 \; = \; m \; \pm \;  \sqrt{|\gamma|^2 - \varepsilon^2} \; ,  \quad E_2 \;  = \; m \; \mp \;  \sqrt{|\gamma|^2 - \varepsilon^2} \; .
\end{equation} 
The respective positive semi-definite metric $\eta_+$  of Case 1 obtained by $\eta \; = \; \eta_0 + \eta_3\,$ is given by (For details see Appendix \ref{xappenda1}!):
\begin{eqnarray} \eta_+ & = & \frac{|N_1|^2+|N_2|^2}{2} \;\frac{|\gamma|}{\sqrt{|\gamma|^2 - \varepsilon^2}}\; \left(\begin{array}{cc} 1  & + i \, \frac{\gamma^\ast \varepsilon}{|\gamma|^2} \\  - i \, \frac{\gamma \, \varepsilon}{|\gamma|^2} & 1 \end{array}\right) \nonumber \\[2mm]
 & \pm & \frac{|N_1|^2-|N_2|^2}{2} \left(\begin{array}{cc} 0  & \frac{\gamma^\ast}{|\gamma|} \\  \frac{\gamma}{|\gamma|} & 0 \end{array}\right) ,
\end{eqnarray}
\subsection{The metric for a two-dimensional Hamilton operator proposed by C.~M.~Bender et al.} 
\noindent Appendix \ref{xappenda2} contains our results for the metric of the following asymmetric two-dimensional Hamilton operator $H$  (with $t$, $s$, $r$, $\theta$ and $\phi$ being real-valued parameters) considered for $t\not=s$ and $\phi\not= 0$ by A.\ Das and L.~Greenwood \cite{Das:2009it}\cite{Das:2011zza} and A.~Ghatak and B.~Mandal \cite{Ghatak:2013wga}, i.~e.,
\begin{equation} H \; = \;  \left(\begin{array}{cc} r\, e^{+i\,\theta}  & s \, e^{+i\,\phi} \\ t \, e^{-i\,\phi} & r\, e^{-i\,\theta} \end{array}\right) \; ,
\end{equation}
which was proposed and discussed before by C.~M.~Bender et al.\ \cite{Bender:2002vv}\cite{Bender:2005tb}\cite{Bender:2019cwm} (see also \cite{Ahmed:2003nn}\cite{Geyer:2007}\cite{Mannheim:2009zj}) in the symmetric limit  $t=s$ and $\phi= 0$ , i.~e., 
\begin{equation} H \; = \;  \left(\begin{array}{cc} r\, e^{+i\,\theta}  & s \\ s  & r\, e^{-i\,\theta} \end{array}\right) \; . \label{hambendparis1}
\end{equation}
\begin{figure}[h]
\begin{center}
\includegraphics[width=29pc]{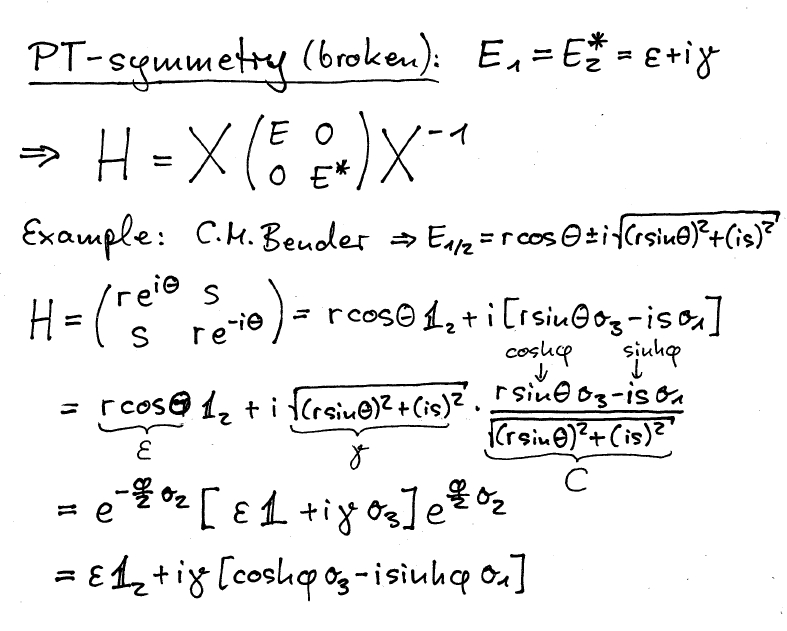}
\end{center}
\caption{The most effective procedure to diagonalize the Hamilton operator Eq.\ (\ref{hambendparis1}) and to determine the ${\cal C}$-operator with ${\cal C}^2=1_2$ had been demonstrated for the case of broken ${\cal PT}$-symmetry in our oral presentation in Paris \cite{Kleefeld:2012} in 2012.}
\end{figure}
Refering to Appendix \ref{xappenda2} for all other cases we want to consider at this place only some results for Case 1, in which the real-valued eigen-values of the asymmetric Hamilton operator ($t\not=s$ and $\phi\not= 0$) are determined by:
\begin{eqnarray}  E_1 & = & r\, \cos\theta  \; \pm \; \sqrt{s\,t - (r\, \sin\theta)^2} \; , \\
 E_2 & = & r\, \cos\theta \; \mp \; \sqrt{s\,t - (r\, \sin\theta)^2} \; .
\end{eqnarray} 
The respective positive semi-definite metric $\eta_+$ found for the asymmetric Hamilton operator ($t\not=s$ and $\phi\not= 0$)  for Case 1  is due to $\eta \; = \; \eta_0 + \eta_3$ given by the following expression (For details see Appendix \ref{xappenda2}!):
\clearpage
\begin{eqnarray}
\eta_+  & = & \frac{|N_1|^2+|N_2|^2}{2} \;\; \sqrt{\frac{s\, t}{s\,t - (r\, \sin\theta)^2}}  \nonumber \\
 & \cdot & \Bigg( \frac{s + t}{2 \, s\, t} \left(\begin{array}{cc} t &  - i \, r \, \sin\theta \;e^{+i\,\phi} \\   i \, r \, \sin\theta \; e^{-i\,\phi} & s \end{array}\right)   \nonumber \\
 &  & \pm \; \frac{s - t}{2 \, s\, t} \; \sqrt{s\,t - (r\, \sin\theta)^2} \;  \left(\begin{array}{cc} 0 &  e^{+i\,\phi} \\  e^{-i\,\phi} & 0 \end{array}\right)  \Bigg) \nonumber \\
 & + & \frac{|N_1|^2-|N_2|^2}{2} \;\; \sqrt{\frac{s\, t}{s\,t - (r\, \sin\theta)^2}}  \nonumber \\ 
 & \cdot & \Bigg( \frac{s - t}{2 \, s\, t} \left(\begin{array}{cc} t &  - i \, r \, \sin\theta \; e^{+i\,\phi} \\   i \, r \, \sin\theta \; e^{-i\,\phi} & s \end{array}\right)   \nonumber \\
 &  & \pm \; \frac{s + t}{2 \, s\, t} \; \sqrt{s\,t - (r\, \sin\theta)^2} \;  \left(\begin{array}{cc} 0 &  e^{+i\,\phi} \\  e^{-i\,\phi} & 0 \end{array}\right)  \Bigg)  . 
\end{eqnarray}
This result simplifies significantly for the symmetric Hamilton operator, i.~e., for $t=s$ and $\phi= 0$:\footnote{The term proportional to $\sigma_1$ has been discovered in \cite{Kleefeld:2009vd} due to a logical inconsistency.}
\begin{eqnarray}
\eta_+  & = & \frac{|N_1|^2+|N_2|^2}{2} \;\; \frac{1}{\sqrt{1 - \left(\frac{r}{s}\, \sin\theta\right)^2}}  \; \left(\begin{array}{cc} 1 &  -\, i \, \frac{r}{s} \, \sin\theta  \\[1mm]   i \, \frac{r}{s} \, \sin\theta \;  & 1 \end{array}\right)   \nonumber \\[1mm]
 & \pm & \frac{|N_1|^2-|N_2|^2}{2} \;\;  \left(\begin{array}{cc} 0 &  1 \\[2mm]  1 & 0 \end{array}\right) \nonumber \\[2mm]
 & = &  \frac{|N_1|^2+|N_2|^2}{2\; \sqrt{1 - \left(\frac{r}{s}\, \sin\theta\right)^2}} \;  \Bigg(  \left(\begin{array}{cc} 1 &  - i \; \frac{r}{s} \, \sin\theta  \\[1mm]   i \; \frac{r}{s} \, \sin\theta  & 1 \end{array}\right) \nonumber \\[1mm]
 & & \pm  \; \frac{|N_1|^2-|N_2|^2}{|N_1|^2+|N_2|^2} \; \sqrt{1 - \left(\frac{r}{s}\, \sin\theta\right)^2} \;  \left(\begin{array}{cc} 0 &  1 \\[1mm]  1 & 0 \end{array}\right) \Bigg) \; . \quad \label{resklee1}
\end{eqnarray}
It should be compared to the corresponding results of Ref.\ \cite{Geyer:2007}, i.\ e.,
\begin{equation} \eta_+ \; = \; \left(\begin{array}{cc} 1 &  \sin \gamma - i \; \frac{r}{s} \, \sin\theta  \\[2mm]  \sin\gamma +  i \; \frac{r}{s} \, \sin\theta  & 1 \end{array}\right) \; \end{equation}
suggesting obviously the following two relations:
\begin{eqnarray}  \frac{|N_1|^2+|N_2|^2}{2\; \sqrt{1 - \left(\frac{r}{s}\, \sin\theta\right)^2}} & = & 1 \; , \label{heissconst1} \\
 \pm  \; \frac{|N_1|^2-|N_2|^2}{|N_1|^2+|N_2|^2} \; \sqrt{1 - \left(\frac{r}{s}\, \sin\theta\right)^2} & = & \sin \gamma \; ,  \label{heissconst2}
\end{eqnarray}
implying $\sin\gamma = \pm \; \frac{|N_1|^2-|N_2|^2}{2}$ and being for $|N_1|^2\,|N_2|^2>0$ consistent with the following constraint imposed by Ref.\ \cite{Geyer:2007}:
\begin{equation} \left( \sin \gamma \right)^2 + \left( \frac{r}{s} \, \sin\theta  \right)^2 < 1 \; . \end{equation} 
It has to be stated nonetheless that  relation  (\ref{heissconst1})  underlying Ref.\  \cite{Geyer:2007} is imposing --- without need --- a constraining upper bound to the values of $|N_1|^2$ and $|N_2|^2$ being not present in our result Eq.\ (\ref{resklee1}).
\subsection{The metric for a general ${\cal PT}$-symmetric Hamilton operator proposed by C.\ M.\ Bender, P.~Meisinger and  Q.\ Wang extended by A.~Mostafazadeh}  
\noindent In Appendix \ref{xappenda3} we present our results for the metric of the following --- due to $u\not=0$  --- asymmetric two-dimensional Hamilton operator $H$  investigated by A.\ Mostafazadeh \cite{Mostafazadeh:2003gz} (see also Q.~Wang et al.\ \cite{Wang:2010br}) (with $r$, $s$, $t$, $u$ and $\phi$ being real-valued parameters):
\begin{equation} H =  \left(\begin{array}{cc} [\,r + (t\, \cos\phi - i \, s\,\sin\phi)\, ] & [\, i\, ( s\, \cos\phi - u) + t \sin\phi\, ] \\[2mm]  [\, i\, ( s\, \cos\phi + u) + t \sin\phi \,] &   [\, r -(t\, \cos\phi - i \, s\,\sin\phi)\,] \end{array}\right) \; , 
\end{equation}
which had been proposed and considered before for in the symmetric limit $u=0$ by C.~M.~Bender, P.~Meisinger and  Q.\ Wang \cite{Bender:2003gu}\cite{Wang:2010br}\cite{Wang:2013oqw}, i.~e.:
\begin{equation} H =  \left(\begin{array}{cc} [\,r + t\, \cos\phi - i \, s\,\sin\phi\, ] & [\, i\, s\, \cos\phi + t \sin\phi\, ] \\[2mm]  [\, i\,  s\, \cos\phi + t \sin\phi \,] &   [\, r -t\, \cos\phi + i \, s\,\sin\phi\,] \end{array}\right) \; .
\end{equation}
Again we present here only some results of Appendix \ref{xappenda3} related to Case 1 for which the real-valued eigen-values of the asymmetric Hamilton operator are denoted in the following way
\begin{equation} E_1 \; = \; r \pm \sqrt{t^2 + u^2 - s^2} \; , \quad E_2 \; = \; r \mp \sqrt{t^2 + u^2 - s^2} \; ,
\end{equation}
and the positive semi-definite metric $\eta_+$ is obtained by $\eta=\eta_0 + \eta_3$ to be the following expression (For details see Appendix \ref{xappenda3}!):
\clearpage
\begin{eqnarray} \eta_+  & = &  \frac{|N_1|^2+|N_2|^2}{2} \;\; \frac{1}{\sqrt{t^2 + u^2 - s^2}}  \nonumber \\[2mm]
 & \cdot &  \frac{1}{\sqrt{\left(\pm \;  \sqrt{t^2 + u^2 - s^2} + t \cos\phi \right)^2 + (s\,\sin\phi)^2}}   \nonumber \\[2mm]
 & \cdot & \Bigg(\left(\begin{array}{cc} t^2 + u^2   &   i \,s\, t\\  -\,  i \, s\,  t  & t^2 + u^2 \end{array}\right)   + s\, u \; \left(\begin{array}{cc} \cos\phi &  \sin\phi \\   \sin\phi & -\cos\phi \end{array}\right) \nonumber \\[2mm]
 &  & \quad \pm \; \cos\phi \;\; \sqrt{t^2 + u^2 - s^2} \;  \left(\begin{array}{cc} t &  i\, s \\  -i\, s & t \end{array}\right)  \Bigg) \; ,\\[2mm]
 & + & \frac{|N_1|^2-|N_2|^2}{2} \;\; \frac{1}{\sqrt{t^2 + u^2 - s^2}}  \nonumber \\[2mm]
 & \cdot &  \frac{1}{\sqrt{\left(\pm \;  \sqrt{t^2 + u^2 - s^2} + t \cos\phi \right)^2 + (s\,\sin\phi)^2}}   \nonumber \\[2mm]
 & \cdot & \Bigg( \cos\phi \;\Bigg(  (t^2- s^2) \left(\begin{array}{cc} \cos\phi &  \sin\phi \\   \sin\phi & -\cos\phi \end{array}\right)  - u\, \left(\begin{array}{cc}  s & i\, t \\  -i\,t & s \end{array}\right) \Bigg)  \nonumber \\[2mm]
 &  & \pm  \; \sqrt{t^2 + u^2 - s^2} \;  \Bigg(  t \left(\begin{array}{cc} \cos\phi &  \sin\phi \\   \sin\phi & -\cos\phi \end{array}\right)  + i\, u\; \left(\begin{array}{cc}  0 & -\,1 \\  1 & 0 \end{array}\right) \Bigg)   \Bigg)  , \nonumber \\
\end{eqnarray}
which simplifies in the symmetric limit $u=0$ considerably:
\begin{eqnarray} \eta_+  & = &   \frac{ t  \;   \pm \; \cos\phi \; \sqrt{t^2  - s^2}}{\sqrt{\left(\pm \;  \sqrt{t^2 - s^2} + t \cos\phi \right)^2 + (s\,\sin\phi)^2}}   \nonumber \\[2mm]
 & \cdot & \Bigg( \; \frac{|N_1|^2+|N_2|^2}{2} \;\frac{1}{\sqrt{t^2  - s^2}} \;  \left(\begin{array}{cc} t &  i\, s \\  -i\, s & t \end{array}\right)  \nonumber \\[2mm]
 &  & \pm \; \frac{|N_1|^2-|N_2|^2}{2} \; \left(\begin{array}{cc} \cos\phi &  \sin\phi \\   \sin\phi & -\cos\phi \end{array}\right) \Bigg)  \,  .
\end{eqnarray}
\noindent This result should be compared to the corresponding metric $\eta_+$ resulting from the expressions for ${\cal C}$ and ${\cal P}$ found in Ref.\ \cite{Bender:2003gu} by invoking Eq.\ (\ref{etaab2}), i.~e., $\eta B = ({\cal CP})^T= {\cal P}^T{\cal C}^T\stackrel{!}{=} {\cal P}\, {\cal C}$:
\begin{eqnarray} \lefteqn{\eta_+\, B \; = \; \frac{1}{\cos\alpha} \; \cdot} \nonumber \\[2mm]
 & \cdot & \left(\begin{array}{cc} \cos\phi & \;\;\;\sin \phi \\ \sin\phi &  -\cos\phi  \end{array}\right) \left(\begin{array}{cc} \cos\phi - i\,\sin\alpha \, \sin\phi & \;\;\;\sin \phi + i\,\sin\alpha \, \cos\phi \\ \sin\phi + i\,\sin\alpha \, \cos\phi &  -\cos\phi+ i\,\sin\alpha \, \sin\phi \end{array}\right)\nonumber \\[2mm]
& = & \frac{1}{\cos\alpha} \left( \left(\begin{array}{cc} 1 & 0 \\ 0 &  1  \end{array}\right) + i \sin\alpha \left(\begin{array}{cc} \cos\phi & \;\;\;\sin \phi \\ \sin\phi &  -\cos\phi  \end{array}\right)\left(\begin{array}{cc}  -\, \sin\phi & \cos\phi \\  \;\;\;\,\cos\phi &  \sin\phi \end{array}\right) \right) \nonumber \\[2mm]
& = & \frac{1}{\cos\alpha} \left(\begin{array}{cc} 1 &  i \sin\alpha \\[1mm] - i \sin\alpha &  1  \end{array}\right) \nonumber \\[2mm]
& = & \frac{t}{\sqrt{t^2  - s^2}} \left(\begin{array}{cc} 1 &  i \,\frac{s}{t} \\[1mm] - \,i \,\frac{s}{t} &  1  \end{array}\right) \; = \;  \frac{1}{\sqrt{t^2  - s^2}} \left(\begin{array}{cc} t &  i \,s \\ - i \,s &  t  \end{array}\right)  \; ,
\end{eqnarray}
implying due to Eq.\ (\ref{cinvol1}) obviously $|N_1|^2=|N_2|^2=1$ and
\begin{equation} B \; = \;  \frac{\sqrt{\left(\pm \;  \sqrt{t^2 - s^2} + t \cos\phi \right)^2 + (s\,\sin\phi)^2}}{ t  \;   \pm \; \cos\phi \; \sqrt{t^2  - s^2}} \; \; 1_2 \; . \end{equation}

\subsection{The metric for the Klein-Gordon Hamilton operator by H.~Feshbach and F.~Villars}
The determination of a positive semi-definite scalar product for the Klein-Gordon theory describing Bosonic fields of zero spin has been an open problem for a long time which has been finally solved for the charged Klein-Gordon field by A.\ Mostafazadeh and F.\ Zamani  between 2003 \cite{Mostafazadeh:2003vy} and 2006 \cite{Mostafazadeh:2006ug}\cite{Mostafazadeh:2006gx} and --- shortly after also in 2006 --- for the neutral Klein-Gordon field by F.\ Kleefeld \cite{Kleefeld:2006bp}. The solution obtained in Ref.\ \cite{Kleefeld:2006bp} for the neutral Klein-Gordon field can be reproduced with the help of the methods presented in this manuscript by starting out with the two-dimensional representation of the Hamilton operator for the Klein-Gordon field suggested by H.~Feshbach and F.~Villars \cite{Feshbach:1958wv} in 1958:
\begin{equation}
H \; = \; (\sigma_3 + i\; \sigma_2) \; \frac{\vec{p}^{\, 2}}{2\, m} \; + m \; \sigma_3 \; = \; \left(\begin{array}{cc} m\; + \;  \frac{\vec{p}^{\, 2}}{2\, m}  &  \frac{\vec{p}^{\, 2}}{2\, m} \\[2mm]  - \, \frac{\vec{p}^{\, 2}}{2\, m} & - \left(m \; + \;  \frac{\vec{p}^{\, 2}}{2\, m}\right) \end{array}\right) \; .
\end{equation}
With the help of Eqs.\ (\ref{xxxident1}) and (\ref{xxxident3}) it is straight forward to determine the expressions for the eigen-value matrices $X$ and $X^+$:
\begin{eqnarray} X & = & X^+ \; =\nonumber \\[2mm]
& = &   \left(\frac{\pm\; \sqrt{\vec{p}^{\, 2} + m^2}  +  m + \frac{\vec{p}^{\, 2}}{2\, m}}{\pm \; 2\, \sqrt{\vec{p}^{\, 2} + m^2}} \right)^{\frac{1}{2}} \left( 1_2 +   \frac{ \frac{\vec{p}^{\, 2}}{2\, m} \left(\begin{array}{cc} 0  & 1 \\ 1 & 0 \end{array}\right)}{\pm\; \sqrt{\vec{p}^{\, 2} + m^2} \; + \; m + \frac{\vec{p}^{\, 2}}{2\, m}} \right) \nonumber \\
& = &   \left(\frac{\pm\; \sqrt{\vec{p}^{\, 2} + m^2}  +  m + \frac{\vec{p}^{\, 2}}{2\, m}}{\pm \; 2\; \sqrt{\vec{p}^{\, 2} + m^2}} \right)^{\frac{1}{2}} \nonumber \\[2mm]
 & \cdot & \left( 1_2 +   \frac{\mp\; \sqrt{\vec{p}^{\, 2} + m^2}  +  m + \frac{\vec{p}^{\, 2}}{2\, m}}{\frac{\vec{p}^{\, 2}}{2\, m}} \; \left(\begin{array}{cc} 0  & 1 \\ 1 & 0 \end{array}\right)\right) \; ,
\end{eqnarray}
yielding the following positive metric $\eta$ for Case 1, i.\ e. a pseudo-Hermitian Hamilton operator ($H^+=\eta\, H \,  \eta^{-1}$) in the ``trivial phase" ($Q=1_2$):
\begin{eqnarray} \eta & = & \eta_0 + \eta_3 \; = \; \frac{|N_1|^2+|N_2|^2}{2} \; \;   X^+ X \; + \; \frac{|N_1|^2-|N_2|^2}{2} \; \;   X^+ \sigma_3 \, X \nonumber \\[2mm]
 & = &  \mbox{\small $\bigcirc\hspace{-3.2mm}\pm$} \;  \Bigg( \;  \frac{|N_1|^2+|N_2|^2}{2} \; \; \frac{1}{2} \left(\begin{array}{cc} \frac{\omega(\vec{p}\,)}{m} + \frac{m}{\omega(\vec{p}\,)} & \frac{\omega(\vec{p}\,)}{m} - \frac{m}{\omega(\vec{p}\,)} \\[2mm] \frac{\omega(\vec{p}\,)}{m} - \frac{m}{\omega(\vec{p}\,)} & \frac{\omega(\vec{p}\,)}{m} + \frac{m}{\omega(\vec{p}\,)} \end{array}\right) \nonumber \\[1mm]
 & & \quad\, \pm\,  \frac{|N_1|^2-|N_2|^2}{2} \; \; \left(\begin{array}{cc} 1  & 0 \\ 0 & -1 \end{array}\right) \Bigg) \nonumber \\[1mm]
 & = &  \mbox{\small $\bigcirc\hspace{-3.2mm}\pm$} \;\left(\; \frac{|N_1|^2+|N_2|^2}{2} \; \;    \frac{\sigma_3 \; H}{\omega(\vec{p}\,)} \; \pm \; \frac{|N_1|^2-|N_2|^2}{2} \; \; \sigma_3 \;\right) \; .
\end{eqnarray}
As pointed out previously, this result for $\eta$ coincides with the result obtained in  \cite{Kleefeld:2006bp} when reexpressing the real-valued Klein-Gordon field $\psi$ in terms of the two components $\varphi$ and $\chi$ of the Feshbach-Villars field by applying the identities $\psi\; =\; \varphi + \chi$ and $i\, \dot{\psi}\; =\; m\, ( \varphi - \chi)\; $  (with $\sqrt{D} \; = \; \omega(\vec{p})\,$):
\begin{eqnarray} \lefteqn{\frac{1}{2\, m}\, \left(\begin{array}{c} D^{1/4} \, \psi \;\; , \;   D^{-1/4} \, \dot{\psi} \end{array}\right) \left(\begin{array}{cc} 1  & -\, i\, a \\[2mm] i\, a & 1 \end{array}\right) \left(\begin{array}{c} D^{1/4} \,  \psi \\[2mm]   D^{-1/4} \,  \dot{\psi} \end{array} \right) \; =} \nonumber \\
 & = &  \frac{1}{2\, m}\, \left(\begin{array}{c}  \psi \;\; , \;  \dot{\psi} \end{array}\right)  \left(\begin{array}{cc} \displaystyle \sqrt{D}  & -\, i\, a \, \\[2mm]  i\, a & \displaystyle \frac{1}{\sqrt{D}} \end{array}\right) \left(\begin{array}{c} \psi \\[2mm]   \dot{\psi} \end{array} \right) \nonumber \\[2mm]
 & = &   \left(\begin{array}{c}  \chi \, ,   \varphi \end{array}\right)  \left[ \frac{1}{2} \left(\begin{array}{cc} \frac{\sqrt{D}}{m} + \frac{m}{\sqrt{D}} & \frac{\sqrt{D}}{m} - \frac{m}{\sqrt{D}} \\[2mm] \frac{\sqrt{D}}{m} - \frac{m}{\sqrt{D}} & \frac{\sqrt{D}}{m} + \frac{m}{\sqrt{D}} \end{array}\right)- a \left(\begin{array}{cc} 1  & 0 \\ 0 & -1 \end{array}\right) \right]  \left(\begin{array}{c} \varphi \\[2mm]   \chi \end{array} \right)  . \nonumber \\[1mm]
\end{eqnarray}

\subsection{The metric for a Hamilton operator related to Wheeler-deWitt  by M.~Znojil}
\noindent Most recently \cite{Znojil:2021ajl} (see also \cite{Znojil:2017oio}) there has been addressed  the following Hamilton operator by M.\ Znojil in the context of the Wheeler-deWitt equation:

 \begin{equation}
H \; = \; \frac{1 + e^{\,2\, \tau}}{2} \; \sigma_1 +  \frac{1 - e^{\,2\, \tau}}{2\,i}\; \sigma_2 \; = \; \left(\begin{array}{cc} 0  &  e^{\,2\, \tau} \\[2mm]  1 & 0 \end{array}\right) \; ,
\end{equation}
possessing the eigen-values $E_1 = \pm \, e^{\,\tau}$ and $E_2 = \mp \, e^{\,\tau}$ and being simultaneously pseudo-Hermitian and anti-pseudo-Hermitian for Im$[\tau]= \ell \, \pi$ (with $\ell$ being integer). Keeping in mind that there obviously holds $n_3=0$ and therefore also
\begin{equation} \frac{E_1-E_2}{2} \; = \; \pm\;  e^{\,\tau} \; \stackrel{!}{=} \;\frac{E_1-E_2}{2} \; (1+n_3) \; , \end{equation}
it is straight forward to derive the matrices $X$ and $X^+$  using Eqs.\ (\ref{xxxident1}) and (\ref{xxxident3}), respectively, i.\ e.: 
\begin{eqnarray} X  & = &  \left(\frac{1}{2} \right)^{\frac{1}{2}} \;  \left(\begin{array}{cc} 1  &   \pm \; e^{\,\tau} \\ \mp\; e^{-\tau}   & 1 \end{array}\right)  \; , \label{zxxxident1} \\[2mm]
 X^+  & = &  \left(\frac{1}{2} \right)^{\frac{1}{2}}  \; \left(\begin{array}{cc} 1 & \mp \; e^{-\tau^\ast} \\   \pm \; e^{\, \tau^\ast} & 1 \end{array}\right)   \; . \label{zxxxident3}
\end{eqnarray}
Hence we can easily calculate the (Hermitian) metric for Case 1, i.\ e.,
\begin{eqnarray} \eta & = & \eta_0 + \eta_3 \; = \; \frac{|N_1|^2+|N_2|^2}{2} \; \;   X^+ X \; + \; \frac{|N_1|^2-|N_2|^2}{2} \; \;   X^+ \sigma_3 \, X \nonumber \\[2mm]
 & = &  \mbox{\small $\bigcirc\hspace{-3.2mm}\pm$} \;  \Bigg( \;  \frac{|N_1|^2+|N_2|^2}{2} \; \; \frac{1}{2} \left(\begin{array}{cc} 1 + e^{- (\tau + \tau^\ast)}  & \pm \left(  e^{\, \tau} \, - \,  e^{- \tau^\ast} \right) \\[2mm] \pm \left( e^{\, \tau^\ast} \, - \,  e^{- \tau} \right) & 1 + e^{\, \tau + \tau^\ast} \end{array}\right) \nonumber \\[2mm]
 & & \quad\, +\,  \frac{|N_1|^2-|N_2|^2}{2} \; \;\frac{1}{2} \left(\begin{array}{cc} 1 - e^{- (\tau + \tau^\ast)}  & \pm \left(  e^{\, \tau} \, +\,  e^{- \tau^\ast} \right) \\[2mm] \pm \left( e^{\, \tau^\ast} \, + \,  e^{- \tau} \right) & -\, 1 + e^{\, \tau + \tau^\ast} \end{array}\right)  \Bigg) \nonumber \\[2mm]
 & = &  \mbox{\small $\bigcirc\hspace{-3.2mm}\pm$} \;  \Bigg( \;  \frac{|N_1|^2}{2} \; \;  \left(\begin{array}{cc} 1  & \pm \;  e^{\, \tau} \\[2mm] \pm \; e^{\, \tau^\ast}  & e^{\, \tau + \tau^\ast} \end{array}\right) +\,  \frac{|N_2|^2}{2} \; \; \left(\begin{array}{cc}  e^{- (\tau + \tau^\ast)}  & \mp \;  e^{- \tau^\ast} \\[2mm] \mp \;   e^{- \tau}  & 1  \end{array}\right)  \Bigg) \; . \nonumber \\
\end{eqnarray}
Since (anti-)pseudo-Hermiticity requires Im$[\tau]= \ell \, \pi$ (with $\ell$ being integer) or, equivalently, 
\begin{equation} e^{\, \tau} \; = \; (-1)^\ell \;  e^{\, \mbox{\tiny Re} [\tau]} \quad \mbox{with $\ell$ being integer} \; ,
\end{equation}
we can rewrite the expression for the metric slightly, i.\ e.:
\begin{eqnarray}
\eta & = &  \mbox{\small $\bigcirc\hspace{-3.2mm}\pm$} \;  \Bigg( \;  \frac{|N_1|^2 \;  e^{\, \mbox{\tiny Re} [\tau]}}{2} \; \;  \left(\begin{array}{cc}  \;  e^{-\, \mbox{\tiny Re} [\tau]}  & \pm \;  (-1)^\ell  \\[2mm] \pm \; (-1)^\ell  & e^{\,  \mbox{\tiny Re} [\tau]} \end{array}\right) \nonumber \\[2mm]
 & & \quad\, +\,  \frac{|N_2|^2 \;  e^{-\, \mbox{\tiny Re} [\tau]}}{2} \; \; \left(\begin{array}{cc}  e^{-\, \mbox{\tiny Re} [\tau]}  & \mp \;  (-1)^\ell  \\[2mm] \mp \;   (-1)^\ell   &    e^{\, \mbox{\tiny Re} [\tau]}  \end{array}\right)  \Bigg) \; . 
\end{eqnarray}
This can be compared to the following expression provided by M.\ Znojil \cite{Znojil:2021ajl}, i.\ e.:
\begin{equation} \eta \; = \; \oplus   \left(\begin{array}{cc}  e^{-\, \mbox{\tiny Re} [\tau]}  & \beta  \\[2mm] \beta  &    e^{\, \mbox{\tiny Re} [\tau]} \; , \end{array}\right) 
\end{equation}
suggesting the following two identities for the particular choice of the normalizations $N_1$ and $N_2$ in \cite{Znojil:2021ajl}:
\begin{eqnarray} \frac{|N_1|^2 \;  e^{\, \mbox{\tiny Re} [\tau]}}{2} \; + \; \frac{|N_2|^2 \;  e^{-\, \mbox{\tiny Re} [\tau]}}{2} \quad & = & 1\; , \\[2mm]
 \pm \;  (-1)^\ell \left( \frac{|N_1|^2 \;  e^{\, \mbox{\tiny Re} [\tau]}}{2} \; - \; \frac{|N_2|^2 \;  e^{-\, \mbox{\tiny Re} [\tau]}}{2} \right)\, & = & \beta \; , 
\end{eqnarray}
or, equivalently,
\begin{eqnarray} |N_1|^2  & = &   \left( 1 \pm \,  (-1)^\ell \, \beta\right) \; e^{-\, \mbox{\tiny Re} [\tau]} \; , \\[2mm]
 |N_2|^2 & = &   \left( 1 \mp \,  (-1)^\ell \, \beta \right) \; e^{\, \mbox{\tiny Re} [\tau]} \;  . 
\end{eqnarray}
Since $|N_1|^2$ and $|N_2|^2$ must be non-negative it gets clear why M.\ Znojil imposes the constraint $|\beta|<1$.
\clearpage
\section{Conclusions and outlook}  \label{xxsec13}
The fundamental result of this work (see Eq.\ (\ref{etaqnx0}) or Eqs.\ (\ref{pseudherm1}) and (\ref{pseudherm2})), allowing to calculate the metric $\eta$ of pseudo-Hermitian (i.\ e.\ $H^+ \; = \; \eta \;  H \;  \eta^{-1}$) or anti-pseudo-Hermitian (i.\ e.\ $H^+ \; = \; -\, \eta \;  H \;  \eta^{-1}\; \Leftrightarrow \; (i\,  H)^+ \; =   \eta \; i \, H \;  \eta^{-1}$) Hamilton operators can be summarized in the following way:
\begin{equation} \eta \; = \;  (N\,X)^+  \,  Q\, N \, X \,  , \label{etaqnx1} \end{equation}
with $Q\in \{ 1_2\, ,  P \}$ for two-dimensional Hamilton operators. 

What is --- to our best knowledge --- new in the  presented work is the investigation of the effect of $Q$ on the metric in the here so-called ``non-trivial phase"  $Q\not=1_2$ relating by permutation the eigen-values of $H$ and $H^+$ or $i H$ and $(i H)^+$ for pseudo-Hermitian or anti-pseudo-Hermitian Hamilton operators, respectively, i.\ e.:
\begin{eqnarray} H^+ \; = \; \eta \;  H \;  \eta^{-1} \;\, & \Rightarrow &  \quad\;\; \left( \begin{array}{c} E^\ast_1 \\[2mm] E^\ast_2 \end{array}\right) \; = \; Q \; \left( \begin{array}{c} E_1 \\[2mm] E_2 \end{array}\right) \; , \\[2mm]
(i\, H)^+ \; = \; \eta \; i\,  H \;  \eta^{-1} & \Rightarrow &  \left( \begin{array}{c} (i \, E_1)^\ast \\[2mm] (i\, E_2)^\ast \end{array}\right) \; = \; Q \; \left( \begin{array}{c} i\, E_1 \\[2mm] i\, E_2 \end{array}\right) \; . 
\end{eqnarray} 
Hence, --- as has been pointed out in Section \ref{implhpseudo1} --- there hold the following identities for the energy eigen-values of  two-dimensional (anti-)pseudo-Hermitian Hamilton operators:
\begin{eqnarray} H^+  =  \eta \;  H \;  \eta^{-1} & \Rightarrow &  \nonumber  \\[2mm]
 Q \, = \, 1_2 \quad & \Rightarrow &   \left( \begin{array}{c} E^\ast_1 \\[2mm] E^\ast_2 \end{array}\right) \;  = \; 1_2  \left( \begin{array}{c} E_1 \\[2mm] E_2 \end{array}\right) \;  = \;  \left( \begin{array}{c} E_1 \\[2mm] E_2 \end{array}\right) , \label{twodimcas1} \\[2mm]
Q \, = \, P \quad & \Rightarrow &  \left( \begin{array}{c} E^\ast_1 \\[2mm] E^\ast_2 \end{array}\right) \; = \; P  \left( \begin{array}{c} E_1 \\[2mm] E_2 \end{array}\right) \;  = \;   \left( \begin{array}{c} E_2 \\[2mm] E_1 \end{array}\right) ,  \label{twodimcas2}\\[5mm]
 (i\, H)^+  = \eta \; i\,  H \;  \eta^{-1} & \Rightarrow &  \nonumber \\[2mm] 
Q \, = \, 1_2 \quad & \Rightarrow &  \left( \begin{array}{c} (i \, E_1)^\ast \\[2mm] (i\, E_2)^\ast \end{array}\right)  \; = \; 1_2  \left( \begin{array}{c} i\, E_1 \\[2mm] i\, E_2 \end{array}\right) \; = \; \left( \begin{array}{c} i\, E_1 \\[2mm] i\, E_2 \end{array}\right),  \label{twodimcas3}\\[2mm]
Q \, = \, P \quad & \Rightarrow &  \left( \begin{array}{c} (i \, E_1)^\ast \\[2mm] (i\, E_2)^\ast \end{array}\right) \; = \; P \left( \begin{array}{c} i\, E_1 \\[2mm] i\, E_2 \end{array}\right)  \; = \; \left( \begin{array}{c} i\, E_2 \\[2mm] i\, E_1 \end{array}\right) . \quad  \label{twodimcas4}
\end{eqnarray} 
As an outlook beyond the scope of this work we would like to point out here merely that the results for (anti-)pseudo-Hermitian two-dimensional Hamilton operators are analogously generalized e.g. to (anti-)pseudo-Hermitian three-dimensional Hamilton operators (see also \cite{Wang:2010br}) by keeping in mind that $Q$ should now be one of the 6 elements $1_3$,  $P_{231}$,  $P_{312}$, $P_{132}$, $P_{321}$, $P_{213}$ of the symmetric group in three dimensions defined in the following way:
\begin{eqnarray} & & \quad 1_3  = \left( \begin{array}{ccc} 1 & 0 & 0 \\[1mm]  0 & 1 & 0 \\[1mm] 0 & 0 & 1 \end{array}\right)   , \; P_{231}  = \left( \begin{array}{ccc} 0 & 1 & 0 \\[1mm]  0 & 0 & 1 \\[1mm] 1 & 0 & 0 \end{array}\right)  , \; P_{312} = \left( \begin{array}{ccc} 0 & 0 & 1 \\[1mm]  1 & 0 & 0 \\[1mm] 0 & 1 & 0 \end{array}\right) , \nonumber \\
 & & P_{132}  = \left( \begin{array}{ccc} 1 & 0 & 0 \\[1mm]  0 & 0 & 1 \\[1mm] 0 & 1 & 0 \end{array}\right)   , \; P_{321}  = \left( \begin{array}{ccc} 0 & 0 & 1 \\[1mm]  0 & 1 & 0 \\[1mm] 1 & 0 & 0 \end{array}\right)  , \; P_{213} = \left( \begin{array}{ccc} 0 & 1 & 0 \\[1mm]  1 & 0 & 0 \\[1mm] 0 & 0 & 1 \end{array}\right) . \nonumber \\
\end{eqnarray} 
Instead of the four cases (see e.\ g.\ Eqs.\ (\ref{twodimcas1}), (\ref{twodimcas2}), (\ref{twodimcas3}), (\ref{twodimcas4})) we have to consider now 12 cases implying the following 12 identities for  energy eigen-values of  three-dimensional (anti-)pseudo-Hermitian Hamilton operators:
\begin{eqnarray} H^+ \; = \; \eta \;  H \;  \eta^{-1} & \Rightarrow &  \nonumber  \\[2mm]
 Q \; = \; 1_3 \quad & \Rightarrow & E^\ast_1 = E_1\;,  \quad  E^\ast_2 = E_2\;,  \quad  E^\ast_3 = E_3\; , \nonumber \\[2mm]
 Q \; = \; P_{231} & \Rightarrow & E^\ast_1 = E_1\; = \;   E^\ast_2 = E_2\; = \;   E^\ast_3 = E_3\; , \nonumber \\[2mm]
 Q \; = \; P_{312} & \Rightarrow & E^\ast_1 = E_1\; = \;   E^\ast_2 = E_2\; = \;   E^\ast_3 = E_3\; , \nonumber \\[2mm]
 Q \; = \; P_{132} & \Rightarrow & E^\ast_1 = E_1\;,  \quad  E^\ast_3 = E_2\; , \nonumber \\[2mm]
 Q \; = \; P_{321} & \Rightarrow & E^\ast_2 = E_2\;,  \quad  E^\ast_3 = E_1\; , \nonumber \\[2mm]
 Q \; = \; P_{213} & \Rightarrow & E^\ast_3 = E_3\;,  \quad  E^\ast_2 = E_1\; , \\[2mm]
(i H)^+ \; =\;\; \; \eta \;\; i H  & \eta^{-1}  & \; \Rightarrow  \nonumber  \\[2mm]
 Q \; = \; 1_3 \quad & \Rightarrow & (i E_1)^\ast_1 = i E_1\;,  \; (i E_2)^\ast = i E_2\;,  \; (i  E_3)^\ast =  i E_3 \;  , \nonumber  \\[2mm]
 Q \; = \; P_{231} & \Rightarrow & (i E_1)^\ast = i E_1\; = \;  (i E_2)^\ast = i E_2\; = \;  (i E_3)^\ast = i E_3\; , \nonumber \\[2mm]
 Q \; = \; P_{312} & \Rightarrow &  (i E_1)^\ast = i E_1\; = \;  (i E_2)^\ast = i E_2\; = \;  (i E_3)^\ast = i E_3\; , \nonumber \\[2mm]
 Q \; = \; P_{132} & \Rightarrow & (i E_1)^\ast = i E_1\;,  \quad (i E_3)^\ast = i E_2\; , \nonumber \\[2mm]
 Q \; = \; P_{321} & \Rightarrow & (i E_2)^\ast = i E_2\;,  \quad (i E_3)^\ast = i E_1\; , \nonumber \\[2mm]
 Q \; = \; P_{213} & \Rightarrow & (i E_3)^\ast = i E_3\;,  \quad (i E_2)^\ast = i E_1\; .
\end{eqnarray}
Even if the consideration and discussion of three-dimensional Hamilton operators  is very interesting, it appears rather academic. It is nonetheless our strong believe that the respective discussion of Hamilton operators respresented by four (or higher even) dimensional matrices will open the door to the understanding of anti-causal causal Quantum Theory as lined out e.~g.~in our Refs.~\cite{Kleefeld:2002au}\cite{Kleefeld:2004jb}\cite{Kleefeld:2004qs}\cite{Kleefeld:2005hf}\cite{Kleefeld:2012fp}. 

As a concrete example we consider in Appendix \ref{xappenda4} the four-dimensional Hamilton operator of the the Fermionic (anti-)causal Harmonic Oscillator presented --- to our best knowledge --- for the first time by T.D.~Lee and C.G.~Wick \cite{Lee:1970iw} (see also \cite{Cotaescu:1983nc}) and rederived in a different context throughout  our doctoral thesis \cite{Kleefeld:1999} (see also \cite{Kleefeld:1998yj}\cite{Kleefeld:1998dg}\cite{Kleefeld:2003zj}\cite{Kleefeld:2003dx}) (with $\Omega\not=\Omega^\ast$):
\begin{equation} h  \; = \; h_H + h_A\; = \;  \frac{\hbar \Omega}{2} \; (d^+  b - b\; d^+) \; + \; \frac{\hbar \Omega^\ast}{2} \; (b^+  d - d\; b^+) \; = h^+ \; . \label{hamleewick0} 
\end{equation}
As shown in Appendix \ref{xappenda4} it is possible to systematically derive the following matrix representation $H$ of the Hamilton operator  (see Eq.\ (\ref{matrephamlw1})):
\begin{eqnarray}
H & = & H_H \, \otimes \, 1_2 + 1_2 \, \otimes \, H_A \nonumber \\[3mm]
& = & \left(\begin{array}{cc} +\, \frac{\hbar \Omega}{2} & 0 \\ 0 & -\, \frac{\hbar \Omega}{2} \end{array}\right)  \otimes  \left(\begin{array}{cc} 1 & 0 \\ 0 & 1 \end{array}\right) + \left(\begin{array}{cc} 1 & 0 \\ 0 & 1 \end{array}\right)  \otimes  \left(\begin{array}{cc} +\, \frac{\hbar \Omega^\ast}{2} & 0 \\ 0 & -\, \frac{\hbar \Omega^\ast}{2} \end{array}\right) \nonumber \\[4mm] 
 & = &  \frac{\hbar \Omega}{2} \; (\bar{D}  \, B - B\, \bar{D}) \; + \; \frac{\hbar \Omega^\ast}{2} \; (\bar{B}  \, D - D\, \bar{B}) \nonumber \\[2mm] 
 & = &  \frac{\hbar \Omega}{2} \; (B^+   B - B\, B^+) \; + \; \frac{\hbar \Omega^\ast}{2} \; (D^+   D - D\, D^+) \; \not= \; H^+ \; ,  \label{matrefx1} \end{eqnarray}
which is shown --- with the chosen normalization of the right and left eigenstates --- to be pseudo-Hermitian ($H^+ \; = \; \eta \; H \; \eta^{-1}$) with respect to the following indefinite four-dimensional metric:
\begin{equation} \eta \; = \; \eta^{-1} \; = \; \eta^+ \; = \; \left(\begin{array}{cccc} - 1 & 0 & 0 & 0 \\ 0 & 0  & 1 & 0 \\ 0 & 1 & 0 & 0 \\  0 & 0 & 0 & 1  \end{array}\right) \; . \label{leewickpseudometric1}
\end{equation}
Morever it is shown in Appendix \ref{xappenda4} that the matrix representation $H$ (see Eq.\ (\ref{matrefx1})) of the Hamilton operator $h$ can be expressed  in terms of the matrix representations of the respective annihilation and creation operators:\footnote{It is remarkable that the matrix $-\,\sigma_3$ (anti-commuting with the Pauli-matrices $\sigma_1$ and $\sigma_2$) occuring in the expressions for $\bar{B}=(-\,\sigma_3)\,\otimes \,\frac{1}{2}(\sigma_1+ \,i \,\sigma_2)$ and $D=(-\,\sigma_3)\,\otimes\, \frac{1}{2}(\sigma_1- \,i\, \sigma_2)$ is responsible for the required anti-commutativity of Fermionic operators (see also Refs.\ \cite{Steeb:1991}\cite{Steeb:2012}\cite{Steeb:2014}\cite{Hardy:2019}). Had we defined instead $\bar{B}=1_2\,\otimes\, \frac{1}{2}(\sigma_1+ \,i \,\sigma_2)$ and $D=1_2\,\otimes\, \frac{1}{2}(\sigma_1- \,i\, \sigma_2)$, the resulting creation and annihilation operators would instead commute with $B$ and $\bar{D}$, while the metric Eq.~(\ref{leewickpseudometric1}) would take the following --- different --- form \cite{Steeb:1991}:\\
\begin{equation} \eta \; = \; \eta^{-1} \; = \; \eta^+ \; = \; \left(\begin{array}{cccc} + 1 & 0 & 0 & 0 \\ 0 & 0  & 1 & 0 \\ 0 & 1 & 0 & 0 \\  0 & 0 & 0 & 1  \end{array}\right)\; .\end{equation}}
\begin{eqnarray} \bar{D} & = & B^+ \; = \; \quad \left(\begin{array}{cc} 0 & 1 \\ 0 & 0 \end{array}\right)  \otimes  \left(\begin{array}{cc} 1 & 0 \\ 0 & 1 \end{array}\right)  ,  \\[2mm]
 \bar{B} & = & D^+ \; = \; \,\left(\begin{array}{cc} -1 & 0 \\ 0 & 1 \end{array}\right)  \otimes  \left(\begin{array}{cc} 0 & 1 \\ 0 & 0 \end{array}\right)  ,  \\[2mm]
 D & = & \bar{B}^+  \; = \; \,\left(\begin{array}{cc} -1 & 0 \\ 0 & 1 \end{array}\right)  \otimes  \left(\begin{array}{cc} 0 & 0 \\ 1 & 0 \end{array}\right)  ,  \\[2mm]
 B & = & \bar{D}^+ \; = \; \;\;\;\left(\begin{array}{cc} 0 & 0 \\ 1 & 0 \end{array}\right)  \otimes  \left(\begin{array}{cc} 1 & 0 \\ 0 & 1 \end{array}\right)  .
\end{eqnarray}
It is shown (see Eq.\ (\ref{phrelations1})) that the metric $\eta$ induces --- by equivalence --- on one hand an interchange of the two annihilation operators, on the other hand also an interchange of the two creation operators: 
\begin{eqnarray}
 D^+ \; = \;  \bar{B} & = & \eta \; \bar{D} \; \eta^{-1} \; , \nonumber \\[2mm]
 B^+ \; = \;  \bar{D} & = & \eta \; \bar{B} \; \eta^{-1} \;,  \nonumber \\[2mm]
 \bar{D}^+ \; = \;  B & = & \eta \; D \; \eta^{-1} \; , \nonumber \\[2mm]
 \bar{B}^+ \; = \;  D & = & \eta \; B \; \eta^{-1} \; . \label{phrelations2}
\end{eqnarray}
The Hamilton operator $h\,=\,h_H + h_A$ (or its matrix representation $H\; = \; H_H \, \otimes 1_2\; + \;1_2 \, \otimes \, H_A$) of T.D.~Lee and C.G.~Wick is a direct sum of a non-Hermitian --- as we call it --- causal (or holomorphic) Hamilton operator $h_H$  (or its matrix representation $H_H \, \otimes 1_2$) and an --- also non-Hermitian --- anti-causal (or anti-holomorphic) Hamilton operator $h_A \; = \; h_H^+$ (or its matrix representation $1_2 \, \otimes \, H_A \; = \; 1_2 \, \otimes \, H_H^+$). It has served e.~g. \ in our Refs.\ \cite{Kleefeld:2004qs}\cite{Kleefeld:2004jb} as one suitable example for a Hamilton operator on the basis of which can be constructed a consistent (anti)causal and analytic Quantum Theory for particle physics. {\em It avoids causality and analyticity violations, as there are no interactions between the causal (or holomorphic) Hamilton operator and the respective anti-causal (or anti-holomorphic) Hamilton operator.} Causality is contained in the Hamilton operator by construction, if {\em all} the eigenvalues of the causal Hamilton operator $h_H$  (or its matrix representation $H_H \, \otimes 1_2$) possess an at least infinitesimal {\em negative} imaginary part (implying here Im$[\Omega]<0$), while {\em all} eigenvalues of the anti-causal Hamilton operator $h_A$  (or its matrix representation $1_2 \; \otimes \; H_A$) possess an at least infinitesimal {\em positive} imaginary part (implying here Im$[\Omega^\ast]>0$). As a consequence the causal Hamilton operator and the anti-causal Hamilton operator have to be manifestly {\em non-Hermitian}. As can been learned from Eqs.\ (\ref{phrelations2}) it is the task of the indefinite metric $\eta$ to perform a interchange of the causal and the anti-causal sector of the theory, while the pseudo-Hermiticity $H^+ \; = \; \eta \; H \; \eta^{-1}$ of the overall Hamilton operator reveals a symmetry due to $h_A = h_h^+$ (or $H_A = H_H^+$) between the causal (or holomorphic) and the anti-causal (or anti-holomorphic) Hamilton operator. Due to non-vanishing imaginary parts of the eigen-values of the causal (or holomorphic) and anti-causal (or anti-holomorphic) Hamilton operators we consider here a system being in the phase of {\em completely} broken ${\cal PT}$-symmetry. 

What will happen, if there are interactions between the causal sector and the anti-causal sector of the theory, has been already discussed in the Introduction with the help of the two-dimensional Hamilton operator Eq.\ (\ref{compghostx1}) describing two complex ghosts with bare mass $m-i \,\varepsilon$ and $m+i\,\varepsilon$ ($m$ and $\varepsilon$ are real-valued and {\em positive}) interacting via a complex coupling $\gamma\,$:
\begin{equation} H \; = \;  \left(\begin{array}{cc} m-i \,\varepsilon  & \gamma^\ast \\ \gamma & m+i \,\varepsilon \end{array}\right) \; . 
\end{equation}
As long as the modulous $|\gamma|$ of the coupling $\gamma$ is sufficently small, i.~e., \mbox{$|\gamma|^2<\varepsilon^2$}, the eigen-values $E_{1,2} = m \mp i\, \sqrt{\varepsilon^2-|\gamma|^2}$ describing the effective mass of the causal and anti-causal complex ghost are complex-valued and causality maintains to be valid (while observing for $\gamma\not=0$ a loss of analyticity due to the fact that the holomorphic and the anti-holomorphic sector of the theory interact). For $|\gamma|^2>\varepsilon^2$ the eigen-values $E_{1,2} = m \,\mp \,\sqrt{|\gamma|^2 - \varepsilon^2}$ of $H$ are real-valued, i.~e., the imaginary parts of the effective ``masses" of the two eigen-states are zero, which would mean the loss of causality. If the interaction $\gamma$ is so strong that there holds $\sqrt{|\gamma|^2 - \varepsilon^2} \; > \;  m$, one of the real eigen-values would get even negative. The appearance of negative energy states would indicate in Hermitian Quantum Physics the breakdown of the probability concept due to the appearance of negative probabilities. Fortunately this pathology can be cured in ${\cal PT}$-symmetric Quantum Theory by identifying some positive metric $\eta_+$. Nonetheless, the existence of this metric does {\em not} cure the manifest loss of causality and analyticity. 

Unfortunately, the majority of scientists educated according to the rules of traditional Hermitian Quantum Theory expect --- in the unjustified hope that causal boundary conditions can be built in the framework consistently afterwards --- the eigen-values of the Hamilton operator (being the operator respresenting the energy as an observable) and the eigen-values of other operators respresenting different (Hermitian) observables to be real-valued implying a very incomplete correspondence between Classical Theory and Quantum Theory. Moreover they are tempted --- keeping in mind the probabilistic structure underlying Quantum Theory --- to prefer a positive semi-definite scalar product provided by the ``trivial phase" $Q=1_2$ in the hope to obtain a meaningful Quantum Theory even for ${\cal PT}$-symmetric Hamiltonians. This is why the focus of the majority of scientists working in the field of non-Hermitian physics lies in the construction of a positive semi-definite metric for ${\cal PT}$-symmetric pseudo-Hermitian Hamilton operators in ``trivial phase" $Q=1_2$ being commonly known as the phase of ``unbroken ${\cal PT}$-symmetry". 
Moreover, the very existence of the indefinite metric in the ``non-trivial" phase $Q=P$ being commonly known as the phase of ``broken ${\cal PT}$-symmetry" is --- according to common interpretation --- associated either with the very existence of sources of energy gain and sinks of energy loss in a physical system or with a mathematical pathology arising due to the presence of unphysical ghosts in a specific region of the parameter space of the Hamilton operator which should be avoided or ``busted"  \cite{Bender:2004sv}\cite{Bender:2007nj}.

As we will show in a forthcoming publication \cite{Kleefeld:2021} the formulation of a \mbox{(anti-)causal} Quantum Theory along the lines illustrated here in considering a non-Hermitian Hamilton operator proposed by T.D.~Lee and C.G.~Wick  {\em will} contain --- despite the underlying indefinite metric --- a meaningful probabilistic framework, as long as the causal and anti-causal sectors of the theory are not at all allowed to interact. Moreover the manifestly non-Hermitian nature of the formalism allows to set up a non-Hermitian Quantum Theory on the basis of non-Hermitian observables which allows a correspondence principle between Quantum and Classical Physics valid in the whole complex plane. This is quite compatible to what has been conjectured alread in 1981 by P.~B.~Burt:\\[2mm]
{\em\small ``Already on the classical level the most general physics is obtained by solving the equations of motion without constraint. In fact, imposing constraints prior to the solution of the equations of motion can lead one to erroneous conclusions. In \mbox{parallel}, in quantum field theory we adopt the premise that non-Hermitian solutions of the field equations are acceptable. The physics contained in the hermiticity assumption will be built in after explicit solutions are constructed." }\\[0mm]
\mbox{} \hfill {\small (Philip Barnes Burt, 1981, p.\ 29 in \cite{Burt:1981})} 

\clearpage
\noindent {\bf Acknowledgements}

\noindent 33 years after leaving with the age of 5 Tansania --- the country where I have been born ---  I had the priviledge to return --- on invitation --- to the African continent to present in November 2005 some of my ideas on non-Hermitian physics --- developed mainly at the Inst. Theor. Phys. III of Univ.\ Erlangen, Germany, under supervision of my doctoral thesis supervisor M.~Dillig and afterwards at the CFIF of the IST in Lisbon, Portugal, under the very inspiring and lasting support of G.\ Rupp ---  orally during the 4th Int. Workshop on ``Pseudo-Hermitian Hamiltonians in Quantum Physics" at the Inst. of Theor. Physics of Stellenbosch University, Matieland, South Africa. During the workshop one of the organizers being simultaneously author of Ref.\  \cite{Geyer:2007} (see also Ref.~\cite{Znojil:2006ugs}) raised to me the question why even the positive semi-definite metric in non-Hermitian Quantum physics is ambiguous. Now, 16 years later, I am very grateful to be able to provide hereby my definite answer to the question confirming to great extend my conjectures made in Refs.\ \cite{Kleefeld:2009vd} and \cite{Kleefeld:2006bp} which had been developed mainly at the Nucl.\ Phys.\ Inst.\ of the Acad.\ of Sciences of Czechian Republic under the kind hospitality and support of M.~Znojil being founding and ongoing organizer of the aforementioned workshop series and one of the authors of Ref.~\cite{Znojil:2006ugs}.
\begin{figure}[b]
\begin{center}
\includegraphics[width=13pc]{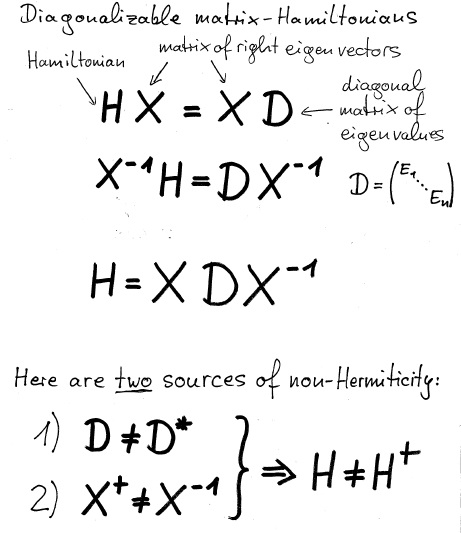}
\end{center}
\caption{The origin of the non-Hermiticity of a Hamilton operator respresented by some matrix had been outlined e.g.\ in our oral presentation in Paris \cite{Kleefeld:2012} in 2012. Although the matrix of the eigenvalues $D$ does not enter the expression for the metric $\eta$ itself, the metric contains at least the matrix $Q$ being responsible for the permutation of the energy eigen-values. Interestingly, the non-trivial role of $Q$ in the metric $\eta$ even persists for $H$ being Hermitian, i.~e. for $D=D^\ast$ and $X^+ = X^{-1}$.}
\end{figure}
\clearpage
\begin{appendix} 
\section{A two-dimensional Hamilton operator for two interacting complex ghosts} \label{xappenda1}
\noindent In this section we will apply our analysis to the following two-dimensional Hamilton operator $H$ (and the respective partners $H^+$, $i H$, $(i H)^+$) describing two complex ghosts with bare mass $m-i \,\varepsilon$ and $m+i\,\varepsilon$ (with $m$ and $\varepsilon$ being real-valued) interacting via a complex coupling $\gamma = \gamma_1 + i  \, \gamma_2$ (with $\gamma_1$ and $\gamma_2$ being real-valued) and the complex conjugate coupling $\gamma^\ast = \gamma_1 - i  \, \gamma_2$:
\begin{eqnarray} & & \;\, H \; = \;  \left(\begin{array}{cc} m-i \,\varepsilon  & \gamma^\ast \\ \gamma & m+i \,\varepsilon \end{array}\right) \; , \;\quad\;\; i \, H \; = \;  \left(\begin{array}{cc} i\, m+ \varepsilon  & i\, \gamma^\ast \\ i\, \gamma & i\, m- \varepsilon \end{array}\right)  , \nonumber \\[2mm]
 & & H^+ \; = \;  \left(\begin{array}{cc} m+i \,\varepsilon  & \gamma^\ast \\ \gamma & m-i \,\varepsilon \end{array}\right) \; , \; (i \, H)^+ \; = \;  \left(\begin{array}{cc} - i\, m+ \varepsilon  & - i\, \gamma^\ast \\ - i\, \gamma & i\, m- \varepsilon \end{array}\right)  . \nonumber \\[2mm]
\end{eqnarray}
A calculation of traces and determinants yields:
\begin{eqnarray}  \mbox{tr}[H] \; = & \mbox{tr}[H^+] & = \; 2\, m \; , \\[2mm] 
 \mbox{tr}[i H] \; = & -\,\mbox{tr}[(i H)^+] & = \;  2 \,m\, i \; , \\[2mm]
 \mbox{det}[H] \; = & \mbox{det}[H^+] & = \; m^2 + \varepsilon^2 - |\gamma^2| \; , \\[2mm] 
 \mbox{det}[i H] \; = & \mbox{det}[(i H)^+] & = \;-  m^2 - \varepsilon^2 + |\gamma^2| \; .
\end{eqnarray}
According to the discussion of Eqs.\ (\ref{consid1}) and (\ref{consid2})  the two-dimensional Hamilton operator $H$ and the conjugate Hamilton operator $H^+$ are pseudo-Hermitian for all real parameters $\gamma_1$, $\gamma_2$, $\varepsilon$, $m$, while $H$ and $H^+$ are anti-pseudo-Hermitian  for all real parameters $\gamma_1$, $\gamma_2$, $\varepsilon$ under the restriction  $m=0$. Obviously there holds:
\begin{equation} \mbox{tr}[H]^2-4\,  \mbox{det}[H] \; = \;4 \, ( |\gamma|^2 - \varepsilon^2) \; , \;\; 4\,  \mbox{det}[H] - \mbox{tr}[H]^2 \; = \; 4\,(  \varepsilon^2 -|\gamma|^2)  \; ,
\end{equation}
and
\begin{eqnarray} \frac{E_1-E_2}{2} \; n_ 3 \; = \;  \frac{H_{11}-H_{22}}{2} & = & -\, i \, \varepsilon \; , \\[2mm]
\frac{E^\ast_1-E^\ast_2}{2} \; n^\ast_ 3 \; = \;  \frac{H^\ast_{11}-H^\ast_{22}}{2} & = & + \, i \, \varepsilon \; .
\end{eqnarray}
Following the discussion of Sections \ref{xmetrici}, \ref{xmetricp} and \ref{xconvrep1}  we obtain the following results for the four different cases under consideration:
\newpage
\begin{itemize} 
\item {\bf Case 1:} $H$ is pseudo-Hermitian ($H^+ = + \eta \, H \, \eta^{-1}\,$), \\
\mbox{} $\qquad\quad\;\;$ the phase of the metric is trivial ($Q=1_2$)
\begin{eqnarray}  |\gamma|^2 \; > & \varepsilon^2 & \quad \Rightarrow \nonumber \\[2mm]
 H  & = &  m \; 1_2 \; \pm \; \sqrt{|\gamma|^2 - \varepsilon^2} \; \vec{n} \cdot \vec{\sigma} \; ,  \\[2mm]
 E_1 & = & m \; \pm \;  \sqrt{|\gamma|^2 - \varepsilon^2} \; ,  \quad E_2 \;  = \; m \; \mp \;  \sqrt{|\gamma|^2 - \varepsilon^2} \; , \\[2mm]
 \frac{E_1-E_2}{2}  & = & \frac{E^\ast_1-E^\ast_2}{2} \; = \;\pm \;  \sqrt{|\gamma|^2 - \varepsilon^2} \;  , \\[2mm]  
\vec{n} \cdot \vec{\sigma} & = & \pm\; \frac{\left(\begin{array}{cc} -i \,\varepsilon  & \gamma^\ast \\ \gamma & +i \,\varepsilon \end{array}\right)}{\sqrt{|\gamma|^2 - \varepsilon^2}} \; , \qquad \\[2mm]
X & = & \left(\frac{\sqrt{|\gamma|^2 - \varepsilon^2} \; \mp i\,\varepsilon}{2\; \sqrt{|\gamma|^2 - \varepsilon^2}} \right)^{\frac{1}{2}} \left( 1_2 \mp   \frac{\left(\begin{array}{cc} 0  & - \gamma^\ast \\ + \gamma & 0 \end{array}\right)}{\sqrt{|\gamma|^2 - \varepsilon^2} \; \mp i\, \varepsilon} \right)  , \qquad \\[2mm]
X^+ & = &  \left(\frac{\sqrt{|\gamma|^2 - \varepsilon^2} \; \pm i\,\varepsilon}{2\; \sqrt{|\gamma|^2 - \varepsilon^2}} \right)^{\frac{1}{2}} \left( 1_2 \pm   \frac{\left(\begin{array}{cc} 0  & - \gamma^\ast \\ + \gamma & 0 \end{array}\right)}{\sqrt{|\gamma|^2 - \varepsilon^2} \; \pm i\, \varepsilon} \right)  , \qquad \\[2mm]
\eta_0 & = & \eta^+_0 \; = \;  \frac{|N_1|^2+|N_2|^2}{2} \; \;   X^+ X \nonumber \\[2mm]
 & = & \mbox{\small $\bigcirc\hspace{-3.2mm}\pm$}\;\frac{|N_1|^2+|N_2|^2}{2} \;\frac{|\gamma|}{\sqrt{|\gamma|^2 - \varepsilon^2}}\; \left(\begin{array}{cc} 1  & + i \, \frac{\gamma^\ast \varepsilon}{|\gamma|^2} \\  - i \, \frac{\gamma \, \varepsilon}{|\gamma|^2} & 1 \end{array}\right)   , \qquad \\[2mm]
\eta_3 & = & \eta^+_3 \; = \;  \frac{|N_1|^2-|N_2|^2}{2} \; \;   X^+\sigma_3 \, X  \nonumber \\[2mm]
 & = & \mbox{\small $\bigcirc\hspace{-3.2mm}\pm$}\;\pm \; \frac{|N_1|^2-|N_2|^2}{2} \left(\begin{array}{cc} 0  & \frac{\gamma^\ast}{|\gamma|} \\  \frac{\gamma}{|\gamma|} & 0 \end{array}\right) . \qquad 
\end{eqnarray}
\item {\bf Case 2:} $H$ is pseudo-Hermitian ($H^+ = + \eta \, H \, \eta^{-1}\,$), \\
\mbox{} $\qquad\quad\;\;$ the phase of the metric is non-trivial ($Q=P$)
\begin{eqnarray} \varepsilon^2 \; > & |\gamma|^2 & \quad \Rightarrow \nonumber \\[2mm]
 H  & = &  m \; 1_2 \; \pm \; i\; \sqrt{\varepsilon^2 -|\gamma|^2} \; \vec{n} \cdot \vec{\sigma} \; ,  \\[2mm]
 E_1 & = & m \; \pm \, i\;  \sqrt{\varepsilon^2 -|\gamma|^2} \; ,  \; E_2 \; = \;  m \; \mp \, i \;  \sqrt{\varepsilon^2 -|\gamma|^2}\;  , \\[2mm]  
 \frac{E_1-E_2}{2}  & = & -\;  \frac{E^\ast_1-E^\ast_2}{2} \; = \;  \pm \, i\;  \sqrt{\varepsilon^2 -|\gamma|^2} \;  , \\[2mm]  
\vec{n} \cdot \vec{\sigma} & = & \mp\, i \; \frac{\left(\begin{array}{cc} -i \,\varepsilon  & \gamma^\ast \\ \gamma & +i \,\varepsilon \end{array}\right)}{\sqrt{\varepsilon^2 -|\gamma|^2}} \; , \qquad \\[2mm] 
X  \; = \;  X^+  & = &    \left( \frac{\sqrt{\varepsilon^2 -|\gamma|^2}\; \mp \varepsilon}{2\; \sqrt{\varepsilon^2 -|\gamma|^2}} \right)^{\frac{1}{2}} \left( 1_2 \pm   \frac{\left(\begin{array}{cc} 0  & - i\, \gamma^\ast \\ + i\, \gamma & 0 \end{array}\right)}{\sqrt{\varepsilon^2 -|\gamma|^2} \; \mp \, \varepsilon} \right)  , \qquad \\[2mm]
\eta_1 & = &  \frac{N^\ast_2\,N_1+N^\ast_1\,N_2}{2} \; \;    X^+ \; \vec{e}\cdot \vec{\sigma}\;   X   \nonumber \\
 & = &  \frac{N^\ast_2\,N_1+N^\ast_1\,N_2}{2} \; \Big( \cos\varphi  \;  X^+ \sigma_1\,   X + \sin\varphi  \;  X^+ \sigma_2\,   X\Big)   \nonumber \\[2mm]
 & = & \mbox{\small $\bigcirc\hspace{-3.2mm}\pm$}\; \;\frac{N^\ast_2\,N_1+N^\ast_1\,N_2}{2} \; \cdot \; \nonumber \\[2mm]
 & \cdot &  \Bigg(  \frac{ \cos\varphi  \;\mbox{Re} [\gamma]+\sin\varphi  \; \mbox{Im} [\gamma]}{|\gamma|} \left(\begin{array}{cc} 0   & \frac{\gamma^\ast}{|\gamma|}  \\ \frac{\gamma}{|\gamma|}  & 0  \end{array}\right) \nonumber \\
 & &  \pm  \; \frac{\sin\varphi  \; \mbox{Re} [\gamma] - \cos\varphi  \;\mbox{Im} [\gamma]}{\sqrt{\varepsilon^2 -|\gamma|^2}}\left(\begin{array}{cc} 1   & + i \, \frac{\varepsilon \, \gamma^\ast}{|\gamma|^2} \\ -\, i \, \frac{\varepsilon\, \gamma}{|\gamma|^2} & 1  \end{array}\right) \Bigg)  , \nonumber \\[2mm]
 & & \\
 \eta_2 & = &  \frac{N^\ast_2\,N_1-N^\ast_1\,N_2}{2\, i} \; \;    X^+ \; \vec{e}_\perp\cdot \vec{\sigma} \; X\nonumber \\
  & = &  \frac{N^\ast_2\,N_1-N^\ast_1\,N_2}{2\, i} \; \Big( \sin\varphi  \;  X^+ \sigma_1\,   X - \cos\varphi  \;  X^+ \sigma_2\,   X\Big)  \nonumber \\[2mm]
 & = & \mbox{\small $\bigcirc\hspace{-3.2mm}\pm$}\; \;\frac{N^\ast_2\,N_1-N^\ast_1\,N_2}{2\, i} \; \cdot \; \nonumber \\[2mm]
 & \cdot &  \Bigg(  \frac{ \sin\varphi  \;\mbox{Re} [\gamma]-\cos\varphi  \; \mbox{Im} [\gamma]}{|\gamma|} \left(\begin{array}{cc} 0   & \frac{\gamma^\ast}{|\gamma|}  \\ \frac{\gamma}{|\gamma|}  & 0  \end{array}\right) \nonumber \\
 & &  \mp  \; \frac{\cos\varphi  \; \mbox{Re} [\gamma] + \sin\varphi  \;\mbox{Im} [\gamma]}{\sqrt{\varepsilon^2 -|\gamma|^2}}\left(\begin{array}{cc} 1   & + i \, \frac{\varepsilon \, \gamma^\ast}{|\gamma|^2} \\ -\, i \, \frac{\varepsilon\, \gamma}{|\gamma|^2} & 1  \end{array}\right) \Bigg) \; . \nonumber \\
  & & 
\end{eqnarray}
\item {\bf Case 3:} $H$ is anti-pseudo-Hermitian ($H^+ = - \,\eta \, H \, \eta^{-1}$) implying\\
\mbox{} $\qquad\quad\;\;$ $i H$ to be pseudo-Hermitian  ($(i H)^+ = + \,\eta \; i H \; \eta^{-1}$), \\
\mbox{} $\qquad\quad\;\;$ the phase of the metric is trivial ($Q=1_2$)

\begin{eqnarray}  \varepsilon^2 \; > & |\gamma|^2 & , \;  m\; = \;  0 \quad \Rightarrow \nonumber \\[2mm]
 H  & = &  \pm \, i\; \sqrt{\varepsilon^2 -|\gamma|^2} \; \vec{n} \cdot \vec{\sigma} \; , \\[2mm]
 E_1 & = & \pm \, i \;  \sqrt{\varepsilon^2 -|\gamma|^2} \; ,  \quad  E_2 \; = \; \mp \, i \;  \sqrt{\varepsilon^2 -|\gamma|^2} \; , \\[2mm] 
 \frac{E_1-E_2}{2}  & = &-\;  \frac{E^\ast_1-E^\ast_2}{2} \; = \;  \pm \, i\;  \sqrt{\varepsilon^2 -|\gamma|^2} \;  , \\[2mm]  
\vec{n} \cdot \vec{\sigma} & = & \mp\, i \; \frac{\left(\begin{array}{cc} -i \,\varepsilon  & \gamma^\ast \\ \gamma & +i \,\varepsilon \end{array}\right)}{\sqrt{\varepsilon^2 -|\gamma|^2}}  \; , \qquad \\[2mm] 
X  =  X^+  & = &   \left( \frac{\sqrt{\varepsilon^2 -|\gamma|^2}\; \mp \varepsilon}{2\; \sqrt{\varepsilon^2 -|\gamma|^2}} \right)^{\frac{1}{2}} \left( 1_2 \pm   \frac{\left(\begin{array}{cc} 0  & - i\, \gamma^\ast \\ + i\, \gamma & 0 \end{array}\right)}{\sqrt{\varepsilon^2 -|\gamma|^2} \; \mp \, \varepsilon} \right)  , \qquad \\[2mm]
\eta_0 & = & \eta^+_0 \; = \;  \frac{|N_1|^2+|N_2|^2}{2} \; \;   X^+ X  \nonumber \\[2mm]
 & = & \mbox{\small $\bigcirc\hspace{-3.2mm}\pm$}\;\mp\; \frac{|N_1|^2+|N_2|^2}{2} \; \frac{1}{\sqrt{|\gamma|^2 - \varepsilon^2}} \left(\begin{array}{cc} \varepsilon  &  i \,\gamma^\ast \\ - i \,\gamma  & \varepsilon \end{array}\right)   , \qquad \\[2mm]
\eta_3 & = & \eta^+_3 \; = \;  \frac{|N_1|^2-|N_2|^2}{2} \; \;   X^+\sigma_3 \, X  \nonumber \\[2mm]
 & = & \mbox{\small $\bigcirc\hspace{-3.2mm}\pm$}\;\frac{|N_1|^2-|N_2|^2}{2} \;   \left(\begin{array}{cc} +1  & 0 \\  0 & -1 \end{array}\right)  . \quad 
\end{eqnarray}
\item {\bf Case 4:} $H$ is anti-pseudo-Hermitian ($H^+ = - \,\eta \, H \, \eta^{-1}$) implying\\
\mbox{} $\qquad\quad\;\;$ $i H$ to be pseudo-Hermitian  ($(i H)^+ = + \,\eta \; i H \; \eta^{-1}$), \\
\mbox{} $\qquad\quad\;\;$ the phase of the metric is non-trivial ($Q=P$)
\begin{eqnarray}   |\gamma|^2 \; > & \varepsilon^2 & , \;  m\; = \;  0 \quad \Rightarrow \nonumber \\[2mm]
 H  & = &  \pm \; \sqrt{|\gamma|^2 - \varepsilon^2} \; \vec{n} \cdot \vec{\sigma} \; ,  \\[2mm]
 E_1 & = & \pm  \;  \sqrt{|\gamma|^2 - \varepsilon^2} \; ,  \quad E_2 \; = \; \mp  \;  \sqrt{|\gamma|^2 - \varepsilon^2} \; , \\[2mm] 
 \frac{E_1-E_2}{2}  & = & \frac{E^\ast_1-E^\ast_2}{2} \; = \;\pm \;  \sqrt{|\gamma|^2 - \varepsilon^2} \;  , \\[2mm]  
 \vec{n} \cdot \vec{\sigma} & = & \pm \; \frac{\left(\begin{array}{cc} -i \,\varepsilon  & \gamma^\ast \\ \gamma & +i \,\varepsilon \end{array}\right)}{\sqrt{|\gamma|^2 - \varepsilon^2}}  \; , \\[2mm] 
X & = & \left(\frac{\sqrt{|\gamma|^2 - \varepsilon^2} \; \mp i\,\varepsilon}{2\; \sqrt{|\gamma|^2 - \varepsilon^2}} \right)^{\frac{1}{2}} \left( 1_2 \mp   \frac{\left(\begin{array}{cc} 0  & -\gamma^\ast \\ + \gamma & 0 \end{array}\right)}{\sqrt{|\gamma|^2 - \varepsilon^2} \; \mp i\, \varepsilon} \right)  , \qquad \\[2mm]
X^+ & = & \left(\frac{\sqrt{|\gamma|^2 - \varepsilon^2} \; \pm i\,\varepsilon}{2\; \sqrt{|\gamma|^2 - \varepsilon^2}} \right)^{\frac{1}{2}} \left( 1_2 \pm   \frac{\left(\begin{array}{cc} 0  & - \gamma^\ast \\ + \gamma & 0 \end{array}\right)}{\sqrt{|\gamma|^2 - \varepsilon^2} \; \pm i\, \varepsilon} \right) , \qquad \\[2mm]
 \eta_1 & = &  \frac{N^\ast_2\,N_1+N^\ast_1\,N_2}{2} \; \;    X^+ \; \vec{e}\cdot \vec{\sigma}\;   X     \nonumber \\[2mm]
  & = &  \frac{N^\ast_2\,N_1+N^\ast_1\,N_2}{2} \; \Big( \cos\varphi  \;  X^+ \sigma_1\,   X + \sin\varphi  \;  X^+ \sigma_2\,   X\Big)  \nonumber \\[2mm]
 & = & \mbox{\small $\bigcirc\hspace{-3.2mm}\pm$}\; \;\frac{N^\ast_2\,N_1+N^\ast_1\,N_2}{2} \; \cdot \; \nonumber \\[2mm]
 & \cdot &  \frac{1}{|\gamma|} \Bigg(  \frac{ \cos\varphi  \;\mbox{Im} [\gamma]-\sin\varphi  \; \mbox{Re} [\gamma]}{\sqrt{|\gamma|^2-\varepsilon^2}} \left(\begin{array}{cc} \varepsilon   & i\,\gamma^\ast  \\ -i \, \gamma  & \varepsilon  \end{array}\right) \nonumber \\
 & & \;\;  \mp  \; \Big(\sin\varphi  \; \mbox{Im} [\gamma] + \cos\varphi  \;\mbox{Re} [\gamma]\Big)\left(\begin{array}{cc} +1   & 0 \\ 0 & -\, 1  \end{array}\right) \Bigg)  ,  \\[2mm] 
 \eta_2 & = &  \frac{N^\ast_2\,N_1-N^\ast_1\,N_2}{2\, i} \; \;    X^+ \; \vec{e}_\perp\cdot \vec{\sigma} \; X\nonumber \\[2mm]
 & = &  \frac{N^\ast_2\,N_1-N^\ast_1\,N_2}{2\, i} \; \Big( \sin\varphi  \;  X^+ \sigma_1\,   X - \cos\varphi  \;  X^+ \sigma_2\,   X\Big)  \nonumber \\[2mm]
 & = & \mbox{\small $\bigcirc\hspace{-3.2mm}\pm$}\; \;\frac{N^\ast_2\,N_1-N^\ast_1\,N_2}{2\, i} \; \cdot \; \nonumber \\[2mm]
 & \cdot &  \frac{1}{|\gamma|} \Bigg(  \frac{ \sin\varphi  \;\mbox{Im} [\gamma]+\cos\varphi  \; \mbox{Re} [\gamma]}{\sqrt{|\gamma|^2-\varepsilon^2}} \left(\begin{array}{cc} \varepsilon   & i\,\gamma^\ast  \\ -i \, \gamma  & \varepsilon  \end{array}\right) \nonumber \\
 & & \;\;  \pm  \; \Big(\cos\varphi  \; \mbox{Im} [\gamma] - \sin\varphi  \;\mbox{Re} [\gamma]\Big)\left(\begin{array}{cc} +1   & 0 \\ 0 & -\, 1  \end{array}\right) \Bigg)  . 
\end{eqnarray}
 \end{itemize}
\newpage
\section{A two-dimensional Hamilton operator considered by C.\ M.\ Bender et al.} \label{xappenda2}
\noindent In this section we will apply our analysis to the following asymmetric two-dimensional Hamilton operator $H$ (and the respective partners $H^+$, $i H$, $(i H)^+$) considered  for $t=s$ and $\phi= 0$ by C.\ M.\ Bender et al.\ \cite{Bender:2002vv}\cite{Bender:2005tb}\cite{Bender:2019cwm}\cite{Geyer:2007} \cite{Mannheim:2009zj}  (see also Z.\ Ahmed \cite{Ahmed:2003nn}) and later for $t\not=s$ and $\phi\not= 0$ by A.\ Das and L.~Greenwood \cite{Das:2009it}\cite{Das:2011zza} and A.~Ghatak and B.~Mandal \cite{Ghatak:2013wga} (with $t$, $s$, $r$, $\theta$ and $\phi$ being real-valued parameters):
\begin{eqnarray} & & \;\, H \; = \;  \left(\begin{array}{cc} r\, e^{+i\,\theta}  & s \, e^{+i\,\phi} \\ t \, e^{-i\,\phi} & r\, e^{-i\,\theta} \end{array}\right) \; , \;\quad\;\; i \, H \; = \;   \left(\begin{array}{cc} i\, r\, e^{+i\,\theta}  & i\, s \, e^{+i\,\phi} \\ i\, t \, e^{-i\,\phi} & i\,  r\, e^{-i\,\theta} \end{array}\right)  , \nonumber \\[2mm]
 & & H^+ \; = \;  \left(\begin{array}{cc} r\, e^{-i\,\theta}  & t \, e^{+i\,\phi} \\ s \, e^{-i\,\phi} & r\, e^{+i\,\theta} \end{array}\right)  \; , \; (i \, H)^+ \; = \;  \left(\begin{array}{cc} - i\, r\, e^{-i\,\theta}  & -i\, t \, e^{+i\,\phi} \\ -i\, s \, e^{-i\,\phi} & -i\,  r\, e^{+i\,\theta} \end{array}\right)  . \nonumber \\[2mm]
\end{eqnarray}
A calculation of traces and determinants yields:
\begin{eqnarray}  \mbox{tr}[H] \; = & \mbox{tr}[H^+] & = \; 2\, r\,\cos\theta \; , \\[2mm] 
 \mbox{tr}[i H] \; = & -\,\mbox{tr}[(i H)^+] & = \;  2 \, i \,r\, \cos\theta \; , \\[2mm]
 \mbox{det}[H] \; = & \mbox{det}[H^+] & = \; r^2 - t \, s \; , \\[2mm] 
 \mbox{det}[i H] \; = & \mbox{det}[(i H)^+] & = \;-  r^2 + t\, s \; .
\end{eqnarray}
According to the discussion of Eqs.\ (\ref{consid1}) and (\ref{consid2})  the two-dimensional Hamilton operator $H$ and the conjugate Hamilton operator $H^+$ are pseudo-Hermitian for all real parameters $t$, $s$, $r$, $\theta$, while $H$ and $H^+$ are anti-pseudo-Hermitian  for all real parameters $t$, $s$ under the restriction  $r\,\cos\theta=0$. For $r\,\cos\theta\not =0$ there holds:
\begin{eqnarray} \mbox{tr}[H]^2-4\,  \mbox{det}[H] & = & 4\, \left(s\,t - (r\, \sin\theta)^2\right) \; , \\[2mm]
 4\,  \mbox{det}[H] - \mbox{tr}[H]^2 & = & 4\, \left((r\, \sin\theta)^2-s\,t\right) \; ,
\end{eqnarray}
and
\begin{eqnarray} \frac{E_1-E_2}{2} \; n_ 3 \; = \;  \frac{H_{11}-H_{22}}{2} & = &  +\, i \, r\, \sin \theta \; , \\[2mm]
\frac{E^\ast_1-E^\ast_2}{2} \; n^\ast_ 3 \; = \;  \frac{H^\ast_{11}-H^\ast_{22}}{2} & = & -\, i \, r\, \sin \theta \; .
\end{eqnarray}

In the case of anti-pseudo-Hermiticity ($r\,\cos\theta=0$) the asymmetric two-dimensional Hamilton operator $H$ (and the respective partners $H^+$, $i H$, $(i H)^+$) take the following simple form:
\begin{eqnarray} & & \;\, H \; = \;  \left(\begin{array}{cc}  (-1)^\ell\,  i\,  r & s \, e^{+i\,\phi} \\ t \, e^{-i\,\phi} & -\, (-1)^\ell \,  i\,  r\end{array}\right)  , \; i \, H \; = \;   \left(\begin{array}{cc} - \,   (-1)^\ell \, r \, e^{+i\,\phi}  & i\, s \, e^{-i\,\phi} \\ i\, t &    (-1)^\ell\, r \end{array}\right)  , \nonumber \\[2mm]
 & & H^+ \; = \;   \left(\begin{array}{cc}  -\, (-1)^\ell\,  i\,  r & t \, e^{+i\,\phi} \\ s \, e^{-i\,\phi} &  (-1)^\ell \,  i\,  r\end{array}\right)  \; , \; (i \, H)^+ \; = \;  \left(\begin{array}{cc} - \,   (-1)^\ell \, r  & -\, i\, t \, e^{+i\,\phi} \\ -\,i\, s \, e^{-i\,\phi} &    (-1)^\ell\, r \end{array}\right)  . \nonumber \\[2mm]
\end{eqnarray}
will $\ell$ being some arbitrary integer. Moreover, anti-pseudo-Hermiticity yields:
\begin{eqnarray} \mbox{tr}[H]^2-4\,  \mbox{det}[H] & = & 4\, \left(s\,t - r^2\right) \; , \\[2mm]
 4\,  \mbox{det}[H] - \mbox{tr}[H]^2 & = & 4\, \left(r^2-s\,t\right) \; ,
\end{eqnarray}
and
\begin{eqnarray} \frac{E_1-E_2}{2} \; n_ 3 \; = \;  \frac{H_{11}-H_{22}}{2} & = &  +\, (-1)^\ell\,  i\,  r \; , \\[2mm]
\frac{E^\ast_1-E^\ast_2}{2} \; n^\ast_ 3 \; = \;  \frac{H^\ast_{11}-H^\ast_{22}}{2} & = & -\, (-1)^\ell\,  i\,  r \; .
\end{eqnarray}
Following the discussion of Sections \ref{xmetrici}, \ref{xmetricp} and \ref{xconvrep1} we obtain the following results for the four different cases under consideration:

\begin{itemize} 
\item {\bf Case 1:} $H$ is pseudo-Hermitian ($H^+ = + \eta \, H \, \eta^{-1}\,$), \\
\mbox{} $\qquad\quad\;\;$ the phase of the metric is trivial ($Q=1_2$)
\begin{eqnarray}  s\,t  & > &  (r\, \sin\theta)^2 \quad \Rightarrow \nonumber \\[2mm]
 H  & = &  r\, \cos\theta \,\; 1_2 \; \pm \; \sqrt{s\,t - (r\, \sin\theta)^2} \; \,\vec{n} \cdot \vec{\sigma} \; ,  \\[2mm]
 E_1 & = & r\, \cos\theta  \; \pm \; \sqrt{s\,t - (r\, \sin\theta)^2} \; , \\[2mm]
 E_2 & = & r\, \cos\theta \; \mp \; \sqrt{s\,t - (r\, \sin\theta)^2} \; , \\[2mm]
 \frac{E_1-E_2}{2}  & = &  \frac{E^\ast_1-E^\ast_2}{2} \; = \; \pm \; \sqrt{s\,t - (r\, \sin\theta)^2} \; , \\[2mm]
\vec{n} \cdot \vec{\sigma} & = & \pm\; \frac{\left(\begin{array}{cc} +i \,r\,\sin\theta  & s \, e^{+i\,\phi} \\ t \, e^{-i\,\phi} & -i \,r\,\sin\theta \end{array}\right)}{\sqrt{s\,t - (r\, \sin\theta)^2}}   , \qquad \\[2mm]
X & = &  \left(\frac{\sqrt{s\,t - (r\, \sin\theta)^2} \; \pm i\,r\,\sin\theta}{2\; \sqrt{s\,t - (r\, \sin\theta)^2}} \right)^{\frac{1}{2}} \nonumber \\[2mm]
 & \cdot & \left( 1_2 \mp   \frac{\left(\begin{array}{cc} 0  & - s  \, e^{+i\,\phi} \\ t  \, e^{-i\,\phi} & 0 \end{array}\right)}{\sqrt{s\,t - (r\, \sin\theta)^2} \; \pm i\,r\,\sin\theta} \right)  , \qquad \\[2mm]
X^+ & = &  \left(\frac{\sqrt{s\,t - (r\, \sin\theta)^2} \; \mp i\,r\,\sin\theta}{2\; \sqrt{s\,t - (r\, \sin\theta)^2}} \right)^{\frac{1}{2}} \nonumber \\[2mm]
 & \cdot & \left( 1_2 \pm   \frac{\left(\begin{array}{cc} 0  & - t  \, e^{+i\,\phi} \\ s  \, e^{-i\,\phi} & 0 \end{array}\right)}{\sqrt{s\,t - (r\, \sin\theta)^2} \; \mp i\,r\,\sin\theta} \right)  , \quad \\[5mm]
\eta_0 & = & \eta^+_0 \; = \;  \frac{|N_1|^2+|N_2|^2}{2} \; \;   X^+ X \nonumber  \\[2mm]
 & = & \mbox{\small $\bigcirc\hspace{-3.2mm}\pm$}\; \frac{|N_1|^2+|N_2|^2}{2} \;\; \sqrt{\frac{s\, t}{s\,t - (r\, \sin\theta)^2}}  \nonumber \\
 & \cdot & \Bigg( \frac{s + t}{2 \, s\, t} \left(\begin{array}{cc} t &  - i \, r \, \sin\theta \;e^{+i\,\phi} \\   i \, r \, \sin\theta \; e^{-i\,\phi} & s \end{array}\right)   \nonumber \\[2mm]
 &  & \pm \; \frac{s - t}{2 \, s\, t} \; \sqrt{s\,t - (r\, \sin\theta)^2} \;  \left(\begin{array}{cc} 0 &  e^{+i\,\phi} \\  e^{-i\,\phi} & 0 \end{array}\right)  \Bigg) \\[35mm]
\eta_3 & = & \eta^+_3 \; = \;  \frac{|N_1|^2-|N_2|^2}{2} \; \;   X^+\sigma_3 \, X  \nonumber \\[2mm]
 & = & \mbox{\small $\bigcirc\hspace{-3.2mm}\pm$}\;\frac{|N_1|^2-|N_2|^2}{2} \;\; \sqrt{\frac{s\, t}{s\,t - (r\, \sin\theta)^2}}  \nonumber \\ 
 & \cdot & \Bigg( \frac{s - t}{2 \, s\, t} \left(\begin{array}{cc} t &  - i \, r \, \sin\theta \; e^{+i\,\phi} \\   i \, r \, \sin\theta \; e^{-i\,\phi} & s \end{array}\right)   \nonumber \\[2mm]
 &  & \pm \; \frac{s + t}{2 \, s\, t} \; \sqrt{s\,t - (r\, \sin\theta)^2} \;  \left(\begin{array}{cc} 0 &  e^{+i\,\phi} \\  e^{-i\,\phi} & 0 \end{array}\right)  \Bigg)  . 
\end{eqnarray}
\item {\bf Case 2:} $H$ is pseudo-Hermitian ($H^+ = + \eta \, H \, \eta^{-1}\,$), \\
\mbox{} $\qquad\quad\;\;$ the phase of the metric is non-trivial ($Q=P$)
\begin{eqnarray}    (r\, \sin\theta)^2 & > &  s\,t \quad \Rightarrow \nonumber \\[2mm]
 H  & = &  r\, \cos\theta \;\, 1_2 \; \pm \; i\; \sqrt{(r\, \sin\theta)^2-s\,t} \;\, \vec{n} \cdot \vec{\sigma} \; ,  \\[2mm]
 E_1 & = & r\, \cos\theta \; \pm \; i\; \sqrt{(r\, \sin\theta)^2-s\,t} \; , \\[2mm] 
 E_2 & = & r\, \cos\theta \; \mp \; i\; \sqrt{(r\, \sin\theta)^2-s\,t}\;  , \\[2mm]  
 \frac{E_1-E_2}{2}  & = & -\;  \frac{E^\ast_1-E^\ast_2}{2} \; = \;  \pm \; i\; \sqrt{(r\, \sin\theta)^2-s\,t} \; , \\[2mm]
\vec{n} \cdot \vec{\sigma} & = & \mp\, i \; \frac{\left(\begin{array}{cc} +i \,r \,\sin\theta  & s \, e^{+i\,\phi} \\ t \, e^{-i\,\phi} & -i \,r \, \sin\theta \end{array}\right)}{\sqrt{(r\, \sin\theta)^2-s\,t}}  \; ,  \\[2mm]
X  & = &  \left( \frac{\sqrt{(r\, \sin\theta)^2-s\,t}\; \pm r\,\sin\theta}{2\; \sqrt{(r\, \sin\theta)^2-s\,t}} \right)^{\frac{1}{2}} \nonumber \\[2mm]
 & \cdot &  \left( 1_2 \pm   \frac{\left(\begin{array}{cc} 0  & - i\, s  \, e^{+i\,\phi} \\ + i\, t  \, e^{-i\,\phi} & 0 \end{array}\right)}{\sqrt{(r\, \sin\theta)^2-s\,t}\; \pm r\,\sin\theta} \right)  , \qquad \\[25mm]
 X^+  & = & \left( \frac{\sqrt{(r\, \sin\theta)^2-s\,t}\; \pm r\,\sin\theta}{2\; \sqrt{(r\, \sin\theta)^2-s\,t}} \right)^{\frac{1}{2}} \nonumber \\[2mm]
 & \cdot &  \left( 1_2 \pm   \frac{\left(\begin{array}{cc} 0  & - i\, t  \, e^{+i\,\phi} \\ + i\, s  \, e^{-i\,\phi} & 0 \end{array}\right)}{\sqrt{(r\, \sin\theta)^2-s\,t}\; \pm r\,\sin\theta} \right)  , \qquad \\[2mm]
\eta_1 & = &  \frac{N^\ast_2\,N_1+N^\ast_1\,N_2}{2} \; \;    X^+ \; \vec{e}\cdot \vec{\sigma}\;   X   \nonumber \\[2mm]
  & = &  \frac{N^\ast_2\,N_1+N^\ast_1\,N_2}{2} \; \Big( \cos\varphi  \;  X^+ \sigma_1\,   X + \sin\varphi  \;  X^+ \sigma_2\,   X\Big)   \nonumber \\
 & = & \mbox{\small $\bigcirc\hspace{-3.2mm}\pm$}\;\; \frac{N^\ast_2\,N_1+N^\ast_1\,N_2}{2} \;   \cdot   \nonumber \\[2mm]
 & \cdot & \Bigg( (\cos\varphi \; \cos\phi \; - \; \sin\varphi \; \sin\phi ) \left(\begin{array}{cc} 0 &  e^{+i\,\phi} \\  e^{-i\,\phi} & 0 \end{array}\right) \nonumber \\[2mm]
 & & \pm \; \frac{\cos\varphi \; \sin\phi \; + \; \sin\varphi \; \cos\phi}{\sqrt{(r\, \sin\theta)^2-s\,t}} \; \cdot \nonumber \\[2mm]
 & & \quad \cdot \; \left(\begin{array}{cc} t &  - i \, r \, \sin\theta \; e^{+i\,\phi} \\   i \, r \, \sin\theta \; e^{-i\,\phi} & s \end{array}\right)  \Bigg) \; , \\[2mm]
 & & \nonumber \\
 \eta_2 & = &  \frac{N^\ast_2\,N_1-N^\ast_1\,N_2}{2\, i} \; \;    X^+ \; \vec{e}_\perp\cdot \vec{\sigma} \; X\nonumber \\[2mm]
 & = &  \frac{N^\ast_2\,N_1-N^\ast_1\,N_2}{2\, i} \; \Big( \sin\varphi  \;  X^+ \sigma_1\,   X - \cos\varphi  \;  X^+ \sigma_2\,   X\Big) \nonumber \\[3mm]
 & = & \mbox{\small $\bigcirc\hspace{-3.2mm}\pm$}\;\; \frac{N^\ast_2\,N_1-N^\ast_1\,N_2}{2\, i} \;   \cdot   \nonumber \\[2mm]
 & \cdot & \Bigg( (\sin\varphi \; \cos\phi \; + \; \cos\varphi \; \sin\phi ) \left(\begin{array}{cc} 0 &  e^{+i\,\phi} \\  e^{-i\,\phi} & 0 \end{array}\right) \nonumber \\[2mm]
 & & \pm \; \frac{\sin\varphi \; \sin\phi \; - \; \cos\varphi \; \cos\phi}{\sqrt{(r\, \sin\theta)^2-s\,t}} \; \cdot \nonumber \\[2mm]
 & & \quad \cdot \; \left(\begin{array}{cc} t &  - i \, r \, \sin\theta \; e^{+i\,\phi} \\   i \, r \, \sin\theta \; e^{-i\,\phi} & s \end{array}\right)  \Bigg) \; .
\end{eqnarray}
\item {\bf Case 3:} $H$ is anti-pseudo-Hermitian ($H^+ = - \,\eta \, H \, \eta^{-1}$) implying\\
\mbox{} $\qquad\quad\;\;$ $i H$ to be pseudo-Hermitian  ($(i H)^+ = + \,\eta \; i H \; \eta^{-1}$), \\
\mbox{} $\qquad\quad\;\;$ the phase of the metric is trivial ($Q=1_2$)
\begin{eqnarray}   r^2 \; > &  s\,t & , \;  r\cos\theta \; = \;  0 \quad \Rightarrow \nonumber \\[2mm]
 H  & = &  \pm \, i\; \sqrt{r^2 -s\,t} \; \; \vec{n} \cdot \vec{\sigma} \; , \\[2mm]
 E_1 & = & \pm \, i \;  \sqrt{r^2 -s\,t} \; ,  \quad  E_2 \; = \; \mp \, i \;  \sqrt{r^2 -s\,t} \; , \\[2mm] 
 \frac{E_1-E_2}{2}  & = &  -\;  \frac{E^\ast_1-E^\ast_2}{2} \; = \; \pm \, i \;  \sqrt{r^2 -s\,t} \; , \\[2mm]
\vec{n} \cdot \vec{\sigma} & = & \mp\, i \; \frac{\left(\begin{array}{cc}  (-1)^\ell\,  i\,  r & s \, e^{+i\,\phi} \\ t \, e^{-i\,\phi} & -\, (-1)^\ell \,  i\,  r\end{array}\right)}{\sqrt{r^2 -s\,t}} , \qquad \\[2mm]
X   & = & \left( \frac{\sqrt{r^2 -s\,t}\; \pm (-1)^\ell\, r}{2\; \sqrt{r^2 -s\,t}} \right)^{\frac{1}{2}} \nonumber \\[2mm]
 & \cdot &  \left( 1_2 \pm   \frac{\left(\begin{array}{cc} 0  & - i\, s  \, e^{+i\,\phi} \\ + i\, t  \, e^{-i\,\phi} & 0 \end{array}\right)}{\sqrt{r^2 -s\,t}\; \pm (-1)^\ell\, r} \right)  , \qquad \\[2mm]
X^+  & = &  \left( \frac{\sqrt{r^2 -s\,t}\; \pm (-1)^\ell\, r}{2\; \sqrt{r^2 -s\,t}} \right)^{\frac{1}{2}} \nonumber \\[2mm]
 & \cdot &  \left( 1_2 \pm   \frac{\left(\begin{array}{cc} 0  & - i\, t  \, e^{+i\,\phi} \\ + i\, s  \, e^{-i\,\phi} & 0 \end{array}\right)}{\sqrt{r^2 -s\,t}\; \pm (-1)^\ell\, r} \right)  , \qquad \\[2mm]
\eta_0 & = & \eta^+_0 \; = \;  \frac{|N_1|^2+|N_2|^2}{2} \; \;   X^+ X  \nonumber \\[2mm]
 & = & \mbox{\small $\bigcirc\hspace{-3.2mm}\pm$}\;\frac{|N_1|^2+|N_2|^2}{2} \; \Bigg(  \frac{s-t}{2\, s \, t} \; \left(\begin{array}{cc} t  & 0 \\ 0 & -s \end{array}\right) \nonumber \\[2mm]
 &  & \pm \;   \frac{s+t}{2\, s \, t} \;  \; \frac{1}{\sqrt{r^2 -s\,t}} \left(\begin{array}{cc}   (-1)^\ell\, r\, t & - i s\, t\; e^{+i\,\phi} \\ + i  \, s\, t \; e^{-i\,\phi} & (-1)^\ell\, r\, s \end{array}\right) \Bigg)  , \qquad \\[2mm]
\eta_3 & = & \eta^+_3 \; = \;  \frac{|N_1|^2-|N_2|^2}{2} \; \;   X^+\sigma_3 \, X \nonumber  \\[2mm]
 & = & \mbox{\small $\bigcirc\hspace{-3.2mm}\pm$}\;\frac{|N_1|^2-|N_2|^2}{2} \;\Bigg(  \frac{s+t}{2\, s \, t} \;  \left(\begin{array}{cc} t  & 0 \\ 0 & -s \end{array}\right)\nonumber \\[2mm]
 & &  \pm \;  \frac{s-t}{2\, s \, t} \;   \frac{1}{\sqrt{r^2 -s\,t}} \left(\begin{array}{cc}  (-1)^\ell\, r\, t & - i \, s\, t \; e^{+i\,\phi} \\ + i \, s\, t \; e^{-i\,\phi}  & (-1)^\ell\, r\, s \end{array} \right) \Bigg) . \quad\quad 
\end{eqnarray}
\item {\bf Case 4:} $H$ is anti-pseudo-Hermitian ($H^+ = - \,\eta \, H \, \eta^{-1}$) implying\\
\mbox{} $\qquad\quad\;\;$ $i H$ to be pseudo-Hermitian  ($(i H)^+ = + \,\eta \; i H \; \eta^{-1}$), \\
\mbox{} $\qquad\quad\;\;$ the phase of the metric is non-trivial ($Q=P$)
\begin{eqnarray}    s\, t \; > &  r^2 & , \;  r\cos\theta\; = \;  0 \quad \Rightarrow \nonumber \\[2mm]
 H  & = &  \pm \; \sqrt{s\, t - r^2} \;\; \vec{n} \cdot \vec{\sigma} \; ,  \\[2mm]
 E_1 & = & \pm  \;  \sqrt{s\, t - r^2} \; ,  \quad E_2 \; = \; \mp  \;  \sqrt{s\, t - r^2} \; , \\[2mm] 
 \frac{E_1-E_2}{2}  & = &  \frac{E^\ast_1-E^\ast_2}{2} \; = \; \pm \; \sqrt{s\,t - r^2} \; , \\[2mm]
 \vec{n} \cdot \vec{\sigma} & = & \pm \; \frac{\left(\begin{array}{cc} +\,(-1)^\ell\,i\, r  & s \, e^{+i\,\phi} \\ t \, e^{-i\,\phi} & - \,(-1)^\ell\,i\, r \end{array}\right)}{\sqrt{s\, t - r^2}}  , \qquad \\[2mm]
X & = &   \left(\frac{\sqrt{s\, t - r^2} \; \pm (-1)^\ell\,i\, r}{2\; \sqrt{s\, t - r^2}} \right)^{\frac{1}{2}} \nonumber \\[2mm] 
 & \cdot & \left( 1_2 \mp   \frac{\left(\begin{array}{cc} 0  & -s  \, e^{+i\,\phi} \\ t  \, e^{-i\,\phi} & 0 \end{array}\right)}{\sqrt{s\, t - r^2} \; \pm (-1)^\ell\,i\, r} \right)  , \qquad \\[2mm]
X^+ & = &  \left(\frac{\sqrt{s\, t - r^2} \; \mp (-1)^\ell\,i\, r}{2\; \sqrt{s\, t - r^2}} \right)^{\frac{1}{2}} \nonumber \\[2mm] 
 & \cdot & \left( 1_2 \pm   \frac{\left(\begin{array}{cc} 0  & -t  \, e^{+i\,\phi} \\ s  \, e^{-i\,\phi} & 0 \end{array}\right)}{\sqrt{s\, t - r^2} \; \mp (-1)^\ell\,i\, r} \right)  , \qquad \\[2mm]
 \eta_1 & = &  \frac{N^\ast_2\,N_1+N^\ast_1\,N_2}{2} \; \;    X^+ \; \vec{e}\cdot \vec{\sigma}\;   X   \nonumber \\[2mm]
  & = &  \frac{N^\ast_2\,N_1+N^\ast_1\,N_2}{2} \; \Big( \cos\varphi  \;  X^+ \sigma_1\,   X + \sin\varphi  \;  X^+ \sigma_2\,   X\Big)   \nonumber \\[2mm]
 & & \nonumber \\
& = & \mbox{\small $\bigcirc\hspace{-3.2mm}\pm$}\;\; \frac{N^\ast_2\,N_1+N^\ast_1\,N_2}{2} \; \sqrt{\frac{s\,t}{s\,t-r^2}} \; \frac{1}{s\, t} \;  \cdot   \nonumber \\[2mm]
 & \cdot & \Bigg( \mp (\cos\varphi \; \cos\phi \; - \; \sin\varphi \; \sin\phi ) \;  \sqrt{s\,t-r^2} \; \left(\begin{array}{cc} t & 0 \\  0 & -s \end{array}\right) \nonumber \\[2mm]
 & & + \; (\cos\varphi \; \sin\phi \; + \; \sin\varphi \; \cos\phi) \; \cdot \nonumber \\[2mm]
 & & \quad \cdot \; \left(\begin{array}{cc}  (-1)^\ell\, r\, t & - i \, s\, t \; e^{+i\,\phi} \\ + i \, s\, t \; e^{-i\,\phi}  & (-1)^\ell\, r\, s \end{array} \right)  \Bigg) \; , \\[2mm]
 & & \nonumber \\
 \eta_2 & = &  \frac{N^\ast_2\,N_1-N^\ast_1\,N_2}{2\, i} \; \;    X^+ \; \vec{e}_\perp\cdot \vec{\sigma} \; X\nonumber \\[2mm]
 & = &  \frac{N^\ast_2\,N_1-N^\ast_1\,N_2}{2\,i} \; \Big( \sin\varphi  \;  X^+ \sigma_1\,   X - \cos\varphi  \;  X^+ \sigma_2\,   X\Big)  \nonumber \\[3mm]
& = & \mbox{\small $\bigcirc\hspace{-3.2mm}\pm$}\;\; \frac{N^\ast_2\,N_1-N^\ast_1\,N_2}{2\, i} \; \sqrt{\frac{s\,t}{s\,t-r^2}} \; \frac{1}{s\, t} \;  \cdot   \nonumber \\[2mm]
 & \cdot & \Bigg( \mp (\sin\varphi \; \cos\phi \; + \; \cos\varphi \; \sin\phi ) \;  \sqrt{s\,t-r^2} \; \left(\begin{array}{cc} t & 0 \\  0 & -s \end{array}\right) \nonumber \\[2mm]
 & & + \; (\sin\varphi \; \sin\phi \; - \; \cos\varphi \; \cos\phi) \; \cdot \nonumber \\[2mm]
 & & \quad \cdot \; \left(\begin{array}{cc}  (-1)^\ell\, r\, t & - i \, s\, t \; e^{+i\,\phi} \\ + i \, s\, t \; e^{-i\,\phi}  & (-1)^\ell\, r\, s \end{array} \right)  \Bigg) \; .
\end{eqnarray}
 \end{itemize}

\newpage
\section{A general ${\cal PT}$-symmetric Hamilton operator proposed by C.\ M.\ Bender, P.~Meisinger and  Q.\ Wang extended by A.~Mostafazadeh}  \label{xappenda3}
\noindent In this section we will apply our analysis to the following asymmetric two-dimensional Hamilton operator $H$ (and the respective partners $H^+$, $i H$, $(i H)^+$) proposed and considered  for the symmetric case $u=0$ by C.\ M.\ Bender, P.~Meisinger and  Q.\ Wang \cite{Bender:2003gu}\cite{Wang:2010br}\cite{Wang:2013oqw} and later generalized for the asymmetric case $u\not=0$ by A.\ Mostafazadeh \cite{Mostafazadeh:2003gz} (see also Q.~Wang et al.\ \cite{Wang:2010br}) (with $r$, $s$, $t$, $u$ and $\phi$ being real-valued parameters):
\begin{eqnarray} H & = &  \left(\begin{array}{cc} [\,r + (t\, \cos\phi - i \, s\,\sin\phi)\, ] & [\, i\, ( s\, \cos\phi - u) + t \sin\phi\, ] \\[2mm]  [\, i\, ( s\, \cos\phi + u) + t \sin\phi ] &   [\, r -(t\, \cos\phi - i \, s\,\sin\phi)\,] \end{array}\right) \; , \nonumber \\[2mm]
 & & \\
  i \, H & = &  \left(\begin{array}{cc} [\,i \, r +(i\, t\, \cos\phi + \, s\,\sin\phi)\, ] & [\, -\, ( s\, \cos\phi - u) + i\, t \sin\phi\, ] \\[2mm]  [\, -\, ( s\, \cos\phi + u) + i\, t \sin\phi ] & [\, i\, r -(i\, t\, \cos\phi + \, s\,\sin\phi)\,] \end{array}\right)  , \nonumber \\[2mm]
 & & \\
  H^+ & = &  \left(\begin{array}{cc}  [\,r + (t\, \cos\phi + i \, s\,\sin\phi)\, ] & [\,- i\, ( s\, \cos\phi + u) + t \sin\phi\, ] \\[2mm]  [\,- i\, ( s\, \cos\phi - u) + t \sin\phi ] &  [\,r - (t\, \cos\phi + i \, s\,\sin\phi)\, ] \end{array}\right)  \; , \nonumber \\[2mm]
 & & \\ 
(i \, H)^+ & = & \nonumber \\[2mm] 
\lefteqn{\left(\begin{array}{cc} [\,-i \, r +(-i\, t\, \cos\phi + \, s\,\sin\phi)\, ] & [\, -\, ( s\, \cos\phi + u) - i\, t \sin\phi\, ] \\[2mm]  [\, -\, ( s\, \cos\phi - u) - i\, t \sin\phi ] &  [\,-i \, r -(-i\, t\, \cos\phi + \, s\,\sin\phi)\, ] \end{array}\right)  . }\nonumber \\[2mm]
\end{eqnarray}
A calculation of traces and determinants yields:
\begin{eqnarray}  \mbox{tr}[H] \; = & \mbox{tr}[H^+] & = \; 2\, r \; , \\[2mm] 
 \mbox{tr}[i H] \; = & -\,\mbox{tr}[(i H)^+] & = \;  2 \, i \,r \; , \\[2mm]
 \mbox{det}[H] \; = & \mbox{det}[H^+] & = \; +(r^2 + s^2) - (t^2 + u^2) \; , \\[2mm] 
 \mbox{det}[i H] \; = & \mbox{det}[(i H)^+] & = \;-(r^2 + s^2) + (t^2 + u^2) \; .
\end{eqnarray}
According to the discussion of Eqs.\ (\ref{consid1}) and (\ref{consid2})  the two-dimensional Hamilton operator $H$ and the conjugate Hamilton operator $H^+$ are pseudo-Hermitian for all real parameters $r$, $s$, $t$, $u$ and $\phi$, while $H$ and $H^+$ are anti-pseudo-Hermitian  for all real parameters $s$, $t$, $u$ and $\phi$ under the restriction  $r=0$. For $r\not =0$ there holds:
\begin{eqnarray} \mbox{tr}[H]^2-4\,  \mbox{det}[H] & = & 4\, \left(t^2 + u^2 - s^2\right) \; , \label{abcident1} \\[2mm]
 4\,  \mbox{det}[H] - \mbox{tr}[H]^2 & = & 4\, \left(s^2 - t^2 - u^2\right) \; , \label{abcident2}
\end{eqnarray}
and
\begin{eqnarray} \frac{E_1-E_2}{2} \; n_ 3 \; = \;  \frac{H_{11}-H_{22}}{2} & = &  t\, \cos\phi - i \, s\,\sin\phi \; ,  \label{abcident3} \; , \\[2mm]
\frac{E^\ast_1-E^\ast_2}{2} \; n^\ast_ 3 \; = \;  \frac{H^\ast_{11}-H^\ast_{22}}{2} & = & t\, \cos\phi + i \, s\,\sin\phi \; . \label{abcident4}
\end{eqnarray}

In the case of anti-pseudo-Hermiticity ($r=0$) the asymmetric two-dimensional Hamilton operator $H$ (and the respective partners $H^+$, $i H$, $(i H)^+$) take the following simple form:
\begin{eqnarray} H & = &  \left(\begin{array}{cc} [\,t\, \cos\phi - i \, s\,\sin\phi\, ] & [\, i\, ( s\, \cos\phi - u) + t \sin\phi\, ] \\[2mm]  [\, i\, ( s\, \cos\phi + u) + t \sin\phi ] &   -\,[\, t\, \cos\phi - i \, s\,\sin\phi\,] \end{array}\right) \; , \nonumber \\[2mm]
 & & \\
  i \, H & = &  \left(\begin{array}{cc} [\,i\, t\, \cos\phi + \, s\,\sin\phi\, ] & [\, -\, ( s\, \cos\phi - u) + i\, t \sin\phi\, ] \\[2mm]  [\, -\, ( s\, \cos\phi + u) + i\, t \sin\phi ] & -\, [\, i\, t\, \cos\phi + \, s\,\sin\phi\,] \end{array}\right)  , \nonumber \\[2mm]
 & & \\
  H^+ & = &  \left(\begin{array}{cc}  [\,t\, \cos\phi + i \, s\,\sin\phi\, ] & [\,- i\, ( s\, \cos\phi + u) + t \sin\phi\, ] \\[2mm]  [\,- i\, ( s\, \cos\phi - u) + t \sin\phi ] & -\, [\,t\, \cos\phi + i \, s\,\sin\phi\, ] \end{array}\right)  \; , \nonumber \\[2mm]
 & & \\ 
(i \, H)^+ & = & \left(\begin{array}{cc} [\,-i\, t\, \cos\phi + \, s\,\sin\phi\, ] & [\, -\, ( s\, \cos\phi + u) - i\, t \sin\phi\, ] \\[2mm]  [\, -\, ( s\, \cos\phi - u) - i\, t \sin\phi ] &  -\, [\,-i\, t\, \cos\phi + \, s\,\sin\phi\, ] \end{array}\right)  . \nonumber \\[2mm]
\end{eqnarray}
It should be stressed that Eqs.\ (\ref{abcident1}), (\ref{abcident2}), (\ref{abcident3}) and (\ref{abcident1}) are also valid for the case of anti-pseudo-Hermiticity, i.\ e., for $r=0$.  

Following the discussion of Sections \ref{xmetrici}, \ref{xmetricp} and \ref{xconvrep1} we obtain the following results for the four different cases under consideration:
\newpage
\begin{itemize} 
\item {\bf Case 1:} $H$ is pseudo-Hermitian ($H^+ = + \eta \, H \, \eta^{-1}\,$), \\
\mbox{} $\qquad\quad\;\;$ the phase of the metric is trivial ($Q=1_2$)
\begin{eqnarray}  t^2 + u^2   & > &  s^2 \quad \Rightarrow \nonumber \\[2mm]
 H  & = &  r \,\; 1_2 \; \pm \; \sqrt{t^2 + u^2 - s^2} \; \,\vec{n} \cdot \vec{\sigma} \; ,  \\[2mm]
 E_1 & = & r  \; \pm \; \sqrt{t^2 + u^2 - s^2} \; , \\[2mm]
 E_2 & = & r \; \mp \; \sqrt{t^2 + u^2 - s^2} \; , \\[2mm]
 \frac{E_1-E_2}{2}  & = &  \frac{E^\ast_1-E^\ast_2}{2} \; = \; \pm \; \sqrt{t^2 + u^2 - s^2} \; , \\[2mm]
\vec{n} \cdot \vec{\sigma}  & = & \pm \; \frac{\small\left(\begin{array}{cc} [\,t\, \cos\phi - i \, s\,\sin\phi\, ] & [\, i\, ( s\, \cos\phi - u) + t \sin\phi\, ] \\[2mm]  [\, i\, ( s\, \cos\phi + u) + t \sin\phi ] &   -\,[\, t\, \cos\phi - i \, s\,\sin\phi\,] \end{array}\right)}{\sqrt{t^2 + u^2 - s^2}}   , \nonumber \\[2mm]
 & & \\
X & = &  \left(\frac{\sqrt{t^2 + u^2 - s^2} \; \pm(t\, \cos\phi - i \, s\,\sin\phi)}{2\; \sqrt{t^2 + u^2 - s^2}} \right)^{\frac{1}{2}}  \nonumber \\[2mm]
\cdot  \Bigg( 1_2 & \mp &  \frac{\small \left(\begin{array}{cc} 0 & - [\, i\, ( s\, \cos\phi - u) + t \sin\phi\, ] \\[2mm]  [\, i\, ( s\, \cos\phi + u) + t \sin\phi ] &   0 \end{array}\right)}{\sqrt{t^2 + u^2 - s^2} \; \pm(t\, \cos\phi - i \, s\,\sin\phi)} \Bigg)  , \nonumber \\[2mm]
 & & \\
X^+ & = &  \left(\frac{\sqrt{t^2 + u^2 - s^2} \; \pm(t\, \cos\phi + i \, s\,\sin\phi)}{2\; \sqrt{t^2 + u^2 - s^2}} \right)^{\frac{1}{2}} \nonumber \\[2mm]
\cdot  \Bigg( 1_2 & \mp &   \frac{\small \left(\begin{array}{cc} 0 &   [\,- i\, ( s\, \cos\phi + u) + t \sin\phi ] \\[2mm] -\,[\,- i\, ( s\, \cos\phi - u) + t \sin\phi\, ] &   0 \end{array}\right)}{\sqrt{t^2 + u^2 - s^2} \; \pm(t\, \cos\phi + i \, s\,\sin\phi)} \Bigg)  , \nonumber \\[2mm]
 & & \end{eqnarray}
\newpage
\begin{eqnarray} \eta_0 & = & \eta^+_0 \; = \;  \frac{|N_1|^2+|N_2|^2}{2} \; \;   X^+ X \nonumber  \\[2mm]
 & = & \mbox{\small $\bigcirc\hspace{-3.2mm}\pm$}\;\; \frac{|N_1|^2+|N_2|^2}{2} \;\; \frac{1}{\sqrt{t^2 + u^2 - s^2}}  \nonumber \\[2mm]
 & \cdot &  \frac{1}{\sqrt{\left(\pm \;  \sqrt{t^2 + u^2 - s^2} + t \cos\phi \right)^2 + (s\,\sin\phi)^2}}   \nonumber \\[2mm]
 & \cdot & \Bigg(\left(\begin{array}{cc} t^2 + u^2   &   i \,s\, t\\[2mm]  -\,  i \, s\,  t  & t^2 + u^2 \end{array}\right)   + s\, u \; \left(\begin{array}{cc} \cos\phi &  \sin\phi \\   \sin\phi & -\cos\phi \end{array}\right) \nonumber \\[5mm]
 &  & \quad \pm \; \cos\phi \;\; \sqrt{t^2 + u^2 - s^2} \;  \left(\begin{array}{cc} t &  i\, s \\  -i\, s & t \end{array}\right)  \Bigg) \; ,\\[2mm]
\eta_3 & = & \eta^+_3 \; = \;  \frac{|N_1|^2-|N_2|^2}{2} \; \;   X^+\sigma_3 \, X  \nonumber \\[2mm]
 & = & \mbox{\small $\bigcirc\hspace{-3.2mm}\pm$}\;\;\frac{|N_1|^2-|N_2|^2}{2} \;\; \frac{1}{\sqrt{t^2 + u^2 - s^2}}  \nonumber \\[2mm]
 & \cdot &  \frac{1}{\sqrt{\left(\pm \;  \sqrt{t^2 + u^2 - s^2} + t \cos\phi \right)^2 + (s\,\sin\phi)^2}}   \nonumber \\[2mm]
 & \cdot & \Bigg( \cos\phi \;\Bigg(  (t^2- s^2) \left(\begin{array}{cc} \cos\phi &  \sin\phi \\   \sin\phi & -\cos\phi \end{array}\right)  - u\, \left(\begin{array}{cc}  s & i\, t \\  -i\,t & s \end{array}\right) \Bigg)  \nonumber \\[2mm]
 &  & \pm  \; \sqrt{t^2 + u^2 - s^2} \; \cdot \nonumber \\[2mm]
 & & \quad \cdot \;  \Bigg(  t \left(\begin{array}{cc} \cos\phi &  \sin\phi \\   \sin\phi & -\cos\phi \end{array}\right)  + i\, u\; \left(\begin{array}{cc}  0 & -\,1 \\  1 & 0 \end{array}\right) \Bigg)   \Bigg)  . 
\end{eqnarray}
\newpage
\item {\bf Case 2:} $H$ is pseudo-Hermitian ($H^+ = + \eta \, H \, \eta^{-1}\,$), \\
\mbox{} $\qquad\quad\;\;$ the phase of the metric is non-trivial ($Q=P$)
\begin{eqnarray}    s^2 & > &   t^2 + u^2  \quad \Rightarrow \nonumber \\[2mm]
 H  & = &  r \;\, 1_2 \; \pm \; i\; \sqrt{s^2 - t^2 - u^2} \;\, \vec{n} \cdot \vec{\sigma} \; ,  \\[2mm]
 E_1 & = & r \; \pm \; i\; \sqrt{s^2 - t^2 - u^2} \; , \\[2mm] 
 E_2 & = & r \; \mp \; i\; \sqrt{s^2 - t^2 - u^2}\;  , \\[2mm]  
 \frac{E_1-E_2}{2}  & = & -\;  \frac{E^\ast_1-E^\ast_2}{2} \; = \;  \pm \; i\; \sqrt{s^2 - t^2 - u^2} \; , \\[2mm]
\vec{n} \cdot \vec{\sigma} & = & \mp\, i \; \frac{\left(\begin{array}{cc} [\,t\, \cos\phi - i \, s\,\sin\phi\, ] & [\, i\, ( s\, \cos\phi - u) + t \sin\phi\, ] \\[2mm]  [\, i\, ( s\, \cos\phi + u) + t \sin\phi ] &   -\,[\, t\, \cos\phi - i \, s\,\sin\phi\,] \end{array}\right)}{\sqrt{s^2 - t^2 - u^2}}  \; ,  \nonumber \\[2mm]
 & & \\
X & = &  \left(\frac{\sqrt{s^2 - t^2 - u^2} \; \mp i\,(t\, \cos\phi - i \, s\,\sin\phi)}{2\; \sqrt{s^2 - t^2 - u^2}} \right)^{\frac{1}{2}}  \nonumber \\[2mm]
\cdot  \Bigg( 1_2 & \pm & i\;  \frac{\small \left(\begin{array}{cc} 0 & - [\, i\, ( s\, \cos\phi - u) + t \sin\phi\, ] \\[2mm]  [\, i\, ( s\, \cos\phi + u) + t \sin\phi ] &   0 \end{array}\right)}{ \sqrt{s^2 - t^2 - u^2} \; \mp \, i \,  (t\, \cos\phi - i \, s\,\sin\phi)} \Bigg)  , \nonumber \\[2mm]
 & & \\
X^+ & = &  \left(\frac{\sqrt{s^2 - t^2 - u^2} \; \pm i\,(t\, \cos\phi + i \, s\,\sin\phi)}{2\; \sqrt{s^2 - t^2 - u^2}} \right)^{\frac{1}{2}}  \nonumber \\[2mm]
\cdot  \Bigg( 1_2 & \mp & i\;  \frac{\small \left(\begin{array}{cc} 0 &  [\,- i\, ( s\, \cos\phi + u) + t \sin\phi ] \\[2mm] - [\,- i\, ( s\, \cos\phi - u) + t \sin\phi\, ]  &   0 \end{array}\right)}{ \sqrt{s^2 - t^2 - u^2} \; \pm \, i \,  (t\, \cos\phi + i \, s\,\sin\phi)} \Bigg)  , \nonumber \\[2mm]
 & & \end{eqnarray}

\begin{eqnarray}\eta_1 & = &  \frac{N^\ast_2\,N_1+N^\ast_1\,N_2}{2} \; \;    X^+ \; \vec{e}\cdot \vec{\sigma}\;   X   \nonumber \\[2mm]
  & = &  \frac{N^\ast_2\,N_1+N^\ast_1\,N_2}{2} \; \Big( \cos\varphi  \;  X^+ \sigma_1\,   X + \sin\varphi  \;  X^+ \sigma_2\,   X\Big)   \nonumber \\
 & = & \mbox{\small $\bigcirc\hspace{-3.2mm}\pm$}\;\; \frac{N^\ast_2\,N_1+N^\ast_1\,N_2}{2} \;\; \frac{1}{\sqrt{s^2 - t^2 - u^2}}  \nonumber \\[2mm]
 & \cdot &  \frac{1}{\sqrt{\left(\pm \;  \sqrt{s^2 - t^2 - u^2} \; - \, s \sin\phi \right)^2 + (t\,\cos\phi)^2}}   \;   \cdot   \nonumber \\[2mm]
 & \cdot & \Bigg( \cos\varphi \; \Bigg[ \; \sin\phi  \; \Bigg( u \; \left(\begin{array}{cc} s &  i\, t \\  -\,i\, t & s \end{array}\right) - (t^2 - s^2) \left(\begin{array}{cc} \cos\phi &  \sin\phi \\  \sin\phi & -\cos\phi \end{array}\right)\Bigg)  \nonumber \\[2mm]
 & & \qquad\quad\;\;\mp \; \sqrt{s^2 - t^2 - u^2} \;  \Bigg( u \; \left(\begin{array}{cc} 1 &  0 \\  0 & 1 \end{array}\right) + s \left(\begin{array}{cc} \cos\phi &  \sin\phi \\  \sin\phi & -\cos\phi \end{array}\right)\Bigg) \Bigg]\nonumber \\[2mm]
 & & -\,  \sin\varphi \; \Bigg[ \;  \left(\begin{array}{cc} t\,s &  i\, (u^2 - s^2) \\  -\, i\, (u^2 - s^2) & t\, s \end{array}\right) + tu \,  \left(\begin{array}{cc} \cos\phi &  \sin\phi \\  \sin\phi & -\cos\phi \end{array}\right) \nonumber \\[2mm]
 & & \qquad\quad\;\; \mp \; \sin\phi \; \sqrt{s^2 - t^2 - u^2} \;   \left(\begin{array}{cc} t &  i\, s \\  -\,i\, s & t \end{array}\right) \; \Bigg] \; \Bigg) \; , \\[2mm]
 \eta_2 & = &  \frac{N^\ast_2\,N_1-N^\ast_1\,N_2}{2\, i} \; \;    X^+ \; \vec{e}_\perp\cdot \vec{\sigma} \; X\nonumber \\[2mm]
 & = &  \frac{N^\ast_2\,N_1-N^\ast_1\,N_2}{2\, i} \; \Big( \sin\varphi  \;  X^+ \sigma_1\,   X - \cos\varphi  \;  X^+ \sigma_2\,   X\Big) \nonumber \\[3mm]
 & = & \mbox{\small $\bigcirc\hspace{-3.2mm}\pm$}\;\; \frac{N^\ast_2\,N_1-N^\ast_1\,N_2}{2\, i} \;\; \frac{1}{\sqrt{s^2 - t^2 - u^2}}  \nonumber \\[2mm]
 & \cdot &  \frac{1}{\sqrt{\left(\pm \;  \sqrt{s^2 - t^2 - u^2} \; - \, s \sin\phi \right)^2 + (t\,\cos\phi)^2}}   \;   \cdot   \nonumber \\[2mm]
 & \cdot & \Bigg( \sin\varphi \; \Bigg[ \; \sin\phi  \; \Bigg( u \; \left(\begin{array}{cc} s &  i\, t \\  -\,i\, t & s \end{array}\right) - (t^2 - s^2) \left(\begin{array}{cc} \cos\phi &  \sin\phi \\  \sin\phi & -\cos\phi \end{array}\right)\Bigg)  \nonumber \\[2mm]
 & & \qquad\quad\;\;\mp \; \sqrt{s^2 - t^2 - u^2} \;  \Bigg( u \; \left(\begin{array}{cc} 1 &  0 \\  0 & 1 \end{array}\right) + s \left(\begin{array}{cc} \cos\phi &  \sin\phi \\  \sin\phi & -\cos\phi \end{array}\right)\Bigg) \Bigg]\nonumber \\[2mm]
 & & +\,  \cos\varphi \; \Bigg[ \;  \left(\begin{array}{cc} t\,s &  i\, (u^2 - s^2) \\  -\, i\, (u^2 - s^2) & t\, s \end{array}\right) + tu \,  \left(\begin{array}{cc} \cos\phi &  \sin\phi \\  \sin\phi & -\cos\phi \end{array}\right) \nonumber \\[2mm]
 & & \qquad\quad\;\; \mp \; \sin\phi \; \sqrt{s^2 - t^2 - u^2} \;   \left(\begin{array}{cc} t &  i\, s \\  -\,i\, s & t \end{array}\right) \; \Bigg] \; \Bigg) \; .
\end{eqnarray}
\newpage
\item {\bf Case 3:} $H$ is anti-pseudo-Hermitian ($H^+ = - \,\eta \, H \, \eta^{-1}$) implying\\
\mbox{} $\qquad\quad\;\;$ $i H$ to be pseudo-Hermitian  ($(i H)^+ = + \,\eta \; i H \; \eta^{-1}$), \\
\mbox{} $\qquad\quad\;\;$ the phase of the metric is trivial ($Q=1_2$)
\begin{eqnarray}   s^2 & > &   t^2 + u^2 \; , \quad r= 0 \quad \Rightarrow \nonumber \\[2mm]
 H  & = &  \pm \, i\; \sqrt{s^2 - t^2 - u^2} \; \; \vec{n} \cdot \vec{\sigma} \; , \\[2mm]
 E_1 & = & \pm \, i \;  \sqrt{s^2 - t^2 - u^2} \; ,  \quad  E_2 \; = \; \mp \, i \;  \sqrt{s^2 - t^2 - u^2} \; , \\[2mm] 
 \frac{E_1-E_2}{2}  & = &  -\;  \frac{E^\ast_1-E^\ast_2}{2} \; = \; \pm \, i \;  \sqrt{s^2 - t^2 - u^2} \; , \\[2mm]
\vec{n} \cdot \vec{\sigma} & = & \mp\, i \; \frac{\left(\begin{array}{cc} [\,t\, \cos\phi - i \, s\,\sin\phi\, ] & [\, i\, ( s\, \cos\phi - u) + t \sin\phi\, ] \\[2mm]  [\, i\, ( s\, \cos\phi + u) + t \sin\phi ] &   -\,[\, t\, \cos\phi - i \, s\,\sin\phi\,] \end{array}\right)}{\sqrt{s^2 - t^2 - u^2}} , \nonumber \\[2mm]
 & & \\
X & = &  \left(\frac{\sqrt{s^2 - t^2 - u^2} \; \mp i\,(t\, \cos\phi - i \, s\,\sin\phi)}{2\; \sqrt{s^2 - t^2 - u^2}} \right)^{\frac{1}{2}}  \nonumber \\[2mm]
\cdot  \Bigg( 1_2 & \pm & i\;  \frac{\small \left(\begin{array}{cc} 0 & - [\, i\, ( s\, \cos\phi - u) + t \sin\phi\, ] \\[2mm]  [\, i\, ( s\, \cos\phi + u) + t \sin\phi ] &   0 \end{array}\right)}{ \sqrt{s^2 - t^2 - u^2} \; \mp \, i \,  (t\, \cos\phi - i \, s\,\sin\phi)} \Bigg)  , \nonumber \\[2mm]
 & & \\
X^+ & = &  \left(\frac{\sqrt{s^2 - t^2 - u^2} \; \pm i\,(t\, \cos\phi + i \, s\,\sin\phi)}{2\; \sqrt{s^2 - t^2 - u^2}} \right)^{\frac{1}{2}}  \nonumber \\[2mm]
\cdot  \Bigg( 1_2 & \mp & i\;  \frac{\small \left(\begin{array}{cc} 0 &  [\,- i\, ( s\, \cos\phi + u) + t \sin\phi ] \\[2mm] - [\,- i\, ( s\, \cos\phi - u) + t \sin\phi\, ]  &   0 \end{array}\right)}{ \sqrt{s^2 - t^2 - u^2} \; \pm \, i \,  (t\, \cos\phi + i \, s\,\sin\phi)} \Bigg)  , \nonumber \\[2mm]
 & & \end{eqnarray}
\newpage
\begin{eqnarray}\eta_0 & = & \eta^+_0 \; = \;  \frac{|N_1|^2+|N_2|^2}{2} \; \;   X^+  X  \nonumber \\[2mm]
 & = & \mbox{\small $\bigcirc\hspace{-3.2mm}\pm$}\;\;\frac{|N_1|^2+|N_2|^2}{2} \;\; \frac{1}{\sqrt{s^2 - t^2 - u^2}}  \nonumber \\[2mm]
 & \cdot &  \frac{1}{\sqrt{\left(\pm \;  \sqrt{s^2 - t^2 - u^2} \; - \, s \sin\phi \right)^2 + (t\,\cos\phi)^2}}   \nonumber \\[2mm]
 & \cdot & \Bigg( \; s\, \left(\begin{array}{cc}  s & i\, t \\  -i\,t & s \end{array}\right)  +  s\, u  \left(\begin{array}{cc} \cos\phi &  \sin\phi \\   \sin\phi & -\cos\phi \end{array}\right)   \nonumber \\[2mm]
 &  & \mp  \; \sqrt{s^2 - t^2 - u^2} \; \cdot \nonumber \\[2mm]
 & & \quad \cdot \;  \Bigg(   \sin\phi \,   \left(\begin{array}{cc}  s & i\, t \\  -i\,t & s \end{array}\right)  +\, u\; \left(\begin{array}{cc}  0 & 1 \\  1 & 0 \end{array}\right) \Bigg)   \Bigg)  , \\[5mm]
\eta_3 & = & \eta^+_3 \; = \;  \frac{|N_1|^2-|N_2|^2}{2} \; \;   X^+\sigma_3 \, X  \nonumber \\[2mm]
 & = & \mbox{\small $\bigcirc\hspace{-3.2mm}\pm$}\;\;\frac{|N_1|^2-|N_2|^2}{2} \;\; \frac{1}{\sqrt{s^2 - t^2 - u^2}}  \nonumber \\[2mm]
 & \cdot &  \frac{1}{\sqrt{\left(\pm \;  \sqrt{s^2 - t^2 - u^2} \; - \, s \sin\phi \right)^2 + (t\,\cos\phi)^2}}   \nonumber \\[2mm]
 & \cdot & \Bigg(   (s^2-t^2-u^2) \; \left(\begin{array}{cc}  1 & 0 \\  0 & -\, 1 \end{array}\right)+ \cos\phi \;\cdot  \nonumber \\[2mm]
 & & \quad  \cdot \; \Bigg( (t^2- s^2) \;  \left(\begin{array}{cc} \cos\phi &  \sin\phi \\   \sin\phi & -\cos\phi \end{array}\right)  - \, u\, \left(\begin{array}{cc}  s & i\, t \\  -i\,t & s \end{array}\right)\Bigg) \nonumber \\[2mm] 
  &  & \quad \pm \, s \; \, \sqrt{s^2 - t^2 - u^2}\; \left(\begin{array}{cc} - \sin\phi &  \cos\phi \\   \;\; \;\cos\phi & \sin\phi \end{array}\right)   \Bigg)  . 
\end{eqnarray}

\item {\bf Case 4:} $H$ is anti-pseudo-Hermitian ($H^+ = - \,\eta \, H \, \eta^{-1}$) implying\\
\mbox{} $\qquad\quad\;\;$ $i H$ to be pseudo-Hermitian  ($(i H)^+ = + \,\eta \; i H \; \eta^{-1}$), \\
\mbox{} $\qquad\quad\;\;$ the phase of the metric is non-trivial ($Q=P$)

\begin{eqnarray}     t^2 + u^2   & > &  s^2 \; , \quad r=0  \quad \Rightarrow \nonumber \\[2mm]
 H  & = &  \pm \; \sqrt{t^2 + u^2 - s^2} \;\; \vec{n} \cdot \vec{\sigma} \; ,  \\[2mm]
 E_1 & = & \pm  \;  \sqrt{t^2 + u^2 - s^2} \; ,  \quad E_2 \; = \; \mp  \;  \sqrt{t^2 + u^2 - s^2} \; , \\[2mm] 
 \frac{E_1-E_2}{2}  & = &  \frac{E^\ast_1-E^\ast_2}{2} \; = \; \pm \; \sqrt{t^2 + u^2 - s^2} \; , \\[2mm]
\vec{n} \cdot \vec{\sigma} & = & \pm \; \frac{\left(\begin{array}{cc} [\,t\, \cos\phi - i \, s\,\sin\phi\, ] & [\, i\, ( s\, \cos\phi - u) + t \sin\phi\, ] \\[2mm]  [\, i\, ( s\, \cos\phi + u) + t \sin\phi ] &   -\,[\, t\, \cos\phi - i \, s\,\sin\phi\,] \end{array}\right)}{\sqrt{t^2 + u^2 - s^2}}  , \nonumber \\[2mm]
 & & \nonumber \\
X & = &  \left(\frac{\sqrt{t^2 + u^2 - s^2} \; \pm(t\, \cos\phi - i \, s\,\sin\phi)}{2\; \sqrt{t^2 + u^2 - s^2}} \right)^{\frac{1}{2}}  \nonumber \\[2mm]
\cdot  \Bigg( 1_2 & \mp &  \frac{\small \left(\begin{array}{cc} 0 & - [\, i\, ( s\, \cos\phi - u) + t \sin\phi\, ] \\[2mm]  [\, i\, ( s\, \cos\phi + u) + t \sin\phi ] &   0 \end{array}\right)}{\sqrt{t^2 + u^2 - s^2} \; \pm(t\, \cos\phi - i \, s\,\sin\phi)} \Bigg)  , \nonumber \\[2mm]
 & & \\
X^+ & = &  \left(\frac{\sqrt{t^2 + u^2 - s^2} \; \pm(t\, \cos\phi + i \, s\,\sin\phi)}{2\; \sqrt{t^2 + u^2 - s^2}} \right)^{\frac{1}{2}} \nonumber \\[2mm]
\cdot  \Bigg( 1_2 & \mp &   \frac{\small \left(\begin{array}{cc} 0 &   [\,- i\, ( s\, \cos\phi + u) + t \sin\phi ] \\[2mm] -\,[\,- i\, ( s\, \cos\phi - u) + t \sin\phi\, ] &   0 \end{array}\right)}{\sqrt{t^2 + u^2 - s^2} \; \pm(t\, \cos\phi + i \, s\,\sin\phi)} \Bigg)  , \nonumber \\[2mm]
 & & \\[10mm]
 \eta_1 & = &  \frac{N^\ast_2\,N_1+N^\ast_1\,N_2}{2} \; \;    X^+ \; \vec{e}\cdot \vec{\sigma}\;   X   \nonumber \\[2mm]
  & = &  \frac{N^\ast_2\,N_1+N^\ast_1\,N_2}{2} \; \Big( \cos\varphi  \;  X^+ \sigma_1\,   X + \sin\varphi  \;  X^+ \sigma_2\,   X\Big)   \nonumber \\[2mm]
 & & \nonumber \\[10mm]
& = & \mbox{\small $\bigcirc\hspace{-3.2mm}\pm$}\;\; \frac{N^\ast_2\,N_1+N^\ast_1\,N_2}{2} \;\; \frac{1}{\sqrt{t^2 + u^2 - s^2}}  \nonumber \\[1mm]
 & \cdot &  \frac{1}{\sqrt{\left(\pm \;  \sqrt{t^2 + u^2 - s^2} + t \cos\phi \right)^2 + (s\,\sin\phi)^2}}  \nonumber \\[1mm]
 & \cdot & \Bigg( \cos\varphi \; \Bigg[ \; u\; \Bigg(\sin\phi   \, \left(\begin{array}{cc} s &  i\, t \\  -\,i\, t & s \end{array}\right) + u\, \left(\begin{array}{cc} 0 &  1 \\  1 & 0 \end{array}\right) \Bigg) \nonumber \\[1mm]
 & & \quad + \Big( \cos\phi \; (t^2 - s^2) \pm \; t \; \sqrt{t^2 + u^2 - s^2}\;\Big) \left(\begin{array}{cc} -\,\sin\phi &  \cos\phi \\  \cos\phi & \sin\phi \end{array}\right)  \Bigg]\nonumber \\[1mm]
 & & -\,  \sin\varphi \; \Bigg[ \;  t\; \left(\begin{array}{cc} s &  i\, t \\  -\, i\, t & s \end{array}\right) + tu \,  \left(\begin{array}{cc} \cos\phi &  \sin\phi \\  \sin\phi & -\cos\phi \end{array}\right) \nonumber \\[1mm]
 & &  \quad \pm\; \sqrt{t^2 + u^2 - s^2} \;  \Bigg(  \cos\phi \left(\begin{array}{cc} s &  i\, t \\  -\,i\, t & s \end{array}\right) + u\, \left(\begin{array}{cc} 1 &  0 \\  0 & -\,1 \end{array}\right) \Bigg) \; \Bigg] \; \Bigg) \; , \nonumber \\[1mm]
 & & \\
 \eta_2 & = &  \frac{N^\ast_2\,N_1-N^\ast_1\,N_2}{2\, i} \; \;    X^+ \; \vec{e}_\perp\cdot \vec{\sigma} \; X\nonumber \\[1mm]
 & = &  \frac{N^\ast_2\,N_1-N^\ast_1\,N_2}{2\,i} \; \Big( \sin\varphi  \;  X^+ \sigma_1\,   X - \cos\varphi  \;  X^+ \sigma_2\,   X\Big)  \nonumber \\[1mm]
& = & \mbox{\small $\bigcirc\hspace{-3.2mm}\pm$}\;\; \frac{N^\ast_2\,N_1-N^\ast_1\,N_2}{2\, i} \;\; \frac{1}{\sqrt{t^2 + u^2 - s^2}}  \nonumber \\[2mm]
 & \cdot &  \frac{1}{\sqrt{\left(\pm \;  \sqrt{t^2 + u^2 - s^2} + t \cos\phi \right)^2 + (s\,\sin\phi)^2}} \nonumber \\[1mm]
 & \cdot & \Bigg( \sin\varphi \; \Bigg[ \; u\; \Bigg(\sin\phi   \, \left(\begin{array}{cc} s &  i\, t \\  -\,i\, t & s \end{array}\right) + u\, \left(\begin{array}{cc} 0 &  1 \\  1 & 0 \end{array}\right) \Bigg) \nonumber \\[1mm]
 & & \quad + \Big( \cos\phi \; (t^2 - s^2) \pm \; t \; \sqrt{t^2 + u^2 - s^2}\;\Big) \left(\begin{array}{cc} -\,\sin\phi &  \cos\phi \\  \cos\phi & \sin\phi \end{array}\right)  \Bigg]\nonumber \\[1mm]
 & & +\,  \cos\varphi \; \Bigg[ \;  t\; \left(\begin{array}{cc} s &  i\, t \\  -\, i\, t & s \end{array}\right) + tu \,  \left(\begin{array}{cc} \cos\phi &  \sin\phi \\  \sin\phi & -\cos\phi \end{array}\right) \nonumber \\[1mm]
 & &  \quad \pm\; \sqrt{t^2 + u^2 - s^2} \;  \Bigg(  \cos\phi \left(\begin{array}{cc} s &  i\, t \\  -\,i\, t & s \end{array}\right) + u\, \left(\begin{array}{cc} 1 &  0 \\  0 & -\,1 \end{array}\right) \Bigg) \; \Bigg] \; \Bigg) \; . \nonumber \\[1mm]
\end{eqnarray}
\end{itemize}

\newpage
\section{The matrix representation and metric for a non-Hermitian Hamilton operator proposed by T.D.~Lee and C.G.~Wick} \label{xappenda4}
In the following we will discuss here  the Hamilton operator for the Fermionic (anti-)causal Harmonic Oscillator presented --- to our best knowledge --- for the first time by T.D.~Lee and C.G.~Wick \cite{Lee:1970iw} (see also \cite{Cotaescu:1983nc}) and rederived in a different context of  our doctoral thesis \cite{Kleefeld:1999}(see also \cite{Kleefeld:1998yj}\cite{Kleefeld:1998dg}\cite{Kleefeld:2003zj}\cite{Kleefeld:2003dx})
\begin{equation} h  \; = \;  \frac{\hbar \Omega}{2} \; (d^+  b - b\; d^+) \; + \; \frac{\hbar \Omega^\ast}{2} \; (b^+  d - d\; b^+) \; = h^+ \; , \label{hamleewick1} \end{equation}
with $\Omega\not=\Omega^\ast$. That the seemingly Hermitian Hamilton operator possessing the eigen-values $E_{m,n} \; = \; \hbar \Omega \, (m - \frac{1}{2}) + \hbar \Omega^\ast \, (n - \frac{1}{2}) $ and being diagonal with respect to the right eigen-states $\big|m,n\big>$ and left eigen-states $\big<\!\big<m,n\big|$, i.~e.,
\begin{equation} h \, \big|m,n\big> \; = \; E_{m,n} \, \big|m,n\big> \; , \quad \big<\!\big<m,n\big| \, h \; = \; \big<\!\big<m,n\big| \, E_{m,n} \; , \end{equation}
with $\big<0\big|0\big>\; = \;1$ and
\begin{eqnarray} & & \big|1,1\big> \; = \; d^+ \, b^+ \big|0\big> \; , \; \big|1,0\big> \; = \; d^+ \big|0\big> \; ,  \; \big|0,1\big> \; = \; b^+ \big|0\big> \; , \; \big|0,0\big> \; = \; \big|0\big> \;  , \nonumber \\[2mm]
 & & \big<\!\big<1,1\big| \; = \; \big<0\big|\, d\, b \; , \; \big<\!\big<1,0\big| \; = \; \big<0\big|\, b \; , \; \big<\!\big<0,1\big| \; = \; \big<0\big|\, d \; , \;\big<\!\big<0,0\big| \; = \; \big<0\big| \, . \qquad \label{leewickstates1}
\end{eqnarray}
is indeed not Hermitian, yet pseudo-Hermitian, can be noted by inspection of the anti-commutation relations  of the annihilation operators $b$, $d$ and creation operators $b^+$, $d^+$, i.~e.,
\begin{eqnarray} \left(\begin{array}{cc} \{ \, d\, , d^+ \} & \{ \, d\, , b^+ \} \\[1mm] \{ \, b\, , d^+ \} & \{ \, b\, , b^+ \} \end{array}\right) & = & \left(\begin{array}{cc} 0 & 1 \\[1mm] 1 & 0 \end{array}\right) \; , \nonumber \\[2mm]
 \left(\begin{array}{cc} \{ \, d\, , d \, \} & \{ \, d\, , b \, \} \\[1mm] \{ \, b\, , d \, \} & \{ \, b\, , b \, \} \end{array}\right) & = & \left(\begin{array}{cc} 0 & 0 \\[1mm] 0 & 0 \end{array}\right) \; , \nonumber  \\[2mm]
 \left(\begin{array}{cc} \{ \, d^+ , d^+ \} & \{ \, d^+ , b^+ \} \\[1mm] \{ \, b^+ , d^+ \} & \{ \, b^+ , b^+ \} \end{array}\right) & = & \left(\begin{array}{cc} 0 & 0 \\[1mm] 0 & 0 \end{array}\right)  \; , \label{anticomrelx1} \end{eqnarray}
containing obviously an indefinite metric. Consequently we obtain with the help of the spectral expansion of the Hamilton operator also a non-Hermitian four-dimensional matrix representation $H_{m^\prime,n^\prime\, ;\,m,n}\; = \big<m^\prime,n^\prime \big|\, h\,\big|m,n\big>$ of the Hamilton operator with respect to the  right eigen-states $\big|m,n\big>$ and left eigen-states $\big<\!\big<m,n\big|$, i.~e.,
\begin{equation} h \; = \; \sum\limits_{m^\prime,n^\prime}\sum\limits_{m,n} \;\big|m^\prime,n^\prime\big> \big<\!\big<m^\prime,n^\prime \big|\, h\,\big|m,n\big> \big<\!\big<m,n\big| \; , \end{equation}
yielding obviously
\begin{eqnarray} H &= & \left(\begin{array}{cccc} \big<\!\big<1,1 \big|\, h\,\big|1,1\big> & \big<\!\big<1,1 \big|\, h\,\big|1,0\big> & \big<\!\big<1,1 \big|\, h\,\big|0,1\big> & \big<\!\big<1,1 \big|\, h\,\big|0,0\big> \\ 
 \big<\!\big<1,0 \big|\, h\,\big|1,1\big> & \big<\!\big<1,0 \big|\, h\,\big|1,0\big> & \big<\!\big<1,0 \big|\, h\,\big|0,1\big> & \big<\!\big<1,0 \big|\, h\,\big|0,0\big> \\  
\big<\!\big<0,1 \big|\, h\,\big|1,1\big> & \big<\!\big<0,1 \big|\, h\,\big|1,0\big> & \big<\!\big<0,1 \big|\, h\,\big|0,1\big> & \big<\!\big<0,1 \big|\, h\,\big|0,0\big> \\ 
\big<\!\big<0,0 \big|\, h\,\big|1,1\big> & \big<\!\big<0,0 \big|\, h\,\big|1,0\big> & \big<\!\big<0,0 \big|\, h\,\big|0,1\big> & \big<\!\big<0,0 \big|\, h\,\big|0,0\big>   \end{array}\right)  \nonumber \\[2mm]
 & = & \left(\begin{array}{cccc} +\,  \mbox{Re}\, [\Omega] & 0 & 0 & 0 \\ 0 & +\,i\, \mbox{Im}\, [\Omega]  & 0 & 0 \\ 0 & 0 & -\, i\, \mbox{Im}\, [\Omega]  & 0 \\  0 & 0 & 0 & -\, \mbox{Re}\, [\Omega]   \end{array}\right) \nonumber \\[2mm]
 & = & \left(\begin{array}{cccc} +\, \frac{\hbar \Omega}{2} +\, \frac{\hbar \Omega^\ast}{2} & 0 & 0 & 0 \\ 0 & +\, \frac{\hbar \Omega}{2} -\, \frac{\hbar \Omega^\ast}{2} & 0 & 0 \\ 0 & 0 & -\, \frac{\hbar \Omega}{2} +\, \frac{\hbar \Omega^\ast}{2} & 0 \\  0 & 0 & 0 & -\, \frac{\hbar \Omega}{2} -\, \frac{\hbar \Omega^\ast}{2} \end{array}\right)  \nonumber \\[2mm]
 & = & \left(\begin{array}{cc} +\, \frac{\hbar \Omega}{2} & 0 \\ 0 & -\, \frac{\hbar \Omega}{2} \end{array}\right)  \otimes  \left(\begin{array}{cc} 1 & 0 \\ 0 & 1 \end{array}\right) + \left(\begin{array}{cc} 1 & 0 \\ 0 & 1 \end{array}\right)  \otimes  \left(\begin{array}{cc} +\, \frac{\hbar \Omega^\ast}{2} & 0 \\ 0 & -\, \frac{\hbar \Omega^\ast}{2} \end{array}\right)  .\nonumber \\ \end{eqnarray}
Analogously we can determine the underlying matrix representation $\bar{D}$, $\bar{B}$ and $D$, $B$ of the respective creation operators $d^+$, $b^+$ and annihilation operators $d$, $b$, i.~e.,
\begin{eqnarray} d^+ & = & \sum\limits_{m^\prime,n^\prime}\sum\limits_{m,n} \;\big|m^\prime,n^\prime\big> \;  \big<\!\big<m^\prime,n^\prime \big|\, d^+\,\big|m,n\big> \; \big<\!\big<m,n\big| \nonumber \\ 
 & = & \sum\limits_{m^\prime,n^\prime}\sum\limits_{m,n} \;\big|m^\prime,n^\prime\big> \;  \bar{D}_{m^\prime\!,n^\prime\,;\, m,n} \;  \big<\!\big<m,n\big| \; , \\[1mm] 
 b^+ & = & \sum\limits_{m^\prime,n^\prime}\sum\limits_{m,n} \;\big|m^\prime,n^\prime\big> \;  \big<\!\big<m^\prime ,n^\prime \big|\, b^+\,\big|m,n\big> \; \big<\!\big<m,n\big| \nonumber \\ 
 & = & \sum\limits_{m^\prime,n^\prime}\sum\limits_{m,n} \;\big|m^\prime,n^\prime\big> \;  \bar{B}_{m^\prime\!,n^\prime\,;\, m,n} \;  \big<\!\big<m,n\big| \; , \\[1mm]
  d & = & \sum\limits_{m^\prime,n^\prime}\sum\limits_{m,n} \;\big|m^\prime,n^\prime\big> \;  \big<\!\big<m^\prime,n^\prime \big|\, d\,\big|m,n\big> \; \big<\!\big<m,n\big| \nonumber \\
 & = & \sum\limits_{m^\prime,n^\prime}\sum\limits_{m,n} \;\big|m^\prime,n^\prime\big> \;  D_{m^\prime\!,n^\prime\,;\, m,n} \;  \big<\!\big<m,n\big| \; , \\[1mm] 
 b & = & \sum\limits_{m^\prime,n^\prime}\sum\limits_{m,n} \;\big|m^\prime,n^\prime\big> \;  \big<\!\big<m^\prime,n^\prime \big|\, b\,\big|m,n\big> \; \big<\!\big<m,n\big| \nonumber \\ 
 & = & \sum\limits_{m^\prime,n^\prime}\sum\limits_{m,n} \;\big|m^\prime,n^\prime\big> \;  B_{m^\prime\!,n^\prime\,;\, m,n} \;  \big<\!\big<m,n\big|  \; , \end{eqnarray}
yielding (see also \cite{Steeb:1991}\cite{Steeb:2012}\cite{Steeb:2014}\cite{Hardy:2019})
\begin{eqnarray} \bar{D} & = & B^+ \; = \left(\begin{array}{cccc} 0 & 0 & +1 & 0 \\ 0 & 0  & 0 & +1 \\ 0 & 0 & 0  & 0 \\  0 & 0 & 0 & 0   \end{array}\right) \; = \; \left(\begin{array}{cc} 0 & 1 \\ 0 & 0 \end{array}\right)  \otimes  \left(\begin{array}{cc} 1 & 0 \\ 0 & 1 \end{array}\right)  ,  \\[2mm]
 \bar{B} & = & D^+ \; = \left(\begin{array}{cccc} 0 & -1 & 0 & 0 \\ 0 & 0  & 0 & 0 \\ 0 & 0 & 0  & + 1 \\  0 & 0 & 0 & 0   \end{array}\right) \; = \; \left(\begin{array}{cc} -1 & 0 \\ 0 & 1 \end{array}\right)  \otimes  \left(\begin{array}{cc} 0 & 1 \\ 0 & 0 \end{array}\right)  ,  \qquad \\[2mm]
 D & = & \bar{B}^+ \; =  \left(\begin{array}{cccc} 0 & 0 & 0 & 0 \\ -1 & 0  & 0 & 0 \\ 0 & 0 & 0  & 0 \\  0 & 0 & +1 & 0   \end{array}\right) \; = \; \left(\begin{array}{cc} -1 & 0 \\ 0 & 1 \end{array}\right)  \otimes  \left(\begin{array}{cc} 0 & 0 \\ 1 & 0 \end{array}\right)  ,  \\[2mm]
 B & = & \bar{D}^+ \; = \left(\begin{array}{cccc} 0 & 0 & 0 & 0 \\ 0 & 0  & 0 & 0 \\ +1 & 0 & 0  & 0 \\  0 & +1 & 0 & 0   \end{array}\right) \; = \; \left(\begin{array}{cc} 0 & 0 \\ 1 & 0 \end{array}\right)  \otimes  \left(\begin{array}{cc} 1 & 0 \\ 0 & 1 \end{array}\right)  .
\end{eqnarray}
The matrix representation of Eq.\ (\ref{hamleewick1}) can be denoted in complete analogy in terms of the matrix representations of the creation and annihilation operators, i.~e.,
\begin{eqnarray} H  & = &  \frac{\hbar \Omega}{2} \; (\bar{D}  \, B - B\, \bar{D}) \; + \; \frac{\hbar \Omega^\ast}{2} \; (\bar{B}  \, D - D\, \bar{B}) \nonumber \\[2mm] 
 & = &  \frac{\hbar \Omega}{2} \; (B^+   B - B\, B^+) \; + \; \frac{\hbar \Omega^\ast}{2} \; (D^+   D - D\, D^+) \; \not= \; H^+ \; . \label{matrephamlw1} \end{eqnarray}
That this Hamilton operator is --- despite the relation $h^+=h$ on the level of operators (see Eq.\ (\ref{hamleewick1})) --- pseudo-Hermitian gets manifest by  performing a basis change from the  right eigen-states $\big|m,n\big>$ and left eigen-states $\big<\!\big<m,n\big|$ of $h$ to the Hermitian conjugate states $\big|m,n\big>\!\big>=\big<\!\big<m,n\big|^+$ and $\big<m,n\big|=\big|m,n\big>^+$ being respectively the right and left eigen-states of $h^+$. In a first step we recall by inspection of Eqs.\ (\ref{leewickstates1}) that there hold the identities:
\begin{eqnarray} & & \big|1,1\big>\!\big> \; = \; b^+ \, d^+ \big|0\big> \; , \; \big|1,0\big>\!\big> \; = \; b^+ \big|0\big> \; ,  \; \big|0,1\big>\!\big> \; = \; d^+ \big|0\big> \; , \; \big|0,0\big>\!\big> \; = \; \big|0\big> \;  , \nonumber \\[2mm]
 & & \big<1,1\big| \; = \; \big<0\big|\, b\, d \; , \; \big<1,0\big| \; = \; \big<0\big|\, d \; , \; \big<0,1\big| \; = \; \big<0\big|\, b \; , \;\big<0,0\big| \; = \; \big<0\big| \, , \qquad
\end{eqnarray}
and the following spectral expansions of $h^+$ and $h$:

\begin{eqnarray} h^+ & = & \left(\; \sum\limits_{\bar{m}^\prime,\bar{n}^\prime} \; \sum\limits_{\bar{m},\bar{n}} \;\big|\bar{m},\bar{n}\big> \; \big<\!\big<\bar{m},\bar{n} \big|\, h\,\big|\bar{m}^\prime,\bar{n}^\prime\big>\;  \big<\!\big<\bar{m}^\prime,\bar{n}^\prime\big| \; \right)^+ \nonumber \\[2mm]
 & = & \sum\limits_{\bar{m}^\prime,\bar{n}^\prime} \; \sum\limits_{\bar{m},\bar{n}} \;\big|\bar{m}^\prime,\bar{n}^\prime\big>\!\big> \;  \big<\bar{m}^\prime,\bar{n}^\prime \big|\, h^+\big|\bar{m},\bar{n}\big>\!\big> \; \big<\bar{m},\bar{n}\big| \nonumber \\[2mm]
 & = & \sum\limits_{\bar{m}^\prime,\bar{n}^\prime} \; \sum\limits_{\bar{m},\bar{n}} \;\big|\bar{m}^\prime,\bar{n}^\prime\big>\!\big>\;  \big<\!\big<\bar{m},\bar{n} \big|\, h\,\big|\bar{m}^\prime,\bar{n}^\prime\big>^\ast\;  \big<\bar{m},\bar{n}\big| \; ,   \\[2mm]
h & = & \sum\limits_{\bar{m}^\prime,\bar{n}^\prime} \; \sum\limits_{\bar{m},\bar{n}} \;\big|\bar{m}^\prime,\bar{n}^\prime\big> \; \big<\!\big<\bar{m}^\prime,\bar{n}^\prime \big|\, h\,\big|\bar{m},\bar{n}\big>\;  \big<\!\big<\bar{m},\bar{n}\big| \; . \end{eqnarray}
The second step is to perform the matrix element $\big<m^\prime,n^\prime \,\big|\,h^+\,\big|m,n\big>\!\big>=\big<m^\prime,n^\prime \big|\,h\,\big|m,n\big>\!\big>$ of the identity $h^+ \; = \; h$ revealing obviously the following pseudo-Hermiticity relation:
\begin{eqnarray} \lefteqn{\big<\!\big<m,n \big|\, h\,\big|m^\prime,n^\prime\big>^\ast \; =} \nonumber \\[2mm]
 & = & \sum\limits_{\bar{m}^\prime,\bar{n}^\prime} \, \sum\limits_{\bar{m},\bar{n}} \;\big<m^\prime,n^\prime \,\big|\bar{m}^\prime,\bar{n}^\prime\big> \; \big<\!\big<\bar{m}^\prime,\bar{n}^\prime \big|\, h\,\big|\bar{m},\bar{n}\big>\;  \big<\!\big<\bar{m},\bar{n}\big|m,n\big>\!\big> \; , \end{eqnarray}
or, equivalently, 
\begin{eqnarray} H_{m,n\,;\, m^\prime,n^\prime}^\ast & = & \sum\limits_{\bar{m}^\prime,\bar{n}^\prime} \, \sum\limits_{\bar{m},\bar{n}} \;\big<m^\prime,n^\prime \,\big|\bar{m}^\prime,\bar{n}^\prime\big> \; H_{\bar{m}^\prime,\bar{n}^\prime\,;\, \bar{m},\bar{n}}\;  \big<\!\big<\bar{m},\bar{n}\big|m,n\big>\!\big> \\
 & = & \sum\limits_{\bar{m}^\prime,\bar{n}^\prime} \, \sum\limits_{\bar{m},\bar{n}} \;\quad\,\eta_{\,m^\prime,n^\prime \,;\, \bar{m}^\prime,\bar{n}^\prime} \; H_{\bar{m}^\prime,\bar{n}^\prime;\, \bar{m},\bar{n}}\;  \,\Big(\eta^{-1}\Big)_{\bar{m},\bar{n}\, , \, m,n} \; . \quad\quad \end{eqnarray}
For our particular Hamilton operator under consideration we have with the chosen normalization of the right and left eigenstates of $h$ and $h^+$:
\begin{eqnarray}
\eta  & = &  \left(\begin{array}{cccc} \big<1,1 \big|1,1\big> & \big<1,1 \big|1,0\big> & \big<1,1 \big|0,1\big> & \big<1,1 \big|0,0\big> \\[1mm] 
 \big<1,0 \big|1,1\big> & \big<1,0 \big|1,0\big> & \big<1,0 \big|0,1\big> & \big<1,0 \big|0,0\big> \\[1mm]  
\big<0,1 \big|1,1\big> & \big<0,1 \big|1,0\big> & \big<0,1 \big|0,1\big> & \big<0,1 \big|0,0\big> \\[1mm] 
\big<0,0 \big|1,1\big> & \big<0,0 \big|1,0\big> & \big<0,0 \big|0,1\big> & \big<0,0 \big|0,0\big>   \end{array}\right)  \\[2mm] 
\eta^{-1} & = &  \left(\begin{array}{cccc} \big<\!\big<1,1 \big|1,1\big>\!\big> & \big<\!\big<1,1 \big|1,0\big>\!\big> & \big<\!\big<1,1 \big|0,1\big>\!\big> & \big<\!\big<1,1 \big|0,0\big>\!\big> \\[1mm] 
 \big<\!\big<1,0 \big|1,1\big>\!\big> & \big<\!\big<1,0 \big|1,0\big>\!\big> & \big<\!\big<1,0 \big|0,1\big>\!\big> & \big<\!\big<1,0 \big|0,0\big>\!\big> \\[1mm]  
\big<\!\big<0,1 \big|1,1\big>\!\big> & \big<\!\big<0,1 \big|1,0\big>\!\big> & \big<\!\big<0,1 \big|0,1\big>\!\big> & \big<\!\big<0,1 \big|0,0\big>\!\big>  \\[1mm] 
\big<\!\big<0,0 \big|1,1\big>\!\big> & \big<\!\big<0,0 \big|1,0\big>\!\big> & \big<\!\big<0,0 \big|0,1\big>\!\big> & \big<\!\big<0,0  \big|0,0\big>\!\big>   \end{array}\right)  \qquad \\[2mm]
\lefteqn{\Rightarrow \;\; \eta \; = \; \eta^{-1} \; = \;  \eta^+ \; = \;  \Big(\eta^{-1}\Big)^+ \; = \; \left(\begin{array}{cccc} - 1 & 0 & 0 & 0 \\ 0 & 0  & 1 & 0 \\ 0 & 1 & 0 & 0 \\  0 & 0 & 0 & 1  \end{array}\right)   , \qquad}  \end{eqnarray}
implying the following pseudo-Hermiticity relation for the matrix representation of the Hamilton operator and --- in complete analogy  --- several equivalence relations relating the underlying annihilation or creation operators:
\begin{eqnarray} H^+ \; = \; \bar{H} & = & \eta \; H \; \eta^{-1} \; , \nonumber \\[2mm] 
 D^+ \; = \;  \bar{B} & = & \eta \; \bar{D} \; \eta^{-1} \; ,\nonumber \\[2mm]
 B^+ \; = \;  \bar{D} & = & \eta \; \bar{B} \; \eta^{-1} \;, \nonumber \\[2mm]
 \bar{D}^+ \; = \;  B & = & \eta \; D \; \eta^{-1} \; , \nonumber\\[2mm]
 \bar{B}^+ \; = \;  D & = & \eta \; B \; \eta^{-1} \; . \label{phrelations1}
\end{eqnarray}
It is interesting to observe how the anti-commutation relations (see also Eqs.\ (\ref{anticomrelx1})) respected by matrices representing creation- and annihilation operators of  pseudo-Fermions \cite{Bagarello:2012ds} displaying an indefinite metric, i.~e.,
\begin{eqnarray} \left(\begin{array}{cc} \{ \, D\, , \bar{D} \} & \{ \, D\, , \bar{B} \} \\[1mm] \{ \, B\, , \bar{D} \} & \{ \, B\, , \bar{B} \} \end{array}\right) & = & \left(\begin{array}{cc} 0 & 1_2 \otimes 1_2 \\[1mm] 1_2 \otimes 1_2 & 0 \end{array}\right) \; , \\
 \left(\begin{array}{cc} \{ \, D\, , D \, \} & \{ \, D\, , B \, \} \\[1mm] \{ \, B\, , D \, \} & \{ \, B\, , B \, \} \end{array}\right) & = & \left(\begin{array}{cc} 0 & 0 \\[1mm] 0 & 0 \end{array}\right) \; , \\
 \left(\begin{array}{cc} \{ \, \bar{D} , \bar{D} \} & \{ \, \bar{D} , \bar{B} \} \\[1mm] \{ \, \bar{B} , \bar{D} \} & \{ \, \bar{B} , \bar{B} \} \end{array}\right) & = & \left(\begin{array}{cc} 0 & 0 \\[1mm] 0 & 0 \end{array}\right)  \; ,\end{eqnarray}
are converted by the non-trivial identities $ D^+ \; = \;  \bar{B}$, $B^+ \; = \;  \bar{D}$, $\bar{D}^+ \; = \;  B$ and $\bar{B}^+ \; = \;  D$ into standard Fermionic anti-commutation relations, i.~e.,
\begin{eqnarray} \left(\begin{array}{cc} \{ \, D\, , D^+ \} & \{ \, D\, , B^+ \} \\[1mm] \{ \, B\, , D^+ \} & \{ \, B\, , B^+ \} \end{array}\right) & = & \left(\begin{array}{cc} 1_2 \otimes 1_2 & 0 \\[1mm] 0 & 1_2 \otimes 1_2 \end{array}\right) \; , \\
 \left(\begin{array}{cc} \{ \, D\, , D \, \} & \{ \, D\, , B \, \} \\[1mm] \{ \, B\, , D \, \} & \{ \, B\, , B \, \} \end{array}\right) & = & \left(\begin{array}{cc} 0 & 0 \\[1mm] 0 & 0 \end{array}\right) \; , \\
 \left(\begin{array}{cc} \{ \, D^+ , D^+ \} & \{ \, D^+ , B^+ \} \\[1mm] \{ \, B^+ , D^+ \} & \{ \, B^+ , B^+ \} \end{array}\right) & = & \left(\begin{array}{cc} 0 & 0 \\[1mm] 0 & 0 \end{array}\right)  \; ,\end{eqnarray}
displaying no indefinite metric, while $H$ (see Eq.\ (\ref{matrephamlw1})) {\em persists} nonetheless to be pseudo-Hermitian and --- due to $\Omega^\ast \not= \Omega$ --- manifestly non-Hermitian.
\end{appendix}

\end{document}